\documentclass[12pt]{article}

\usepackage{graphicx}
\usepackage[dvipsnames]{xcolor}
\usepackage{lineno}
\usepackage[font=footnotesize,labelfont=bf,labelsep=period]{caption}

\usepackage{scicite}

\usepackage{times}

\newenvironment{sciabstract}{%
\begin{quote} \bf}
{\end{quote}}
\newcommand{\edit}[1]{\noindent \color{black} #1\normalcolor}

\renewcommand\refname{References and Notes}

\newcounter{lastnote}

\title{Computational Discovery of Microstructured Composites with Optimal \edit{Stiffness}-Toughness Trade-Offs} 

\author
{Beichen Li$^{1,2}$, Bolei Deng$^{1,2,3}$, Wan Shou$^{1,2,4\ast}$, Tae-Hyun Oh$^{5}$, \\
Yuanming Hu$^{1,2}$, Yiyue Luo$^{1,2}$, Liang Shi$^{1,2}$, Wojciech Matusik$^{1,2,3\ast}$\\
\\
\normalsize{$^{1}$Computer Science and Artificial Intelligence Laboratory,}\\
\normalsize{Massachusetts Institute of Technology, Cambridge, MA 02139, USA}\\
\normalsize{$^{2}$Department of Electrical Engineering and Computer Science,}\\
\normalsize{Massachusetts Institute of Technology, Cambridge, MA 02139, USA}\\
\normalsize{$^{3}$Department of Mechanical Engineering,}\\
\normalsize{Massachusetts Institute of Technology, Cambridge, MA 02139, USA}\\
\normalsize{$^{4}$Department of Mechanical Engineering,}\\
\normalsize{University of Arkansas, Fayetteville, AR 72701, USA}\\
\normalsize{$^{5}$Department of Electrical Engineering and Graduate School of AI,}\\
\normalsize{POSTECH, Pohang-si, Gyeongsangbuk-do 37673, Korea}\\
\\
\normalsize{$^\ast$Correspondence to: wanshou@mit.edu or wshou@uark.edu (W.S.);}\\
\normalsize{\qquad\,\, wojciech@csail.mit.edu (W.M.).}
}

\date{}

\begin{document} 


\baselineskip24pt


\maketitle 


\pagenumbering{gobble}


\begin{sciabstract}
The conflict between \edit{stiffness} and toughness is a fundamental problem in engineering materials design. However, the systematic discovery of microstructured composites with optimal \edit{stiffness}-toughness trade-offs has never been demonstrated, \edit{hindered by the discrepancies between simulation and reality and the lack of data-efficient exploration of the entire Pareto front. We introduce a generalizable pipeline that integrates physical experiments, numerical simulations, and artificial neural networks to address both challenges. Without any prescribed expert knowledge of material design, our approach implements a nested-loop proposal-validation workflow to bridge the simulation-to-reality gap and discover microstructured composites that are stiff and tough with high sample efficiency. Further analysis of Pareto-optimal designs allows us to automatically identify existing toughness enhancement mechanisms, which were previously discovered through trial-and-error or biomimicry. On a broader scale, our method provides a blueprint for computational design in various research areas beyond solid mechanics, such as polymer chemistry, fluid dynamics, meteorology, and robotics.}
\end{sciabstract}
\textbf{Teaser}: Experiments informed computational discovery of microstructured composites with optimal \edit{stiffness}-toughness trade-offs.

\subsection* {Introduction}
\edit{Stiffness---the ability to resist deformation in response to an applied force},
and toughness---the ability to resist cracks, are two quintessential properties in most engineering materials, since these materials must resist non-recoverable deformation and prevent catastrophic failure under external loading in structural applications. Unfortunately, \edit{stiffness} and toughness are often mutually exclusive because, in order to be tough, a material must be ductile enough to tolerate long cracks and absorb more energy before fracturing \cite{ritchie2011conflicts}. Although a few exceptions have been discovered among microstructured composites through trial-and-error approaches or biomimicry \cite{mayer2005rigid, sanchez2005biomimetism, dimas2013tough, meyers2013structural,yin2019impact, dudukovic2021cellular}, there is no systematic way to design and fabricate such materials.
Recent development in additive manufacturing allows materials to be arranged combinatorially in space, enabling designs that were impossible to physically realize before, including high-performance structural materials \cite{schaedler2011ultralight, zheng2014ultralight, pham2019damage} and functional objects \cite{wehner2016integrated, dudukovic2021cellular}. Furthermore, computational methods (e.g., structural optimization) allow for efficient exploration of optimal designs inside a parameterized design space with the help of numerical simulation \cite{aage2017giga, chen2018computational, butler2018machine, gu2018bioinspired, sigmund2013topology, wu2021topology}.
However, all reliable computational explorations require simulation models to accurately match the corresponding experimental measurements over the entire design space. This is still very difficult for toughness prediction as current simulations---despite harnessing advanced fracture theories---are unable to fully model the fracture of structured materials \cite{yang2021intrinsic, shaikeea2022toughness, surjadi2022design}.
Moreover, the computational process for discovering microstructured composites that are both \edit{stiff and tough} is also challenging, since their conflicting nature dictates that there is not a single best solution but a set of {\it Pareto-optimal} solutions. To obtain the entire Pareto front where one can make simple trade-off decisions among optimal designs, existing structural optimization algorithms typically execute many single-objective optimization routines or rely on evolutionary strategies \cite{zavala2014survey}. Both of these strategies incur a large number of simulations.
Therefore, finding Pareto-optimal designs using these approaches imposes contradictory demands on the simulator: the simulator should model all physical details to be accurate, but it must also run extremely fast to reduce computational cost. Such simulators have to undergo careful design by field experts, which is normally unrealistic, especially for microstructured composites with complicated geometries and abundant material interfaces.

\begin{figure}[tbh!]
    \centering
    \includegraphics[width=\linewidth]{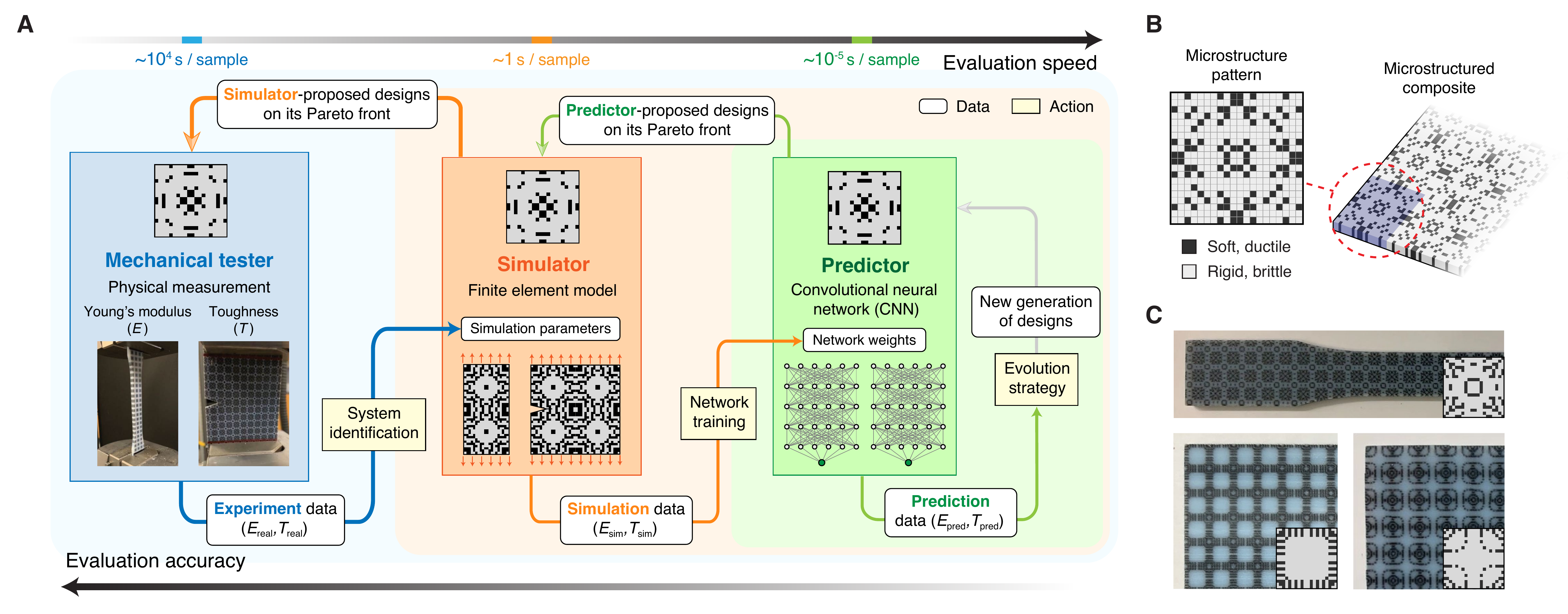}
    \caption{\textbf{Schematic of our approach.} ({\bf A}) Workflow of the proposed nested-loop pipeline. Our system integrates three distinctive evaluators: a mechanical tester, an FEM-based simulator, and a CNN-based predictor, which vary in evaluation speed and accuracy. Data flow directions are indicated by colored arrows. Arrows from slower, more accurate evaluators to faster, less accurate evaluators illustrate the propagation of microstructure performance data for improving the latter's accuracy. Arrows pointing in the opposite direction represent proposed Pareto-optimal designs from faster evaluators to slower ones for validation. The self-loop at the predictor depicts an evolutionary strategy for finding its Pareto front. ({\bf B}) Illustration of a microstructure and the corresponding microstructured composites. The microstructure is defined by a 2D pattern that demonstrates the spatial arrangement of two base materials with contrasting properties. ({\bf C}) Sample photographs of manufactured microstructured composites, annotated by the source microstructure patterns.}
    \label{fig:pipeline}
\end{figure}

Here, we present an efficient and expert-knowledge-free approach for the automatic discovery of microstructured composite designs with optimal \edit{stiffness}-toughness trade-offs \edit{(where stiffness is measured using Young's modulus)}, tackling all of the aforementioned challenges simultaneously. Our approach employs three evaluators (Fig.~\ref{fig:pipeline}A): (1) a mechanical tester that conducts physical measurements, (2) a finite element method (FEM) -based simulator that performs virtual mechanical testing in moderate complexity, and (3) a convolutional neural network (CNN) -based predictor that executes machine learning inference. All evaluators take an arbitrary microstructure design as input and then measure or predict its Young's modulus and toughness as output, with varying evaluation speed and accuracy. The mechanical tester runs very slowly due to labor-intensive specimen fabrication and testing ($\sim10^4$~s/sample), but this provides ground truth performance values for a microstructure design. At the opposite end of the spectrum we have the predictor, which runs extremely fast ($\sim10^{-5}$~s/sample) but yields relatively inaccurate results. In between, we have the simulator, which runs reasonably fast ($\sim1$~s/sample) given its moderate complexity and delivers intermediate accuracy. The full pipeline executes in a nested-loop fashion. Faster evaluators, acting as surrogate models, conduct multi-objective structural optimization and propose microstructure designs on the Pareto front to slower evaluators (from right to left, top Fig.~\ref{fig:pipeline}A). Slower evaluators, being more accurate, validate the performance of these designs and use them as additional training data to improve the accuracy of the faster evaluators (from left to right, bottom Fig.~\ref{fig:pipeline}A). As faster evaluators become increasingly accurate throughout the process, they are able to propose higher-quality designs to the slower evaluators. Overall, our nested-loop pipeline effectively improves sample efficiency in discovering microstructure designs with optimal \edit{stiffness}-toughness trade-offs. The inner loop between the simulator and the predictor reduces the number of numerical simulations for finding the simulator's Pareto front by an order of magnitude compared with other state-of-the-art structural optimization algorithms. The outer loop between the mechanical tester and the simulator simultaneously closes the sim-to-real gap and discovers experimentally verified Pareto-optimal microstructure designs based on only a few dozen physical measurements. Remarkably, this is accomplished with a physics-based simulator that does not incorporate sophisticated or advanced modeling of material inelasticity or fracture. Despite the limited modeling capability, the simulator autonomously learns to match the behavior of Pareto-optimal microstructures with physical experiments.

\subsection*{Results}
The microstructured composites in this study consist of two acrylic-based photopolymers, one rigid and one soft, that contrast in stiffness and ductility (Table S2). The unit cell of a microstructure, which we call a microstructure pattern (Fig.~\ref{fig:pipeline}B and Fig.~S1), is constructed by digitally arranging the base materials in a 20$\times$20 regular grid through 3D printing (Fig.~\ref{fig:pipeline}C). The Young's modulus and toughness of a microstructure are measured on fabricated specimens via tensile mechanical testing. To obtain reliable physical measurements that are comparable in horizontal and vertical directions, we limit the volume fraction of the rigid material in a microstructure pattern between 50\% and 85\%, and further enforce symmetry along axes and diagonals (see Note S1). Even under these constraints, the design space of microstructures is sufficiently large to render exhaustive search methods infeasible.

\begin{figure}[!htbp]
    \centering
    \includegraphics[width=0.8\linewidth]{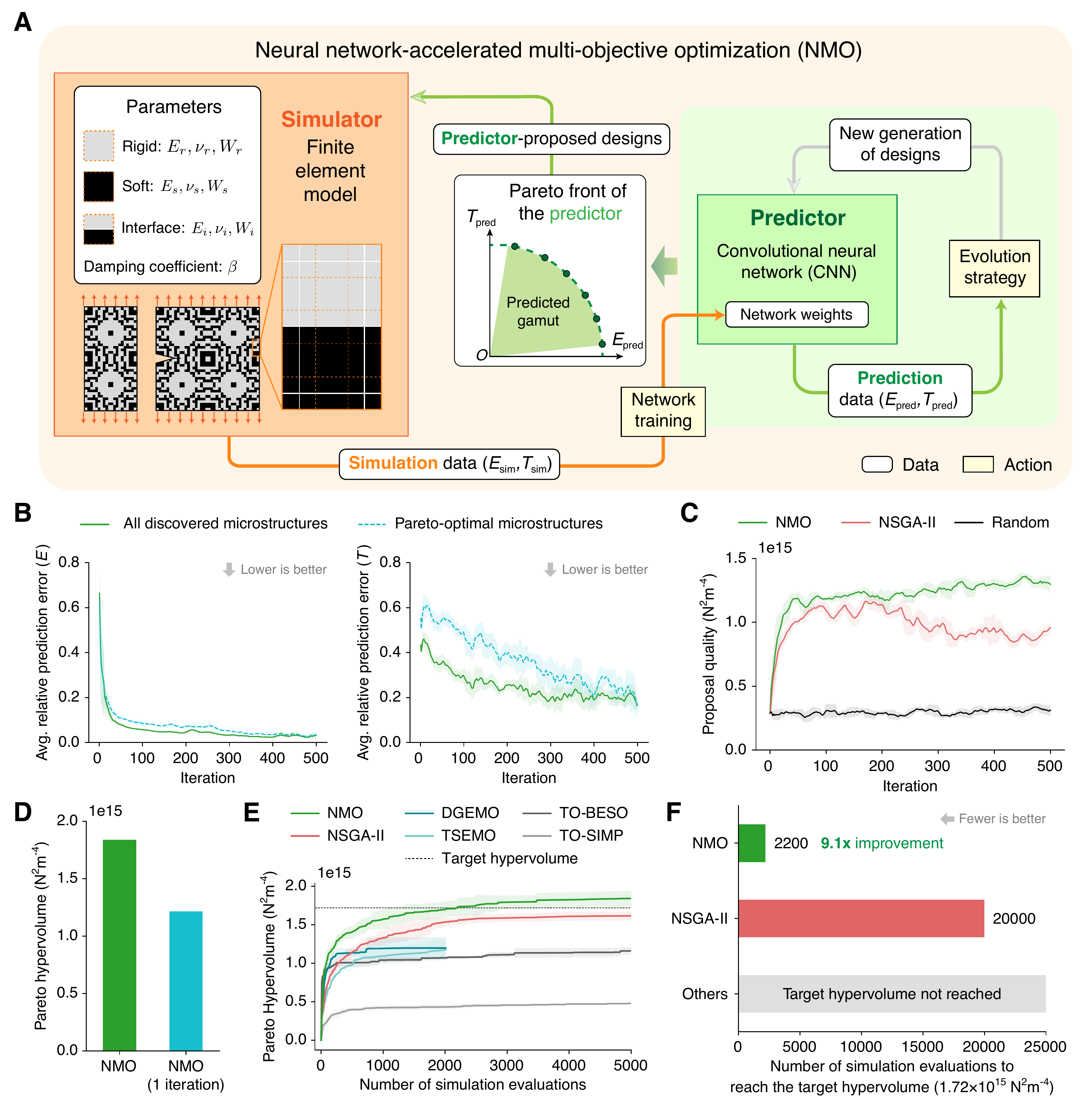}
    \caption{\textbf{The inner loop of the proposed workflow with neural network-accelerated multi-objective optimization (NMO).} ({\bf A}) Workflow of NMO, illustrated by a zoomed-in snapshot of Fig.~\ref{fig:pipeline}A. The simulator has 10 exposed parameters, including material model parameters of rigid, soft, and interface base materials plus a global damping coefficient. ({\bf B}) Average prediction errors of Young's modulus and toughness in NMO over 500 iterations, calculated for all discovered microstructures and specifically those on the simulation Pareto front. Shaded regions indicate standard deviations estimated from adjacent data points. ({\bf C}) Evolution of design proposal quality, as characterized by the Pareto hypervolume of 10 proposed designs in each iteration, over 500 iterations. NMO is compared with NSGA-II and a random sampling strategy. ({\bf D}), Comparison of the final Pareto hypervolumes from NMO and its simplified alternative (NMO one iteration) that only trains the predictor with 5,000 random designs and proposes designs back to the simulator once. ({\bf E}) Comparison between NMO and other multi-objective optimization algorithms in Pareto hypervolume growth within a budget of 5,000 simulation evaluations. The baselines comprise our modified NSGA-II algorithm, topology optimization (TO) (e.g., BESO \cite{querin1998evolutionary, huang2008topology} and SIMP \cite{bendsoe1989optimal}), and multi-objective Bayesian optimization (MOBO) (e.g., DGEMO \cite{lukovic2020diversity} and TSEMO \cite{bradford2018efficient}).
    MOBO algorithms are stopped at 2,000 simulations due to exceeding a time limit of 24 hours. Each solid curve is an average of repeats using five random seeds and the colored region around each curve indicates standard deviation. ({\bf F}) Number of evaluations required for NMO and other baseline algorithms to reach a target hypervolume, marked by the dashed black line in ({\bf E}).}
    \label{fig:inner-loop}
\end{figure}

At the foundation of our approach, the FEM-based simulator is constructed with simplified specimen geometry and material modeling. The FEM grid comprises three types of elements: soft, rigid, and interface elements (see left-hand side of Fig.~\ref{fig:inner-loop}A). To simulate mechanical testing of a microstructured composite, each element type is characterized by three essential material parameters: Young's modulus, Poisson's ratio, and maximal strain energy density (the energy density beyond which an element fails and becomes void). Above them, a global damping coefficient is added to capture energy dissipation and suppress numerical oscillations, resulting in a total of 10 exposed parameters.
The exposed parameters are initially optimized to match experimental measurements of base materials, namely, homogeneous composites fabricated with soft, rigid, and interface materials, respectively \cite{hossain2014effective}.
\edit{To minimize the expert knowledge required, advanced material parameters that control the fracture behavior, such as strength surface and critical energy release rate \cite{kumar2020phase, bourdin2008variational, jia2023controlling}, are omitted from the simulator. This precludes tedious implementation and allows for substantial evaluation throughput.} With a customized high-performance explicit solver for dynamic finite-element analysis (see Note S3), it takes around 0.015 s for every Young’s modulus evaluation and 0.2 s for toughness at maximum throughput (Fig.~S6). 

The inner loop of our nested-loop pipeline explores the maximal simulated performance range of microstructured composites (i.e., the simulated performance gamut) and finds designs on the simulation Pareto front. Since the complex, discontinuous nature of fracture makes it almost impossible to reach theoretically Pareto-optimal microstructures without exhaustive search, the goal of the inner loop is to discover the largest Pareto front possible using a limited number of simulations. Unlike traditional methods that conduct structural optimization using the simulator exclusively, the inner loop of our approach employs the CNN-based predictor as a surrogate model to guide the optimization process (see Note S3 for details), hence named neural network-accelerated multi-objective optimization (NMO) (Fig.~\ref{fig:inner-loop}A). Starting from a dataset of 10 random microstructures (Fig.~S15), NMO repeats the following steps for 500 iterations in total (Fig.~\ref{fig:inner-loop}A). First, the two neural networks of the predictor, one for Young's modulus and one for toughness, are trained using simulated properties of all microstructures in the current dataset.
Then, the Pareto front of the trained predictor is computed using an evolutionary multi-objective optimization algorithm (right part of Fig.~\ref{fig:inner-loop}A; in our case, we use a modified NSGA-II algorithm \cite{deb2002fast}; see Materials and Methods). Each of these Pareto front computations incurs a negligible time cost since the predictor runs orders of magnitude faster than the simulator.
Finally, the predictor proposes 10 Pareto-optimal designs to the simulator for validation.
The validated designs are used to augment the microstructure dataset, to serve as additional training data in the next iteration.
 
In early iterations, the predictor is very inaccurate due to limited training data. Predictions for Pareto-optimal designs are particularly inaccurate since they are far away from the microstructures in the initial dataset.
Nonetheless, as the algorithm proceeds the predictor becomes more accurate by virtue of accumulating training data from the simulator.
This is evidenced by the decreasing prediction errors in both Young's modulus and toughness over iterations (Fig.~\ref{fig:inner-loop}B). Interestingly, the prediction errors of Pareto-optimal microstructures decline even faster (the blue lines in Fig.~\ref{fig:inner-loop}B), indicating that the predictor improves its inference near the simulation Pareto front considerably after retraining.

Meanwhile, the self-learning predictor is able to propose increasingly better microstructure designs to the simulator, which we quantify using the Pareto hypervolume (namely, the area enclosed by the Pareto front) of 10 validated designs in each NMO iteration (Fig.~\ref{fig:inner-loop}C). Compared with a direct application of NSGA-II and random sampling, our approach shows consistently higher design proposal quality over the entire algorithm.
Eventually, the repetitive proposal-validation workflow allows NMO to reach a Pareto hypervolume of 1.84$\times10^{15}$ N$^2$m$^{-4}$ after 500 iterations (5,000 simulation evaluations excluding random initial designs). This is notably better than simply training the predictor via 5,000 random samples and validating Pareto optimal designs once (NMO one iteration in Fig.~\ref{fig:inner-loop}D), which only achieves 66.1\% of our achieved hypervolume.

Most importantly, NMO outperforms state-of-the-art optimization algorithms in terms of exploration efficiency within the design space, including NSGA-II \cite{deb2002fast}, topology optimization (TO) \cite{huang2008topology, sigmund200199}, and other surrogate-model-based algorithms such as multi-objective Bayesian optimization (MOBO) \cite{lukovic2020diversity, bradford2018efficient}. As shown in Fig.~\ref{fig:inner-loop}E, NMO covers the largest Pareto hypervolume within a budget of 5,000 simulations, surpassing TO and MOBO algorithms by a notable margin.
This demonstrates promising opportunities for NMO to tackle structural design problems involving complex, non-linear mechanical properties, where TO exhibits severe limitations.
To further distinguish NMO in sample efficiency, we run both NMO and NSGA-II until reaching a target hypervolume of 1.72$\times10^{15}$ N$^2$m$^{-4}$, and then observe the number of simulations required (Fig.~\ref{fig:inner-loop}F). Here, NMO takes an order of magnitude fewer simulation evaluations to achieve the target hypervolume, demonstrating evidently superior sample efficiency. Aside from the modified NSGA-II in this study, NMO is also compatible with other optimization algorithms for finding the predictor's Pareto front. In practice, it is recommended to choose appropriate algorithms based on problem characteristics to maximize the advantage of the predictor.

\begin{figure}[htbp!]
    \centering
    \includegraphics[width=\linewidth]{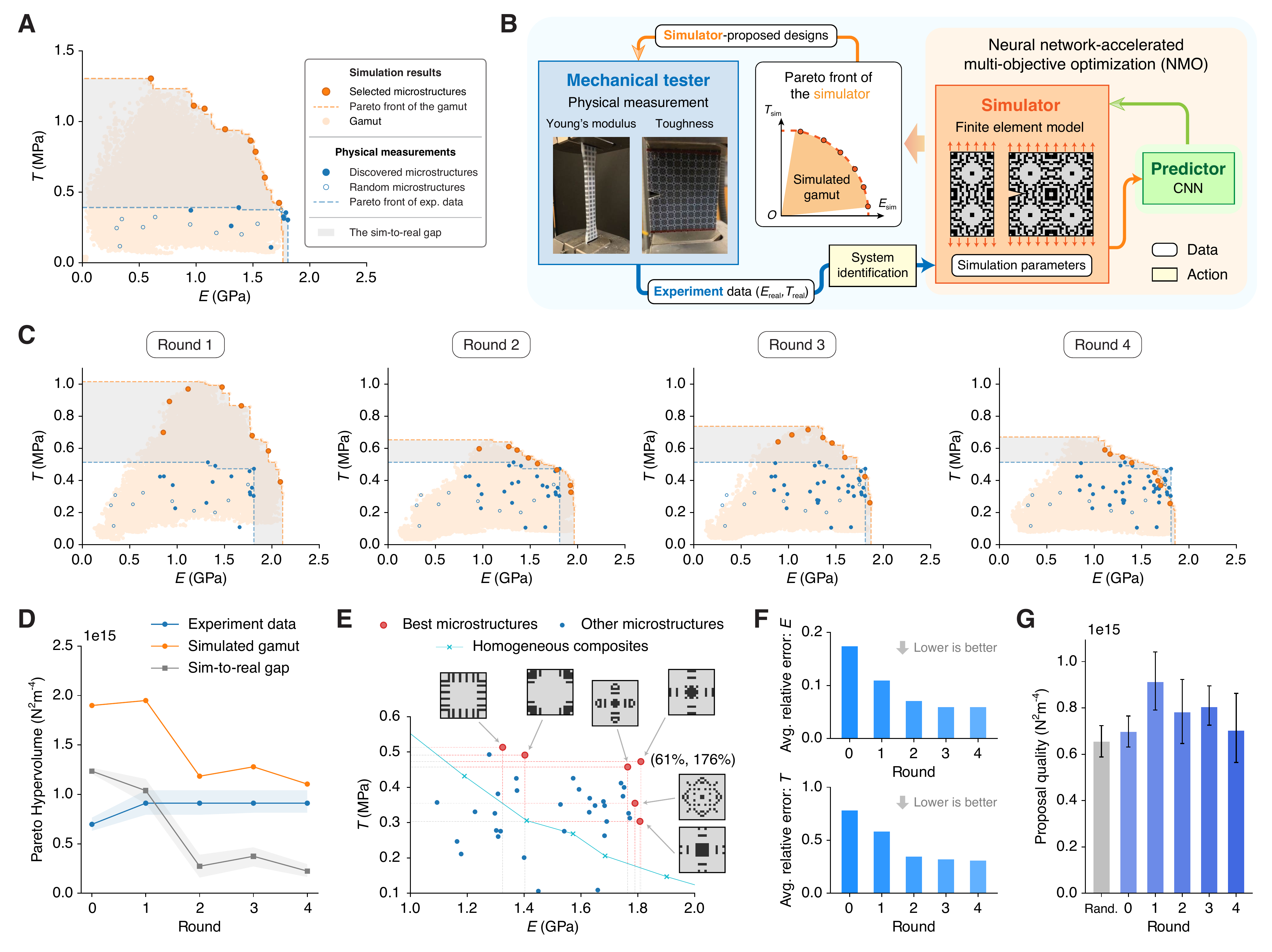}
    \caption{\textbf{The outer loop of the proposed workflow.} ({\bf A}) Experiment data of discovered microstructures and the simulation gamut at the end of Round 0. The sim-to-real gap is defined as the symmetric difference between the Pareto hypervolumes of the experimental Pareto front and the simulation Pareto front. ({\bf B}) Workflow illustration of the outer loop as simplified from Fig.~\ref{fig:pipeline}A. ({\bf C}) Evolution of experiment data and the simulation gamut within four rounds of the outer loop, where the sim-to-real gap shrinks substantially. ({\bf D}) Pareto hypervolume of experiment data and the simulation gamut, and the area of the sim-to-real gap over four rounds of the outer loop. ({\bf E}) Representative microstructures discovered by the pipeline with optimal trade-offs. Numbers in parentheses indicate the improvement of a design in Young’s modulus and toughness compared to homogeneous composites with various volume fractions of the rigid material. ({\bf F}) Average relative simulation error on Young's modulus and toughness in each round of the outer loop. The error is calculated for all 50 discovered microstructures to showcase the improvement in simulation accuracy. ({\bf G}) Quality of microstructure designs proposed by the simulator over the entire outer loop. Proposal quality is evaluated by the experimental Pareto hypervolume of 8 proposed designs in each round. The quality of the 10 randomly chosen microstructures before Round 1 (Rand.) is used as a reference.}
    \label{fig:outer-loop}
\end{figure}

Following the inner loop, we select 8 Pareto-optimal microstructure designs from the simulation Pareto front, fabricate specimens, and mechanically test their actual performance, which concludes Round 0 of the outer loop.
However, since our simulator does not incorporate sophisticated constitutive models or advanced fracture mechanics, the simulated properties of these microstructure designs deviate notably from their physical measurements. Unpredictable manufacturing flaws introduced by the 3D printer and measurement errors due to, e.g., boundary effects, further contribute to the sim-to-real gap. At this stage, the less accurate simulator normally tends to overestimate the performance of microstructures on its Pareto front since the optimization algorithm tries to exploit the discrepancies between simulation and reality. Thus, the experimental Pareto front of the selected microstructure designs is much smaller than the simulation Pareto front (Fig.~\ref{fig:outer-loop}A).

To close the sim-to-real gap, the outer loop of our pipeline improves the accuracy of the simulator through system identification (Fig.~\ref{fig:outer-loop}B), where the exposed parameters of the simulator are optimized to match simulation outputs with physical measurements for all discovered microstructure designs (Fig.~S7). Rather than setting simulation parameters using physical measurements of base materials directly, our approach computes parameters that holistically match the properties of a diverse set of microstructures. The sim-to-real gap is effectively treated as a black box, and system identification automatically absorbs all sources of errors into the exposed parameters of the simulator\edit{, including modeling, fabrication, and measurement errors. This effectively compensates for the omission of advanced material models and fracture mechanisms.} We choose Bayesian optimization (BO) as the solver for system identification, an algorithm frequently used to optimize complex, non-convex functions with many local optima. The BO solver is instantiated from multiple random seeds for better robustness. Moreover, by assigning different weights to the source microstructures, we are able to prioritize matches for near-Pareto-optimal designs so that the simulator generalizes better to other designs with desirable performance (see Note S3). This allows the simulator to propose designs that more likely expand the experimental Pareto front instead of spending its modeling capacity on designs far from optimal.

The outer loop uses the same proposal-validation workflow as the inner loop. Here, the simulator serves as a surrogate model for the mechanical tester to search for experimental Pareto-optimal microstructure designs with high sample efficiency. After Round 0, we assemble a dataset of 18 discovered microstructures, including 8 proposed Pareto-optimal designs from Round 0 and 10 additional random designs for regularization in system identification (hollow blue markers in Fig.~\ref{fig:outer-loop}A). In each subsequent round, our pipeline updates the simulator via system identification, searches for the simulation Pareto front through the inner loop (NMO), proposes 8 Pareto-optimal microstructure designs to the mechanical tester (Fig.~\ref{fig:outer-loop}B), validates their performance by physical measurements, and finally adds them to the dataset of discovered microstructures. We run the outer loop for 4 rounds and visualize the evolution of experiment data and simulated gamut in Fig.~\ref{fig:outer-loop}C. For a quantitative comparison, the advancement of their Pareto hypervolumes is illustrated in Fig.~\ref{fig:outer-loop}D. As the pipeline proceeds, the Pareto front of experiment data gradually converges to the simulation Pareto front and the sim-to-real gap ultimately decreases by an order of magnitude. Among 50 microstructures discovered by the pipeline in total, we select 6 designs on or near the experimental Pareto front as the best examples and present their patterns in Fig.~\ref{fig:outer-loop}E. Compared with homogeneous composites fabricated by mixing base materials at various ratios during 3D printing, they achieve up to 175.6\% higher toughness under the same Young’s modulus, or 61.0\% higher Young’s modulus at the same toughness, demonstrating much better trade-offs between \edit{stiffness} and toughness.
 
Intuitively, the outer loop parallels the inner loop as an automated self-learning process for the simulator \cite{schmidt2009distilling, shalev2011online, xie2020self}. Since the microstructure dataset is augmented by 8 new microstructures per round, the simulator becomes more accurate over time as depicted by its decreasing error from ground truth physical measurements (Fig.~\ref{fig:outer-loop}F). With higher accuracy, the simulator is able to propose increasingly better microstructure designs in each round, as characterized by the improving Pareto hypervolume of their physically validated properties (Fig.~\ref{fig:outer-loop}G). Therefore, even with our relatively simple simulator---which omits complex, advanced modeling techniques---our pipeline simultaneously bridges the sim-to-real gap and discovers microstructures with experimentally verified Pareto-optimal performance.

\begin{figure}[!htbp]
    \centering
    \includegraphics[width=0.9\linewidth]{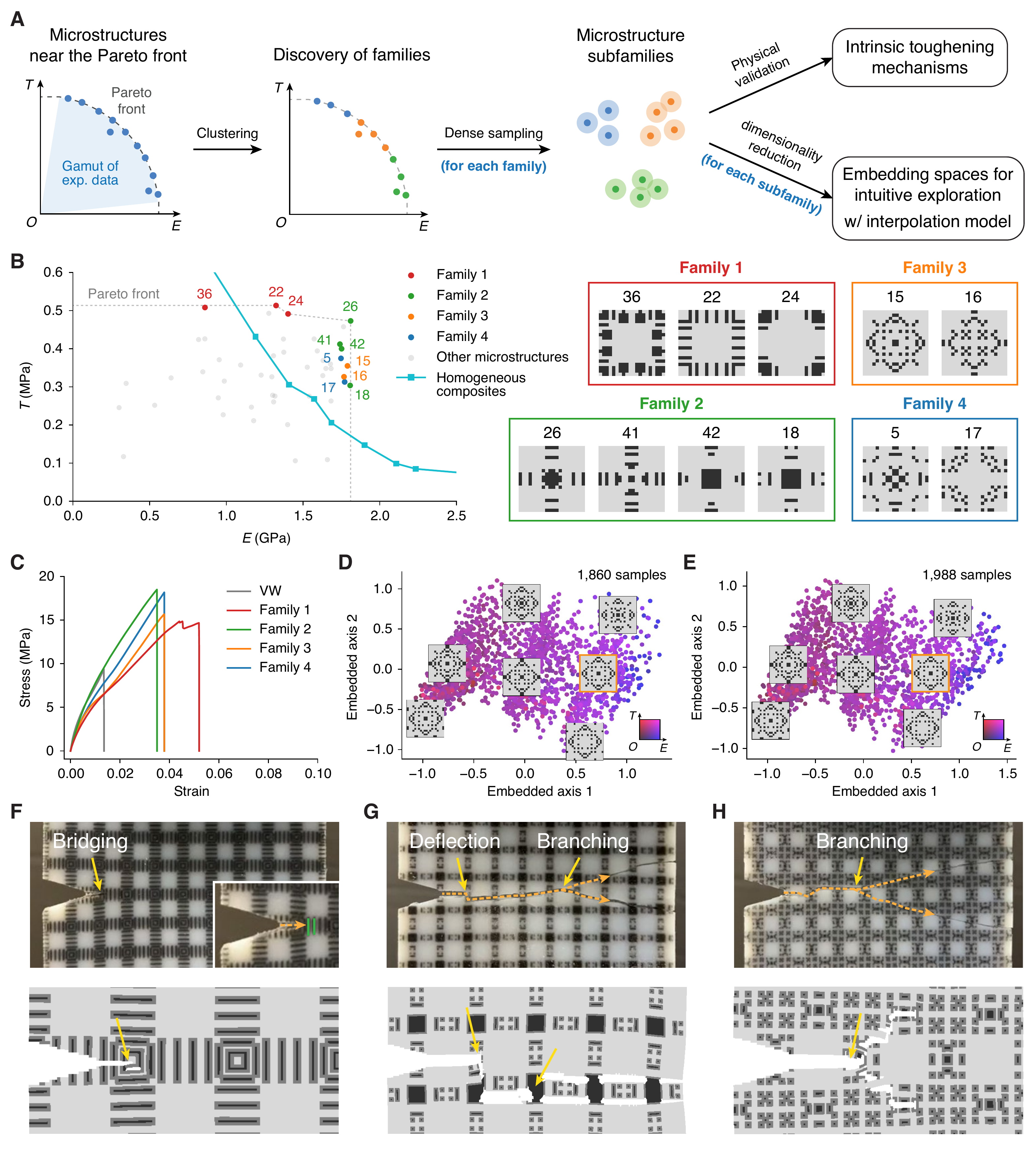}
    \caption{\textbf{Analysis of families and intrinsic toughening mechanisms using discovered Pareto-optimal microstructures.} ({\bf A}) Schematic of the analysis workflow. Microstructures near the Pareto front are grouped into families and further split into subfamilies (see Note S4 and Fig.~S17). The mechanical performance of each family is verified, while the pattern variation in each subfamily is studied in a low-dimensional embedding space. ({\bf B}) Seed microstructures of four families and their mechanical performance compared with homogeneous composites. ({\bf C}) Representative stress-strain curves of microstructures in each family collected from mechanical testing. ({\bf D} and {\bf E}) The Isomap embedding space of an example subfamily before ({\bf D}) and after ({\bf E}) interpolation. Some representative patterns are illustrated. Microstructure properties are encoded in colors (Young’s modulus: the blue channel; toughness: the red channel). Seed patterns are highlighted in orange boxes. ({\bf F}-{\bf H}) Toughening mechanisms observed in experiment and simulation: bridging, deflection, and branching. Snapshots are taken from validation microstructures in Family 1 ({\bf F}), 2 ({\bf G}), and 3 ({\bf H}).}
    \label{fig:analysis}
\end{figure}

From a methodology perspective, our data-driven computational approach inverts the standard design process: instead of implementing handcrafted or bioinspired toughening mechanisms to fabricate synthetic composites \cite{dimas2013tough, wegst2015bioinspired, yin2019impact}, we can interpret these mechanisms from automatically discovered Pareto-optimal designs even without any prior knowledge (Fig.~\ref{fig:analysis}A). To that end, we first notice that microstructures near the experimental Pareto front can be clustered into four major families, where designs in the same family are structurally similar \cite{chen2018computational} (Fig.~\ref{fig:analysis}B). The representative stress-strain curve from each family exemplifies how these designs improve fracture resistance considerably over the rigid base material while maintaining comparable Young's moduli, resulting in optimal \edit{stiffness}-toughness trade-offs (Fig.~\ref{fig:analysis}C). Therefore, the microstructure families allow for extracting and interpreting toughening mechanisms by groups rather than individual designs. To verify the discovered families, we treat the physically measured microstructures as seeds and generate a local performance gamut for each family through dense sampling, from which three validation samples near the Pareto front of each family are selected and validated by the mechanical tester (Fig.~S16 and Table S13). Moreover, to intuitively explore microstructures with similar appearance and performance, we take advantage of our well-trained simulator to generate a local gamut for each seed design near the experimental Pareto front and compute a low-dimensional embedding space using Isomap \cite{balasubramanian2002isomap} (Fig.~\ref{fig:analysis}D). We additionally build an interpolation model for neighboring patterns in the embedding space (see Note S4 and Fig.~S16) to generate patterns that are potentially missed in dense sampling, which eventually refines the embedding space (Fig.~\ref{fig:analysis}E).
 
By analyzing tested specimens and simulation videos, the discovered microstructure families allow us to successfully identify some well-known mechanisms of toughness enhancement, such as bridging, deflection, and branching \cite{yang2021intrinsic, bao1992, gu2017hierarchically, suresh1983micromechanisms, flores1999enhanced} (Fig.~\ref{fig:analysis}F-H). These mechanisms resist crack growth and prevent the formation of clean cleavage as observed in homogeneous composites (Fig.~S33). Their prominence varies depending on the distribution of soft material. Bridging is mostly observed in Family 1, where parallel bars of soft material absorb more energy through elastic deformation and introduce additional stress resistance by crack blunting to slow down crack propagation (Fig.~\ref{fig:analysis}F and Fig.~S22). Deflection is witnessed in both Family 1 and 2, as characterized by chunks of soft material that withstand a large strain and disturb the cracks (Fig.~\ref{fig:analysis}G and Figs.~S22-23). Conversely, branching effects are predominant in Family 3 and 4, which feature scattered elements of soft material. In this case, the propagation energy of the cracks dissipates in the form of branched cracks due to the trapping effect from soft material \cite{ramulu1985mechanics} (Fig.~\ref{fig:analysis}H and Figs.~S24-25). In addition, we also observed other toughing mechanisms that do not obviously fall into the three prevailing categories. For example, some isolated soft material elements are assigned to the edge of a microstructure pattern (Fig.~S24), either near the notch tip or on the path of crack propagation. Despite not being organized into shapes like bars or circles, such elements still effectively dissipate energy and delay crack propagation \cite{ritchie2011conflicts}. Overall, it is common among discovered families that soft material elements enhance toughness through crack blunting while stiff material elements are integrally connected to maintain Young's modulus (see Note S4). \edit{Most importantly, our pipeline automatically discovers the critical soft and stiff elements that contribute to balanced Young’s modulus and toughness without prescribed knowledge of existing toughening mechanisms. These structural features are generated using a fully computational method instead of biomimics or experimental trial-and-error. Therefore, they may pave the way for novel physical insights, provided that further characterization and analysis are conducted by experts in fracture mechanics.}
 
In addition, our computational pipeline has adequate scalability to solve Pareto-optimal microstructure discovery problems in higher design resolutions. For example, we applied our pipeline to discover the Pareto front between stiffness and porosity in isotropic porous materials (see Note S5 for details about fabrication, experiment, and simulation). The microstructure patterns in this case study are represented by 182$\times$104 grids, constituting a considerably larger design space. Nonetheless, our pipeline still maintains excellent sample efficiency in Pareto front discovery (Fig.~S29). Moreover, our approach manages to identify designs with theoretically optimal performance as prescribed by the Hashin-Shtrikman upper bound \cite{hashin1963theory, hashin1963variational} (Fig.~S28).

\subsection*{Discussion}
In conclusion, we have presented a generalizable, scalable, and efficient computational approach that does not require expert knowledge to automatically discover microstructured composites with optimal trade-offs between \edit{stiffness} and toughness. We integrate three distinct evaluation methods---experimental measurement, numerical simulation, and machine learning inference---in a data-driven proposal-validation framework. The essence of our approach is a nested-loop workflow that improves the sample efficiency of slow, accurate evaluators by harnessing fast, approximate evaluators as surrogate models and progressively improving their accuracy. Our pipeline achieves superior sample efficiency when discovering the simulation Pareto front, running an order of magnitude fewer simulations than other state-of-the-art structural optimization algorithms. With a data-driven self-learning simulator, we discover the experimental Pareto front of Young's modulus and toughness using only a few dozen physical measurements. Furthermore, our automated strategy is able to identify prevailing toughening mechanisms that were initially discovered by field experts based on intuition or biomimicry and through experimental trial and error. This is an inversion of traditional composite design approaches since toughening mechanisms stem from Pareto-optimal designs. Most importantly, our pipeline circumvents the time-consuming, labor-intensive development of advanced simulation models and eliminates the need for any prerequisite knowledge of toughness enhancement. Therefore, our pipeline is readily applicable to various trade-off problems well beyond the field of solid mechanics, especially where simulation struggles to match reality, such as polymer chemistry \cite{shields2021bayesian}, computational biology \cite{jumper2021highly,baek2021accurate}, fluid dynamics \cite{kochkov2021machine,raissi2020hidden}, meteorology \cite{bauer2015quiet,schneider2017climate}, and robotics \cite{hwangbo2019learning}.

\section*{Materials and Methods}
\setcounter{figure}{0}

\subsection*{Specimen Manufacturing}
All homogeneous and microstructured composites were manufactured on a Stratasys Object 260 Connex multi-material 3D printer with a nominal resolution of 300 dpi, 600 dpi, and 800 dpi in $x$, $y$, and $z$ directions, respectively. VeroWhitePlus (VW+) and TangoBlackPlus (TB+), two acrylic-based photopolymers, were used as base materials. The model orientation is visualized by the coordinate axes in Fig.~S2. Specifically, microstructure patterns are located on the $x$-$y$ plane and sweep along the $z$ direction, where the longest dimension of the model aligns with the $y$ direction. Regarding the scale of printed specimens, each pixel in a microstructure pattern is calculated to be a 260 $\mu$m square, a reasonable approximation to the least common multiple of the droplet dimensions in $x$ and $y$ directions.

To enable droplet-level control, we employed a voxel printing approach supported by the printer. In the voxel printing mode, droplet configurations were precomputed locally and sent to the printer in real-time. This allows for printing homogeneous composites mixed with base materials at a given ratio. For any volume fraction of the rigid base material, denoted by $\phi$, a printing model was created by probabilistically assigning base materials to droplets, where each droplet was assigned with the rigid material at a probability of $\phi$. Since the base materials are miscible, homogeneous composites were printed by mixing the droplets on the fly. Homogeneous composite materials with $\phi$ = 0.5, 0.6, 0.65, 0.7, 0.75, 0.8, 0.85, and 0.9 were printed as benchmarks.

\subsection*{Physical Measurement}
For Young's modulus measurements, all testing specimens were printed following the dimensions recommended in ASTM D638. For toughness measurements, as there is no standard reference for heterogeneous composites, we devised our specimens shown in Fig.~S2. The dimensions of a toughness specimen are 104$\times$52$\times$3.2 mm. A notch with a length of 10.4 mm was printed, equivalent to 20\% of the specimen's width. After that, a triangle cut was added using a milling machine at a depth of 7.8 mm, which is 15\% of the specimen’s width. The length of the gripper area on either side is 25 mm, leaving a 54~mm gauge area in the middle.

After printing, all specimens were carefully cleaned to remove support material and dried at room temperature. An ultraviolet (UV) light post-curing process was conducted on both the top and bottom sides of the sample to get more uniform curing. This step was performed in a Fusion UV system with a constant belt speed of around 2 cm/s. Additionally, the printed notch of each toughness specimen was cleaned by a new razor blade to avoid the influence of residual support material at the tip.
 
All specimens were tested on an Instron 5984 universal testing machine with a maximum load of 150~kN. Tensile tests for Young’s modulus measurements were conducted according to ASTM D638. Four specimens were tested for each composite and the results were averaged. On the toughness side, there is no existing standard targeting heterogeneous composites. Thus, we employed the following tensile test method. The specimen was pulled at a rate of 2 mm/min. Data acquisition stopped once the crack propagated entirely through the specimen. Because crack propagation in microstructured composites is nonlinear, the toughness is defined as the energy absorbed and dissipated per unit volume; namely, the area under the stress-strain curve \cite{mayer2005rigid}.
At least four specimens were tested for each composite, where at least three specimens that manifested the most common and consistent fracture behaviors were considered valid. These valid toughness measurements were averaged, from which a representative stress-strain curve was selected. The same procedure was followed when testing homogeneous composites for consistency. Measured stress-strain curves from the toughness test of several representative microstructures are provided in Fig.~S8. For each microstructure design, four specimens were fabricated and measured (distinguished by color). The stress-strain curves of the same microstructure pattern are generally consistent, with standard deviations consistently less than 5\% of the mean for Young's modulus and less than 20\% for toughness.

\subsection*{Interface Engineering}
Interfaces between the rigid and soft base materials are engineered by applying crafted droplet patterns in interface areas, i.e., half-pixel-wide areas around shared edges and vertices of adjacent pixels (Fig.~S3). The droplet patterns approximate a 50/50 mixture of the base materials and vary depending on the orientation of shared edges. Based on the coordinates defined in Fig.~S3, droplets are arranged in alternate columns for edges along the $x$ direction (vertices included), and alternate rows for edges along the $y$ direction (vertices excluded; see Note S1). The droplet patterns remain consistent in the $z$ direction during 3D printing. Fig.~S4 compares the performance of 5 random microstructures manufactured with engineered interfaces, non-engineered interfaces, and default interfaces to the printer. The results demonstrate that our interface engineering method yields the best trade-off between Young’s modulus and toughness in printed specimens. In addition, the engineered interfaces are modeled using an interface material in simulation and count as the third base material.

\subsection*{Simulation}
The simulator comprises a pair of finite element method (FEM)-based virtual testers, which predict Young's modulus, toughness, and non-linear stress-strain response given a microstructure pattern. The Neo-Hookean material model is used for all three base materials (VW+, TB+, and interface material) in a 2D setting (see Supplementary for details). The FEM grids represent simplified geometries of manufactured specimens. To model the interfaces between the rigid and soft materials, microstructure units in the FEM grids are upsampled by two times, and elements located at interfaces are assigned with the interface material (Fig.~S5). Dirichlet boundary conditions are enforced on displacements in the pulling direction. An explicit solver is used for time integration as it is usually preferred for dynamic simulations and it is simple to implement. Material viscosity is modeled using numerical damping and controlled by a global damping coefficient. For toughness, crack initiation and propagation are modeled by thresholding the strain energy density of each element as derived from the Neo-Hookean model, where the element is removed upon exceeding a material-specific strain energy density threshold. The simulation stops when the gauge stress drops below 20\% of the ultimate strength, and toughness is calculated using the area beneath the stress-strain curve. For Young’s modulus, a small constant strain is applied, and the grid is allowed to converge to an equilibrium within a fixed number of timesteps. Young’s modulus is then obtained from the measured gauge stress. Algorithmic details and parameter values of the virtual testers can be found in Supplementary.

Our simulator was implemented using the high-performance Taichi programming language \cite{hu2019taichi} to harness the tremendous acceleration power of graphics processing units (GPU). To further increase the throughput, multiple microstructures were batched and simulated simultaneously to enable maximum GPU utilization (Fig.~S6).
 
\subsection*{System Identification}
The simulated behaviors of a microstructure are contingent on several tunable parameters in the simulator, including Young’s moduli, Poisson’s ratios, and strain energy density thresholds of the base materials, plus a global damping coefficient. These parameters are optimized in system identification for a close match between the simulation output and the physical measurements.
\edit{Here, the objective function accounts for multiple sources of errors, including modeling (the Neo-Hookean constitutive model and a simplified fracture mechanism), fabrication (from 3D printing and specimen preparation), and measurement (inaccuracies of the equipment and other random influences from the environment).}
For Young’s modulus, the mean relative prediction error is minimized over a collection of composites, referred to as the system identification dataset. For toughness, the minimization objective is a combination of mean relative prediction errors on both the toughness value and the stress-strain curve, where the relative error of the curve is defined as the area between simulated and experimental stress-strain curves divided by experimental toughness.
To fully utilize the modeling capabilities of the virtual testers, we keep parameter values separated between Young’s modulus and toughness simulation. Furthermore, microstructures in the dataset are weighted to reflect priority in system identification (Fig.~S9).
First, microstructures closer to the experimental Pareto front have larger weights since we mainly focus on near-Pareto-optimal designs. Second, to alleviate possible bias from an uneven performance distribution in the dataset, weight penalties are inflicted upon clusters of microstructures whose physical measurements are too close to one another. A more detailed mathematical formulation of system identification is provided in Supplementary.
We used batch Bayesian optimization (BO) to solve the minimization problem due to its excellent data efficiency and compatibility with our batch simulation mechanism. The Bayesian optimization algorithm was repeated for 5 times using different random seeds, after which the best solution was selected (see Note 3 for implementation details).

\subsection*{Neural Network-Accelerated Multi-Objective Optimization (NMO)}
We developed the NMO algorithm to explore the design space of microstructures and discover Pareto-optimal samples in high sample efficiency. The exploration started from randomly generated microstructures as the first generation. Unlike NSGA-II algorithms which compute offsprings from mutation and crossover operations, NMO employs the predictor, which contains a pair of residual CNNs \cite{he2016deep}, as a surrogate model to guide the proposal of the next generation (see Supplementary for details). Both networks are instantiated from the same architecture template, but the network for toughness is much larger (Fig.~S10). This allows for greater learning capacity to capture the complex mapping from microstructure patterns to toughness. In each iteration, the networks were trained on the simulation results of all discovered microstructures. Based on their predictions, a modified NSGA-II algorithm was run to propose Pareto-optimal samples that likely expand the hypervolume of the current gamut. The proposed samples became the next generation of microstructures once their performance was validated by the simulator.

The modified NSGA-II algorithm within NMO has a custom mutation operator and an additional hash-table-based deduplication mechanism. The custom mutation operator substitutes the original mutation and crossover operators in offspring generation. It changes a microstructure pattern at varied scales probabilistically, including flipping a single pixel, drawing a rasterized line, or overwriting a rectangular area of pixels. Symmetry and structure constraints are subsequently enforced on the mutated pattern. The hash-table-based deduplication mechanism prevents microstructures from being added to the next generation if they have already been discovered, and is a faster alternative to exhaustive match in existing patterns. The NMO algorithm stores discovered microstructures in a global hash table, while each individual run of the modified NSGA-II algorithm uses a separate hash table. More implementation details are provided in Supplementary.

We note that the validation of the inner loop and the outer loop were conducted separately. While NMO was evaluated for the inner loop in Fig.~\ref{fig:inner-loop}, we directly used the modified NSGA-II to verify the outer loop and generate the results in Fig.~\ref{fig:outer-loop}. Despite being evaluated independently, NMO is readily applicable to the computational discovery pipeline since our modularized implementation switches between optimization algorithms very easily.

\subsection*{Sample selection for physical validation}
 
In each round of the outer loop, several microstructures near the simulation Pareto front were selected for physical validation (Fig.~S14). The selection algorithm has two stages. First, to identify microstructures that are sufficiently close to the Pareto front, we associated each microstructure with a rank induced by Pareto dominance. By our definition, the 1st rank comprises microstructures on the Pareto front of the whole gamut, and the $i$-th rank contains microstructures on the Pareto front of a partial gamut where those in higher ranks are excluded. Then, a microstructure qualified as a selection candidate if the following three criteria were satisfied: (1) it is in the 3rd rank or higher; (2) its simulated performance trade-off is better than the actual performance trade-off of homogeneous composites; (3) its toughness is higher than the microstructure with the largest Young’s modulus in the gamut. Second, we chose 8 microstructures from the candidates to cover a wide range of mechanical performance and improve the diversity among discovered microstructures. To that end, the candidate microstructures were divided into 8 radial bins according to their normalized mechanical performance. Then, we solved a combinatorial optimization problem, where one microstructure was chosen in each bin such that the selected microstructures were maximally distinguished from the discovered ones and from each other. The optimization problem was tackled using a beam search algorithm that finds an approximated best solution in a few seconds. In Supplementary Materials, we provide a more detailed description of the selection algorithm, including a mathematical formulation of the combinatorial optimization problem.

\subsection*{Analysis of microstructure families and toughening mechanisms}
After the nested-loop pipeline terminated, eleven microstructures with near-optimal trade-offs between experimental Young’s modulus and toughness were categorized into four families (Fig.~\ref{fig:analysis}B). We refer to these microstructures as seed microstructures. They were used to generate more pattern variations in each family and verify that the similarity in mechanical performance is preserved. This procedure breaks down into the following steps. To start, a family-specific simulator was obtained in each family from system identification on all near-Pareto-optimal microstructures. In this case, the seed microstructures were assigned with large identical weights, while others outside the family were assigned with small identical weights for regularization purposes. Such family-specific simulators have much lower prediction errors on the seed microstructures in the family compared to the global simulator in Round 4 (Table~S7). Then, based on simulation results from the family-specific simulators, we ran a dense evolutionary sampling algorithm around seed microstructures to generate a local gamut for each family. The sampling algorithm is akin to the modified NSGA-II algorithm but limits the mutated patterns within a maximum of 16-pixel difference from seed microstructures. There is neither tournament selection of parents nor competition among the population, hence the target is simply to generate a gamut as dense as possible. Lastly, three near-Pareto-optimal microstructures were chosen from the resulting local gamut for physical validation (Fig.~S16), whose physical measurements were demonstrated to be comparable to seed microstructures (Table~S13).  Therefore, we confirmed that the discovered families contain microstructures with similar patterns and mechanical performance.

To visualize the pattern variation in each microstructure family, we divided each family into subfamilies by extracting microstructures within a 16-pixel radius from each seed microstructure, allowing any microstructure to occur in multiple subfamilies. For every subfamily, a 2D embedding space was computed for all microstructure patterns using Isomap \cite{balasubramanian2002isomap}.
The first Wasserstein distance \cite{villani2003topics}, i.e. the earth mover’s distance (EMD), was used as the distance metric between microstructure patterns instead of the Euclidean distance (Fig.~S18) since EMD takes into account pixel distributions in microstructure patterns and better reflects differences in geometric shapes. Figs.~S17 and B1 illustrate the visualized embedding spaces of 11 subfamilies across 4 families. Furthermore, we built an interpolation model for neighboring microstructures in the Isomap embedding space to generate more microstructures that might be missing in our dense sampling (Fig.~S17). Our interpolation method is based on Wasserstein barycenters from optimal transport theory \cite{agueh2011barycenters, kantorovich1942translocation}, which has been successfully demonstrated as a robust, intuitive interpolation scheme among voxelized shapes \cite{solomon2015convolutional}. In this way, we generated around 5\% more microstructure patterns in each subfamily on average and used them to refine the Isomap embedding spaces. A comparison between the embedding spaces before and after interpolation is illustrated in Figs.~S21, S31, and S32. A mathematical formulation of EMD and our interpolation scheme can be found in Note S4.

The identification of microstructure families simplifies the analysis of intrinsic toughening mechanisms leading to near-optimal trade-offs between Young’s modulus and toughness. As microstructures have similar patterns and performance in each family, they typically share common structural features that enhance fracture resistance. Thus, we used video recordings of seed microstructures and validation microstructures in each family to observe and interpret several predominant toughening mechanisms. For each microstructure, the simulation video produced by family-specific simulators was validated against actual footage from mechanical testing to verify that the mechanisms were properly captured by the simulator. The representative mechanisms of each microstructure family are thoroughly discussed in Supplementary.

\nocite{gu2017printing,mirzaali2020mechanics,sifakis2012fem,kelly2013solid,gonzalez2016batch,bendsoe1988generating,bendsoe2003topology,knowles2006parego,kingma2014adam,dosovitskiy2020image,xia2018topology,russ2019topology,hu2019difftaichi,xia2017evolutionary,mao2020designing}

\bibliography{main}

\begin{thebibliography}{10}

\bibitem{ritchie2011conflicts}
R.~O. Ritchie, The conflicts between strength and toughness.
\newblock {\it Nature materials\/} {\bf 10}, 817--822 (2011).

\bibitem{mayer2005rigid}
G.~Mayer, Rigid biological systems as models for synthetic composites.
\newblock {\it Science\/} {\bf 310}, 1144--1147 (2005).

\bibitem{sanchez2005biomimetism}
C.~Sanchez, H.~Arribart, M.~M.~G. Guille, Biomimetism and bioinspiration as
  tools for the design of innovative materials and systems.
\newblock {\it Nature materials\/} {\bf 4}, 277--288 (2005).

\bibitem{dimas2013tough}
L.~S. Dimas, G.~H. Bratzel, I.~Eylon, M.~J. Buehler, Tough composites inspired
  by mineralized natural materials: computation, 3d printing, and testing.
\newblock {\it Advanced Functional Materials\/} {\bf 23}, 4629--4638 (2013).

\bibitem{meyers2013structural}
M.~A. Meyers, J.~McKittrick, P.-Y. Chen, Structural biological materials:
  critical mechanics-materials connections.
\newblock {\it Science\/} {\bf 339}, 773--779 (2013).

\bibitem{yin2019impact}
Z.~Yin, F.~Hannard, F.~Barthelat, Impact-resistant nacre-like transparent
  materials.
\newblock {\it Science\/} {\bf 364}, 1260--1263 (2019).

\bibitem{dudukovic2021cellular}
N.~A. Dudukovic, E.~J. Fong, H.~B. Gemeda, J.~R. DeOtte, M.~R. Cer{\'o}n, B.~D.
  Moran, J.~T. Davis, S.~E. Baker, E.~B. Duoss, Cellular fluidics.
\newblock {\it Nature\/} {\bf 595}, 58--65 (2021).

\bibitem{schaedler2011ultralight}
T.~A. Schaedler, A.~J. Jacobsen, A.~Torrents, A.~E. Sorensen, J.~Lian, J.~R.
  Greer, L.~Valdevit, W.~B. Carter, Ultralight metallic microlattices.
\newblock {\it Science\/} {\bf 334}, 962--965 (2011).

\bibitem{zheng2014ultralight}
X.~Zheng, H.~Lee, T.~H. Weisgraber, M.~Shusteff, J.~DeOtte, E.~B. Duoss, J.~D.
  Kuntz, M.~M. Biener, Q.~Ge, J.~A. Jackson, {\it et~al.\/}, Ultralight,
  ultrastiff mechanical metamaterials.
\newblock {\it Science\/} {\bf 344}, 1373--1377 (2014).

\bibitem{pham2019damage}
M.-S. Pham, C.~Liu, I.~Todd, J.~Lertthanasarn, Damage-tolerant architected
  materials inspired by crystal microstructure.
\newblock {\it Nature\/} {\bf 565}, 305--311 (2019).

\bibitem{wehner2016integrated}
M.~Wehner, R.~L. Truby, D.~J. Fitzgerald, B.~Mosadegh, G.~M. Whitesides, J.~A.
  Lewis, R.~J. Wood, An integrated design and fabrication strategy for entirely
  soft, autonomous robots.
\newblock {\it Nature\/} {\bf 536}, 451--455 (2016).

\bibitem{aage2017giga}
N.~Aage, E.~Andreassen, B.~S. Lazarov, O.~Sigmund, Giga-voxel computational
  morphogenesis for structural design.
\newblock {\it Nature\/} {\bf 550}, 84--86 (2017).

\bibitem{chen2018computational}
D.~Chen, M.~Skouras, B.~Zhu, W.~Matusik, Computational discovery of extremal
  microstructure families.
\newblock {\it Science advances\/} {\bf 4}, eaao7005 (2018).

\bibitem{butler2018machine}
K.~T. Butler, D.~W. Davies, H.~Cartwright, O.~Isayev, A.~Walsh, Machine
  learning for molecular and materials science.
\newblock {\it Nature\/} {\bf 559}, 547--555 (2018).

\bibitem{gu2018bioinspired}
G.~X. Gu, C.-T. Chen, D.~J. Richmond, M.~J. Buehler, Bioinspired hierarchical
  composite design using machine learning: simulation, additive manufacturing,
  and experiment.
\newblock {\it Materials Horizons\/} {\bf 5}, 939--945 (2018).

\bibitem{sigmund2013topology}
O.~Sigmund, K.~Maute, Topology optimization approaches.
\newblock {\it Structural and Multidisciplinary Optimization\/} {\bf 48},
  1031--1055 (2013).

\bibitem{wu2021topology}
J.~Wu, O.~Sigmund, J.~P. Groen, Topology optimization of multi-scale
  structures: a review.
\newblock {\it Structural and Multidisciplinary Optimization\/} {\bf 63},
  1455--1480 (2021).

\bibitem{yang2021intrinsic}
Y.~Yang, Z.~Song, G.~Lu, Q.~Zhang, B.~Zhang, B.~Ni, C.~Wang, X.~Li, L.~Gu,
  X.~Xie, {\it et~al.\/}, Intrinsic toughening and stable crack propagation in
  hexagonal boron nitride.
\newblock {\it Nature\/} {\bf 594}, 57--61 (2021).

\bibitem{shaikeea2022toughness}
A.~J.~D. Shaikeea, H.~Cui, M.~O’Masta, X.~R. Zheng, V.~S. Deshpande, The
  toughness of mechanical metamaterials.
\newblock {\it Nature materials\/} pp. 1--8 (2022).

\bibitem{surjadi2022design}
J.~U. Surjadi, Y.~Lu, Design criteria for tough metamaterials.
\newblock {\it Nature Materials\/} pp. 1--2 (2022).

\bibitem{zavala2014survey}
G.~R. Zavala, A.~J. Nebro, F.~Luna, C.~A. Coello~Coello, A survey of
  multi-objective metaheuristics applied to structural optimization.
\newblock {\it Structural and Multidisciplinary Optimization\/} {\bf 49},
  537--558 (2014).

\bibitem{hossain2014effective}
M.~Hossain, C.-J. Hsueh, B.~Bourdin, K.~Bhattacharya, Effective toughness of
  heterogeneous media.
\newblock {\it Journal of the Mechanics and Physics of Solids\/} {\bf 71},
  15--32 (2014).

\bibitem{kumar2020phase}
A.~Kumar, O.~Lopez-Pamies, The phase-field approach to self-healable fracture
  of elastomers: A model accounting for fracture nucleation at large, with
  application to a class of conspicuous experiments.
\newblock {\it Theoretical and Applied Fracture Mechanics\/} {\bf 107}, 102550
  (2020).

\bibitem{bourdin2008variational}
B.~Bourdin, G.~A. Francfort, J.-J. Marigo, The variational approach to
  fracture.
\newblock {\it Journal of elasticity\/} {\bf 91}, 5--148 (2008).

\bibitem{jia2023controlling}
Y.~Jia, O.~Lopez-Pamies, X.~S. Zhang, Controlling the fracture response of
  structures via topology optimization: From delaying fracture nucleation to
  maximizing toughness.
\newblock {\it Journal of the Mechanics and Physics of Solids\/} {\bf 173},
  105227 (2023).

\bibitem{querin1998evolutionary}
O.~M. Querin, G.~P. Steven, Y.~M. Xie, Evolutionary structural optimisation
  (eso) using a bidirectional algorithm.
\newblock {\it Engineering computations\/}  (1998).

\bibitem{huang2008topology}
X.~Huang, Y.~Xie, Topology optimization of nonlinear structures under
  displacement loading.
\newblock {\it Engineering structures\/} {\bf 30}, 2057--2068 (2008).

\bibitem{bendsoe1989optimal}
M.~P. Bends{\o}e, Optimal shape design as a material distribution problem.
\newblock {\it Structural optimization\/} {\bf 1}, 193--202 (1989).

\bibitem{lukovic2020diversity}
M.~K. Lukovic, Y.~Tian, W.~Matusik, Diversity-guided multi-objective bayesian
  optimization with batch evaluations.
\newblock {\it 34th Conference on Neural Information Processing Systems
  (NeurIPS 2020)\/}  (2020).

\bibitem{bradford2018efficient}
E.~Bradford, A.~M. Schweidtmann, A.~Lapkin, Efficient multiobjective
  optimization employing gaussian processes, spectral sampling and a genetic
  algorithm.
\newblock {\it Journal of global optimization\/} {\bf 71}, 407--438 (2018).

\bibitem{deb2002fast}
K.~Deb, A.~Pratap, S.~Agarwal, T.~Meyarivan, A fast and elitist multiobjective
  genetic algorithm: Nsga-ii.
\newblock {\it IEEE transactions on evolutionary computation\/} {\bf 6},
  182--197 (2002).

\bibitem{sigmund200199}
O.~Sigmund, A 99 line topology optimization code written in matlab.
\newblock {\it Structural and multidisciplinary optimization\/} {\bf 21},
  120--127 (2001).

\bibitem{schmidt2009distilling}
M.~Schmidt, H.~Lipson, Distilling free-form natural laws from experimental
  data.
\newblock {\it Science\/} {\bf 324}, 81--85 (2009).

\bibitem{shalev2011online}
S.~Shalev-Shwartz, {\it et~al.\/}, Online learning and online convex
  optimization.
\newblock {\it Foundations and trends in Machine Learning\/} {\bf 4}, 107--194
  (2011).

\bibitem{xie2020self}
Q.~Xie, M.-T. Luong, E.~Hovy, Q.~V. Le, {\it Proceedings of the IEEE/CVF
  Conference on Computer Vision and Pattern Recognition\/} (2020), pp.
  10687--10698.

\bibitem{wegst2015bioinspired}
U.~G. Wegst, H.~Bai, E.~Saiz, A.~P. Tomsia, R.~O. Ritchie, Bioinspired
  structural materials.
\newblock {\it Nature materials\/} {\bf 14}, 23--36 (2015).

\bibitem{balasubramanian2002isomap}
M.~Balasubramanian, E.~L. Schwartz, J.~B. Tenenbaum, V.~de~Silva, J.~C.
  Langford, The isomap algorithm and topological stability.
\newblock {\it Science\/} {\bf 295}, 7--7 (2002).

\bibitem{bao1992}
G.~Bao, Z.~Suo, {Remarks on Crack-Bridging Concepts}.
\newblock {\it Applied Mechanics Reviews\/} {\bf 45}, 355-366 (1992).

\bibitem{gu2017hierarchically}
G.~X. Gu, M.~Takaffoli, M.~J. Buehler, Hierarchically enhanced impact
  resistance of bioinspired composites.
\newblock {\it Advanced Materials\/} {\bf 29}, 1700060 (2017).

\bibitem{suresh1983micromechanisms}
S.~Suresh, Micromechanisms of fatigue crack growth retardation following
  overloads.
\newblock {\it Engineering Fracture Mechanics\/} {\bf 18}, 577--593 (1983).

\bibitem{flores1999enhanced}
K.~M. Flores, R.~H. Dauskardt, Enhanced toughness due to stable crack tip
  damage zones in bulk metallic glass.
\newblock {\it Scripta materialia\/} {\bf 41}, 937--943 (1999).

\bibitem{ramulu1985mechanics}
M.~Ramulu, A.~Kobayashi, Mechanics of crack curving and branching—a dynamic
  fracture analysis.
\newblock {\it Dynamic fracture\/} pp. 61--75 (1985).

\bibitem{hashin1963theory}
Z.~Hashin, {\it Theory of Mechanical Behavior of Heterogeneous Media\/} (Towne
  School of Civil and Mechanical Engineering, University of Pennsylvania,
  1963).

\bibitem{hashin1963variational}
Z.~Hashin, S.~Shtrikman, A variational approach to the theory of the elastic
  behaviour of multiphase materials.
\newblock {\it Journal of the Mechanics and Physics of Solids\/} {\bf 11},
  127--140 (1963).

\bibitem{shields2021bayesian}
B.~J. Shields, J.~Stevens, J.~Li, M.~Parasram, F.~Damani, J.~I.~M. Alvarado,
  J.~M. Janey, R.~P. Adams, A.~G. Doyle, Bayesian reaction optimization as a
  tool for chemical synthesis.
\newblock {\it Nature\/} {\bf 590}, 89--96 (2021).

\bibitem{jumper2021highly}
J.~Jumper, R.~Evans, A.~Pritzel, T.~Green, M.~Figurnov, O.~Ronneberger,
  K.~Tunyasuvunakool, R.~Bates, A.~{\v{Z}}{\'\i}dek, A.~Potapenko, {\it
  et~al.\/}, Highly accurate protein structure prediction with alphafold.
\newblock {\it Nature\/} {\bf 596}, 583--589 (2021).

\bibitem{baek2021accurate}
M.~Baek, F.~DiMaio, I.~Anishchenko, J.~Dauparas, S.~Ovchinnikov, G.~R. Lee,
  J.~Wang, Q.~Cong, L.~N. Kinch, R.~D. Schaeffer, {\it et~al.\/}, Accurate
  prediction of protein structures and interactions using a three-track neural
  network.
\newblock {\it Science\/} {\bf 373}, 871--876 (2021).

\bibitem{kochkov2021machine}
D.~Kochkov, J.~A. Smith, A.~Alieva, Q.~Wang, M.~P. Brenner, S.~Hoyer, Machine
  learning--accelerated computational fluid dynamics.
\newblock {\it Proceedings of the National Academy of Sciences\/} {\bf 118}
  (2021).

\bibitem{raissi2020hidden}
M.~Raissi, A.~Yazdani, G.~E. Karniadakis, Hidden fluid mechanics: Learning
  velocity and pressure fields from flow visualizations.
\newblock {\it Science\/} {\bf 367}, 1026--1030 (2020).

\bibitem{bauer2015quiet}
P.~Bauer, A.~Thorpe, G.~Brunet, The quiet revolution of numerical weather
  prediction.
\newblock {\it Nature\/} {\bf 525}, 47--55 (2015).

\bibitem{schneider2017climate}
T.~Schneider, J.~Teixeira, C.~S. Bretherton, F.~Brient, K.~G. Pressel,
  C.~Sch{\"a}r, A.~P. Siebesma, Climate goals and computing the future of
  clouds.
\newblock {\it Nature Climate Change\/} {\bf 7}, 3--5 (2017).

\bibitem{hwangbo2019learning}
J.~Hwangbo, J.~Lee, A.~Dosovitskiy, D.~Bellicoso, V.~Tsounis, V.~Koltun,
  M.~Hutter, Learning agile and dynamic motor skills for legged robots.
\newblock {\it Science Robotics\/} {\bf 4}, eaau5872 (2019).

\bibitem{hu2019taichi}
Y.~Hu, T.-M. Li, L.~Anderson, J.~Ragan-Kelley, F.~Durand, Taichi: a language
  for high-performance computation on spatially sparse data structures.
\newblock {\it ACM Transactions on Graphics (TOG)\/} {\bf 38}, 1--16 (2019).

\bibitem{he2016deep}
K.~He, X.~Zhang, S.~Ren, J.~Sun, {\it Proceedings of the IEEE conference on
  computer vision and pattern recognition\/} (2016), pp. 770--778.

\bibitem{villani2003topics}
C.~Villani, {\it Topics in optimal transportation\/}, no.~58 (American
  Mathematical Soc., 2003).

\bibitem{agueh2011barycenters}
M.~Agueh, G.~Carlier, Barycenters in the wasserstein space.
\newblock {\it SIAM Journal on Mathematical Analysis\/} {\bf 43}, 904--924
  (2011).

\bibitem{kantorovich1942translocation}
L.~V. Kantorovich, {\it Dokl. Akad. Nauk. USSR (NS)\/} (1942), vol.~37, pp.
  199--201.

\bibitem{solomon2015convolutional}
J.~Solomon, F.~De~Goes, G.~Peyr{\'e}, M.~Cuturi, A.~Butscher, A.~Nguyen, T.~Du,
  L.~Guibas, Convolutional wasserstein distances: Efficient optimal
  transportation on geometric domains.
\newblock {\it ACM Transactions on Graphics (TOG)\/} {\bf 34}, 1--11 (2015).

\bibitem{gu2017printing}
G.~X. Gu, F.~Libonati, S.~D. Wettermark, M.~J. Buehler, Printing nature:
  Unraveling the role of nacre's mineral bridges.
\newblock {\it Journal of the mechanical behavior of biomedical materials\/}
  {\bf 76}, 135--144 (2017).

\bibitem{mirzaali2020mechanics}
M.~Mirzaali, A.~H. de~la Nava, D.~Gunashekar, M.~Nouri-Goushki, R.~Veeger,
  Q.~Grossman, L.~Angeloni, M.~Ghatkesar, L.~Fratila-Apachitei, D.~Ruffoni,
  {\it et~al.\/}, Mechanics of bioinspired functionally graded soft-hard
  composites made by multi-material 3d printing.
\newblock {\it Composite Structures\/} {\bf 237}, 111867 (2020).

\bibitem{sifakis2012fem}
E.~Sifakis, J.~Barbic, Fem simulation of 3d deformable solids: a practitioner's
  guide to theory, discretization and model reduction.
\newblock {\it ACM SIGGRAPH 2012 Courses\/} (2012), pp. 1--50.

\bibitem{kelly2013solid}
P.~Kelly, Solid mechanics part i: An introduction to solid mechanics.
\newblock {\it Solid mechanics lecture notes\/} pp. 241--324 (2013).

\bibitem{gonzalez2016batch}
J.~Gonz{\'a}lez, Z.~Dai, P.~Hennig, N.~Lawrence, {\it Artificial intelligence
  and statistics\/} (PMLR, 2016), pp. 648--657.

\bibitem{bendsoe1988generating}
M.~P. Bends{\o}e, N.~Kikuchi, Generating optimal topologies in structural
  design using a homogenization method.
\newblock {\it Computer methods in applied mechanics and engineering\/} {\bf
  71}, 197--224 (1988).

\bibitem{bendsoe2003topology}
M.~P. Bendsoe, O.~Sigmund, {\it Topology optimization: theory, methods, and
  applications\/} (Springer Science \& Business Media, 2003).

\bibitem{knowles2006parego}
J.~Knowles, Parego: A hybrid algorithm with on-line landscape approximation for
  expensive multiobjective optimization problems.
\newblock {\it IEEE Transactions on Evolutionary Computation\/} {\bf 10},
  50--66 (2006).

\bibitem{kingma2014adam}
D.~P. Kingma, J.~Ba, Adam: A method for stochastic optimization.
\newblock {\it arXiv preprint arXiv:1412.6980\/}  (2014).

\bibitem{dosovitskiy2020image}
A.~Dosovitskiy, L.~Beyer, A.~Kolesnikov, D.~Weissenborn, X.~Zhai,
  T.~Unterthiner, M.~Dehghani, M.~Minderer, G.~Heigold, S.~Gelly, {\it
  et~al.\/}, An image is worth 16x16 words: Transformers for image recognition
  at scale.
\newblock {\it arXiv preprint arXiv:2010.11929\/}  (2020).

\bibitem{xia2018topology}
L.~Xia, D.~Da, J.~Yvonnet, Topology optimization for maximizing the fracture
  resistance of quasi-brittle composites.
\newblock {\it Computer Methods in Applied Mechanics and Engineering\/} {\bf
  332}, 234--254 (2018).

\bibitem{russ2019topology}
J.~B. Russ, H.~Waisman, Topology optimization for brittle fracture resistance.
\newblock {\it Computer Methods in Applied Mechanics and Engineering\/} {\bf
  347}, 238--263 (2019).

\bibitem{hu2019difftaichi}
Y.~Hu, L.~Anderson, T.-M. Li, Q.~Sun, N.~Carr, J.~Ragan-Kelley, F.~Durand,
  Difftaichi: Differentiable programming for physical simulation.
\newblock {\it arXiv preprint arXiv:1910.00935\/}  (2019).

\bibitem{xia2017evolutionary}
L.~Xia, F.~Fritzen, P.~Breitkopf, Evolutionary topology optimization of
  elastoplastic structures.
\newblock {\it Structural and Multidisciplinary Optimization\/} {\bf 55},
  569--581 (2017).

\bibitem{mao2020designing}
Y.~Mao, Q.~He, X.~Zhao, Designing complex architectured materials with
  generative adversarial networks.
\newblock {\it Science advances\/} {\bf 6}, eaaz4169 (2020).

\end{thebibliography}

\bibliographystyle{main}

\noindent{\bf Acknowledgments:} The authors thank B. von Vacano, R. Konradi, K. Stoll, H. H. Goetzke, R. de Oliveira, L. Chen, and F. Zhang from BASF, A. Lesser from University of Massachusetts Amherst, Y. C. Yang from University of Maine, R. Bai from Northeastern University, Y. Mao and S. Lin from MIT for discussions. We thank B. Zhu from Dartmouth College, D. Chen from Inkbit, M. Foshey, L. Makatura, M. Guo, and R. Su from MIT for discussions and proofreading. We thank S. Hudson from MIT for the support of mechanical testing equipment. We thank M. K. Lukovi\'{c} and Y. Tian from MIT for implementing the DGEMO algorithm. We also thank R. White and F. Raton for the administration of the project.

\vspace{0.5em}
\noindent{\bf Funding:} B.L., W.S., and Y.H. were financially supported by BASF.

\vspace{0.5em}
\noindent{\bf Author Contributions:} B.L., W.S., B.D., and W.M. conceived the idea. B.L. implemented the proposed framework, processed experiment data, performed algorithm-related evaluation, and conducted computational analysis on Pareto-optimal designs. W.S. manufactured test specimens, conducted physical measurements, and interpreted the toughening mechanisms in Pareto-optimal microstructure families. Y.H. prototyped the FEM simulator. T.-H.O., Y.L., and L.S. assisted with the design of the proposed framework. B.L. and B.D. conducted the case study of the stiffness-porosity trade-off, which involved additional computational infrastructure, specimen fabrication, and physical measurements. B.L. and B.D. led the writing of the manuscript. W.M. supervised the work. All authors discussed, edited, and reviewed the manuscript. 

\vspace{0.5em}
\noindent{\bf Competing Interests:} W.S., B.L., and W.M. are inventors on a pending patent related to this work filed by MIT (No. US20220374569A1, filed on May 20, 2022, and published on November 24, 2022). The authors declare no other competing interests. 

\vspace{0.5em}
\noindent{\bf Data and Materials Availability:} All data needed to evaluate the conclusions in the paper are present in the paper and/or the Supplementary Materials.

\section*{Supplementary Materials}
This PDF file includes: \\
Notes S1-S5 \\
Figs. S1-S33 \\
Table S1-S15 \\

\end{document}


\flushbottom
\maketitle
%
%
\thispagestyle{empty}

\renewcommand{\figurename}{Fig.}
\renewcommand{\thefigure}{S\arabic{figure}}
\renewcommand{\tablename}{Table}
\renewcommand{\thetable}{S\arabic{table}}

\section{\textcolor{black}{Definition of Microstructures}}


\begin{figure}[ht]
    \centering
    \includegraphics[width=0.7\textwidth]{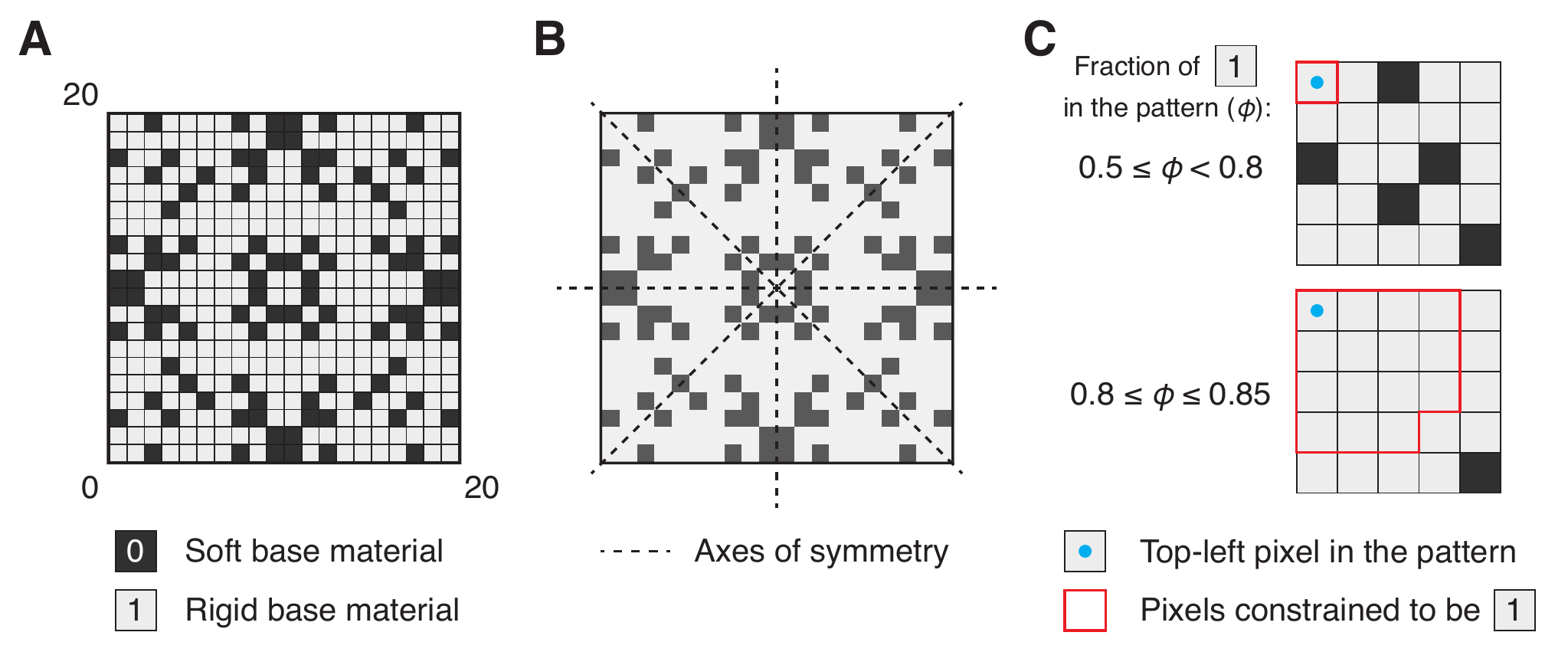}
    \caption{{\bf Definition of microstructures.} ({\bf A}) An example microstructure. ({\bf B}) Symmetry constraints on microstructure patterns. ({\bf C}) Structural constraints on microstructure patterns depending on volume fractions of the rigid base material.}
    \label{fig:design_space}
\end{figure}

As the structural unit of microstructured composites in this work, a microstructure is digitally represented as a 20$\times$20 binary matrix that uniquely defines the spatial arrangement of two base materials in a composite (Fig.~\ref{fig:design_space}A). The matrix is also called a {\it microstructure pattern}. The soft base material is indicated by a value of 0 in a microstructure pattern while the rigid material corresponds to 1. As a microstructure pattern is conceptually similar to an image, we refer to its elements as {\it pixels} to avoid confusion with those in the finite element method (FEM).

To construct the design space for computational exploration, we additionally pose two types of constraints on microstructure patterns. Symmetry constraints (Fig.~\ref{fig:design_space}B) require all microstructures to be invariant to reflections and 90-degree rotations so that they have identical mechanical properties in horizontal and vertical directions. Structural constraints (Fig.~\ref{fig:design_space}C) limit the volume fraction of the rigid material, denoted by $\phi$, between 0.5 and 0.85. Also, pixels around the corner of a pattern are restricted to 1 depending on $\phi$. If $0.5 \leq \phi < 0.8$, only the corner pixel is restricted; otherwise ($0.8 \leq \phi \leq 0.85$), the rule extends to every pixel whose center lies within a 4-pixel-wide radius from the corner vertex. 
The upper bound $\phi \leq 0.85$ ensures reliable and consistent toughness measurements in mechanical testing because the fracture behaviors of very stiff microstructures ($\phi > 0.85$) tend to be highly stochastic and result in low toughness ({\it 59}). 
The lower bound $\phi \geq 0.5$ excludes designs with inferior Young's modulus and toughness due to a predominant influence from the soft material. All constraints aggregated, the design space of microstructures contains more than $10^{16}$ possibilities.

\edit{Our choice of pattern solution is primarily constrained by the 3D printer. A test specimen must contain a sufficient number of microstructure units to achieve proper homogenization and minimize the boundary effect (Fig.~\ref{fig:YM_exp}B). If the resolution is too high, the fine features of a microstructure pattern will be completely blurred during 3D printing as the size of each individual pixel becomes comparable to a material droplet. However, if the resolution is too low, it will take a substantial number of microstructure units to realize such homogenization (otherwise, the large interfaces between base materials will become flaws and weaken the mechanical performance). This makes the resulting specimens virtually indistinguishable from single-phase materials. Thus, we used a 20$\times$20 resolution with a pixel size of 260 $\mu$m to create a meaningful design space while preserving visible pixel boundaries in printed specimens.}

\section{Interface Engineering}

\begin{figure}[ht]
    \centering
    \includegraphics[width=0.6\textwidth]{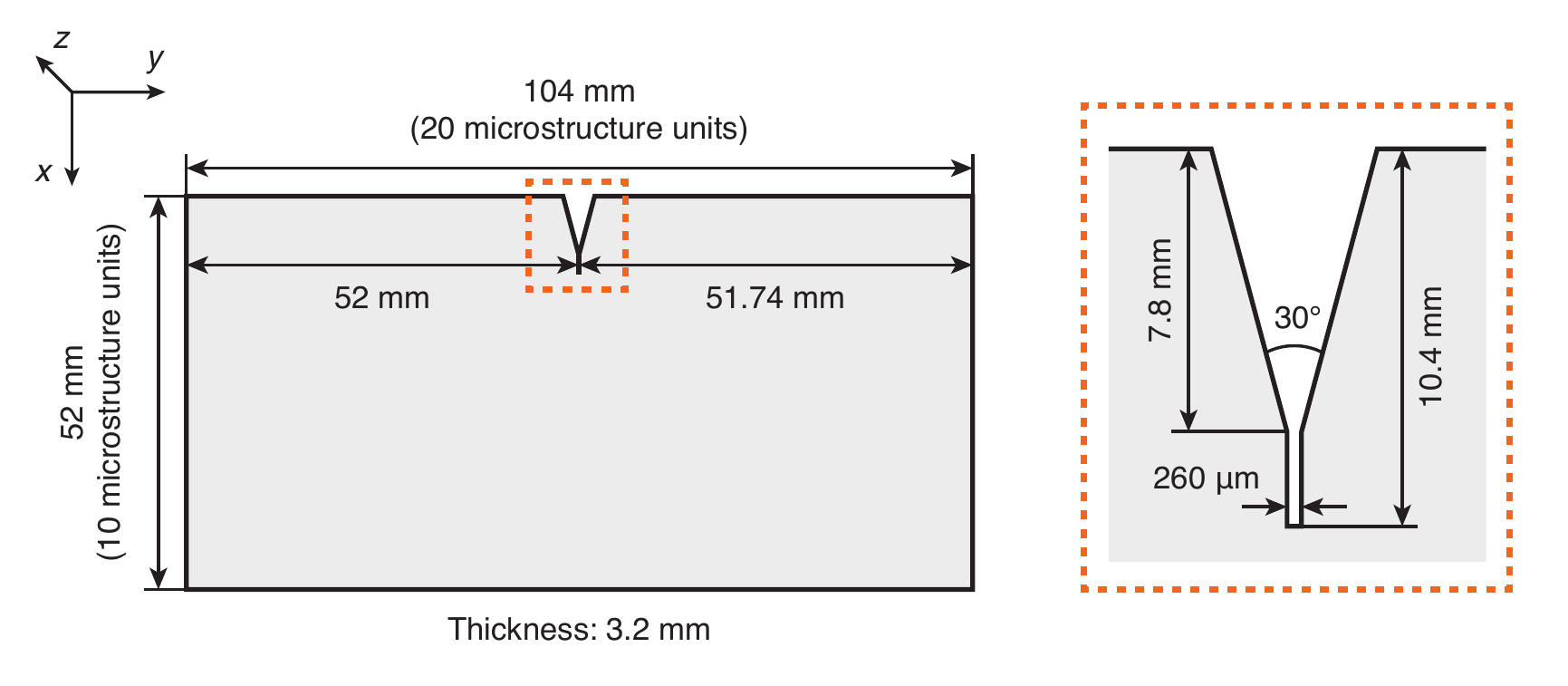}
    \caption{{\bf Dimensions of a toughness specimen with the area around the notch zoomed in.} Coordinate axes indicate the orientation of the specimen in 3D printing.}
    \label{fig:toughness_specimen}
\end{figure}

\begin{figure}[ht]
    \centering
    \includegraphics[width=0.7\textwidth]{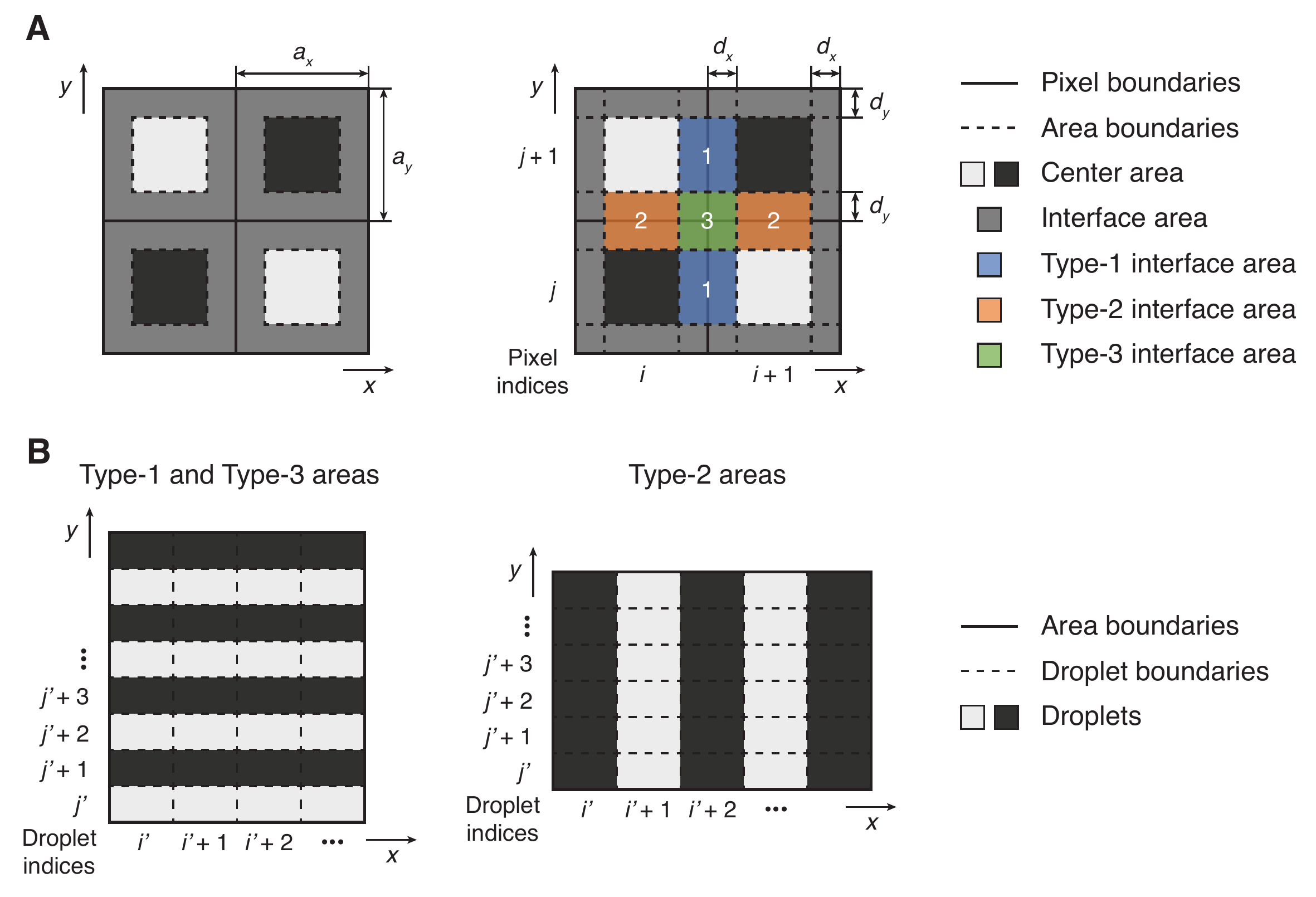}
    \caption{{\bf Schematics of interface engineering.} ({\bf A}) Definition and classification of interface areas. ({\bf B}) Droplet arrangements in interface areas. All coordinate axes conform with the definitions in Fig.~\ref{fig:toughness_specimen}.}
    \label{fig:interface_engineering}
\end{figure}

\begin{table}[ht]
    \centering
    \begin{tabular}{PP|PP|PP}
        \toprule
        \multicolumn{2}{c|}{\textbf{Pixel dimensions}} & \multicolumn{2}{c|}{\textbf{The center area}} & \multicolumn{2}{c}{\textbf{The interface area}} \\
        $x$ ($a_x$) & $y$ ($a_y$) & $x$ & $y$ & $x$ ($d_x$) & $y$ ($d_y$) \\
        \midrule
        3 & 5 & 1 & 3 & 1 & 1 \\
        3 & 6 & 1 & 2 & 1 & 2 \\
        4 & 5 & 2 & 3 & 1 & 1 \\
        4 & 6 & 2 & 2 & 1 & 2 \\
        \bottomrule
    \end{tabular}
    \caption{{\bf Possible dimensions of a pixel and its subdivided areas after quantization into droplets for voxel printing.} All units are droplets.}
    \label{tab:pixel_dimensions}
\end{table}

\begin{figure}[!ht]
    \centering
    \includegraphics[width=0.8\textwidth]{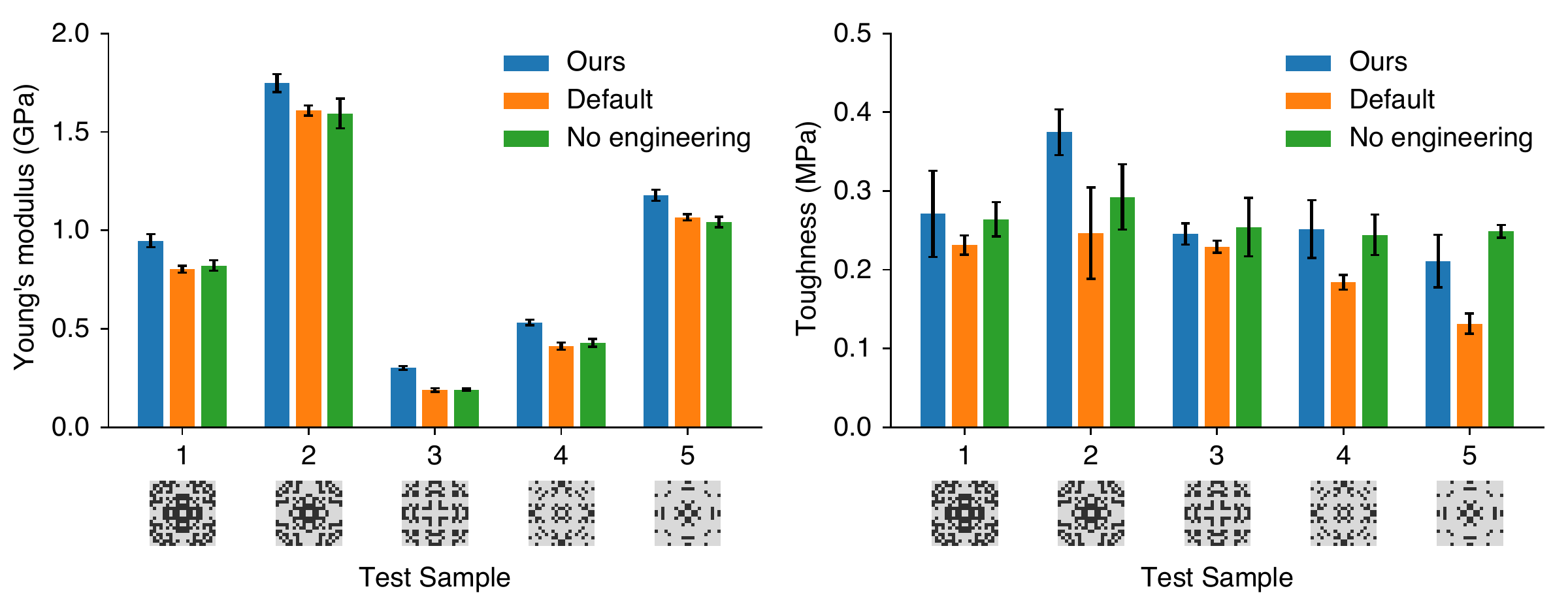}
    \caption{{\bf Comparison of microstructure properties under different interface configurations.} The tested configurations include: default to the printer (default), not engineered (no engineering), and engineered (ours). Standard deviations from multiple measurements are reflected in error bars.}
    \label{fig:interface_comp}
\end{figure}

Upon examination of printed specimens, we noticed that the interface between base materials is relatively strong but can still be the weakest spot during crack propagation ({\it 60}). 
Thus, we engineered the interface to enhance its strength and improve the toughness of microstructures. This was done by applying manually designed droplet patterns in {\it interface areas}. As illustrated in Fig.~\ref{fig:interface_engineering}, interface areas are located around shared edges or vertices of adjacent pixels, where one base material transitions into the other. There are three types of interface areas in total. Let $(i, j)$ be the $x$ and $y$ indices of a pixel in an entire specimen model. Type-1 areas connect two neighboring pixels $(i, j)$ and $(i + 1, j)$ along the $x$ direction, i.e., the direction of crack propagation. Type-2 areas connect two neighboring pixels $(i, j)$ and $(i, j + 1)$ along the $y$ direction, i.e., the pulling direction in tensile testing. Type-3 areas join all four surrounding pixels. Whenever an interface area covers two pixels of different material assignments, droplets in that area are specially arranged to approximate a 50/50 mixture of base materials. Similarly, let $(i', j')$ be the $x$ and $y$ indices of a droplet in the specimen model. Droplets appear in alternating rows in Type-1 and Type-3 areas, where each droplet $(i', j')$ follows the parity of $i + j'$. In contrast, droplets appear in alternating columns in Type-2 areas, each following the parity of $j + i'$. Such patterns remain constant in the $z$ direction.

In the context of voxel printing, a pixel may contain 3$\times$5, 3$\times$6, 4$\times$5, or 4$\times$6 droplets after quantization. Table~\ref{tab:pixel_dimensions} provides the quantized dimensions of a pixel’s internal area contingent on the pixel size. Quantized dimensions of interface areas are then derived from the pixels they intersect with. Type-1 and type-3 areas may contain 2$\times$2 or 2$\times$3 droplets, while type-2 areas vary among 1$\times$2, 1$\times$3, 2$\times$2, and 2$\times$3 droplets in size.

We compared our method of engineering the interface between base materials against two baseline approaches. By default, the 3D printer applies prescribed droplet configurations in interface areas that are invisible to the user. Conversely, the interface is completely user-specified in voxel printing mode. Therefore, it is possible to perform no engineering and make transitions at pixel boundaries as sharp as possible. To test these interface options, we randomly drew 5 samples from microstructures with $0.5 \leq \phi \leq 0.75$ and compared the mechanical properties of their printed specimens (Fig.~\ref{fig:interface_comp}).

\section{Computational Pipeline}

\subsection{Simulation}\label{sec:sim}

The simulator estimates the performance of microstructures without conducting physical measurements. Given an input microstructure pattern, it simulates the actual mechanical testing process on printed specimens using FEM. Young’s modulus and toughness are then computed from recorded stress-strain data. The two virtual testers inside the simulator use the same material model but differ in implementation details and parameter settings.

The nonlinear stress-strain response of the base materials is approximated using a 2D Neo-Hookean model. We followed the strain energy density function in Sifakis et al. ({\it 61}): 
\begin{linenomath}
\begin{ceqn}
\begin{equation}
    W = \frac{\mu}{2}\left(I_1 - 2 - \ln J\right) + \frac{\lambda}{2}\left(\ln J\right)^2,
\end{equation}
\end{ceqn}
\end{linenomath}
where $\mu$ and $\lambda$ are the Lam\'{e} parameters, $J = \mathrm{det}(F)$ is the determinant of the deformation gradient ($F$), and $I_1 = \mathrm{tr}(C)$ is the first invariant of the right Cauchy-Green deformation tensor ($C = F^\mathrm{T} F$). $\mu$ and $\lambda$ are derived from Young's modulus ($E$) and Poisson's ratio ($\nu$) as follows
\begin{linenomath}
\begin{ceqn}
\begin{align}
\begin{split}
    \mu &= \frac{E'}{2(1 + \nu')}, \\
    \lambda &= \frac{E'\nu'}{(1 + \nu')(1 - 2\nu')},
\end{split}
\end{align}
\end{ceqn}
\end{linenomath}
where $E'$ and $\nu'$ are converted from $E$ and $\nu$ as follows ({\it 62}) 
\begin{linenomath}
\begin{ceqn}
\begin{align}
\begin{split}
    \nu' &= \frac{\nu}{1 + \nu}, \\
    E' &= \frac{E(1 + 2\nu)}{(1 + \nu)^2}.
\end{split}
\end{align}
\end{ceqn}
\end{linenomath}
The Young’s moduli and Poisson’s ratios of the base materials were regarded as optimizable parameters in system identification (Section~\ref{sec:sid}).

\begin{figure}[ht]
    \centering
    \includegraphics[width=0.75\textwidth]{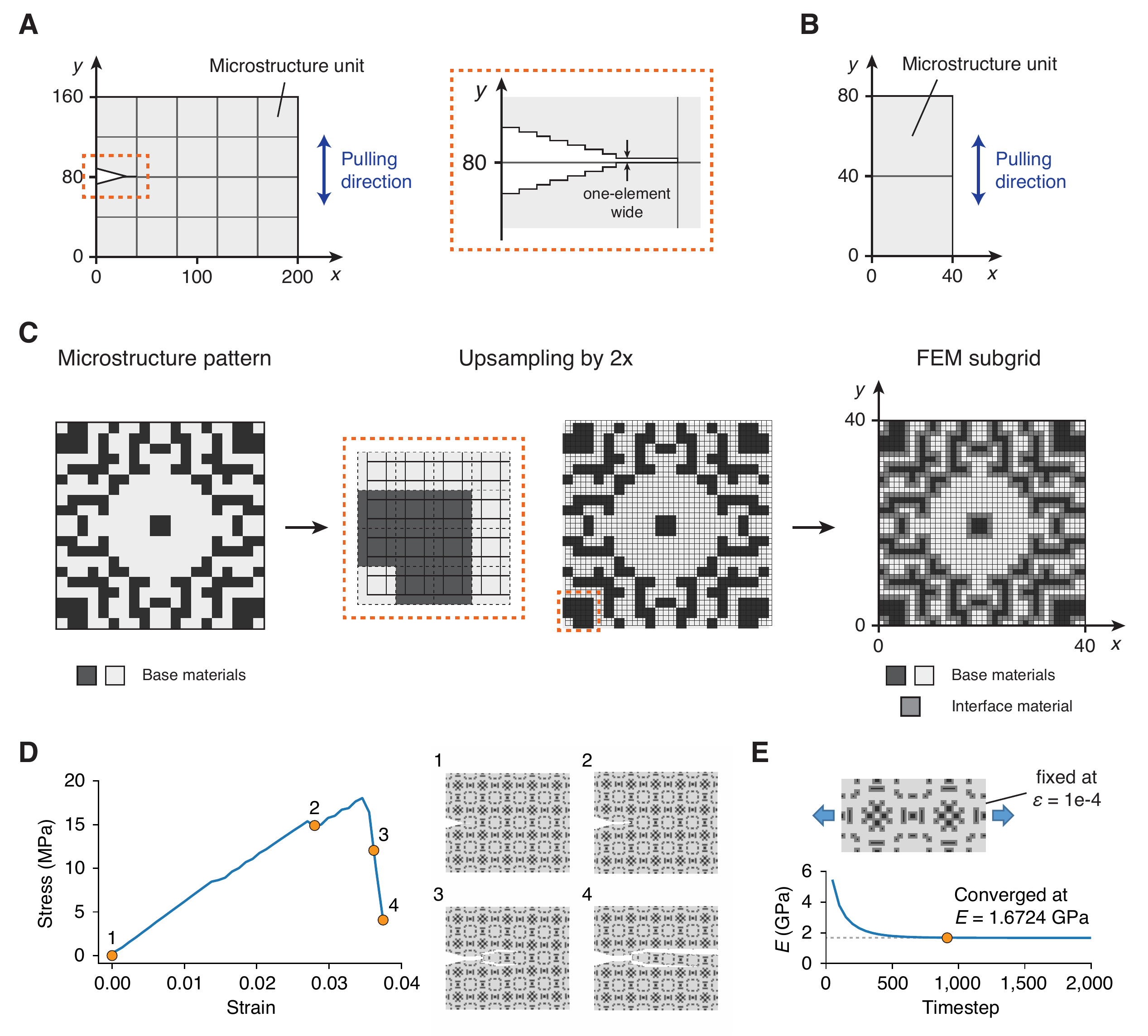}
    \caption{{\bf Configuration of the FEM simulator.} ({\bf A}) Dimensions of the FEM grid in the toughness virtual tester. ({\bf B}) Dimensions of the FEM grid in the Young’s modulus virtual tester. ({\bf C}) upsampling from a microstructure pattern to a 40$\times$40 FEM subgrid. ({\bf D}) An example simulation process of the toughness virtual tester. ({\bf E}) An example simulation process of the Young's modulus virtual tester.}
    \label{fig:simulation_config}
\end{figure}

The virtual tester for toughness operates on a regular grid of 200$\times$160 quadrilateral finite elements (Fig.~\ref{fig:simulation_config}A) where each node has four degrees of freedom. The grid contains 5$\times$4 tiles of microstructure units where each unit covers a 40$\times$40 sub-grid. Notably, a microstructure unit doubles the resolution of the pattern due to modeling the interface between base materials. The interface was modeled as a third homogeneous material, referred to as the {\it interface material}. Finite elements of the interface material and the base materials are equally sized. This is evidenced by our microscopic observation of printed specimens, which indicates that the width of the interface is typically around 100 $\mu$m and comparable to half of the pixel size. The computation of material assignment in a sub-grid is analogous to image upsampling. As shown in Fig.~\ref{fig:simulation_config}C, the FEM subgrid is superimposed on the original pattern where the center of the bottom-left element aligns with the center of the bottom-left pixel. An element inherits the corresponding pixel value if it is located entirely within a pixel; otherwise, the material assignment depends on the values of surrounding pixels in the same way as interface engineering. To mimic the pre-cracking of real specimens, a single-element-wide notch and a triangular cut were applied at the grid position of $(0, 80)$. The ``removed'' elements were assigned with a void material and do not generate nodal forces under deformation. Overall, this configuration correlates well with a printed specimen despite a simpler geometry to reduce computational cost. Dirichlet boundary conditions were enforced on the displacement in the $y$-direction during testing. Nodal forces in each element were computed from the Neo-Hookean model using four Gaussian quadrature points. Nodal velocities and displacements were then obtained from explicit time integration. Crack initiation and propagation were modeled on a per-element basis. To that end, strain energy densities were computed at the quadrature points together with nodal force contributions. An element was considered failed and reassigned with a void material once the maximum strain energy density at quadrature points exceeded a certain threshold specific to each material. Simulation terminated when the global strain went beyond 0.3 or the measured stress fell below 20\% of the ultimate tensile strength, with the only exception being the soft material for which the maximum strain was unlimited. In the end, toughness was calculated by integrating the stress-strain curve.

While it is possible to compute nodal velocities and displacements using an implicit solver, we chose an explicit solver since explicit solvers are generally preferred for nonlinear FEM simulations that involve dynamic failure. In particular, the virtual tester must operate at a sufficiently small time step (around $10^{-7}$ s) to precisely capture the crack propagation in microstructures and maintain numerical stability. Under such a strict requirement, explicit solvers are more efficient due to simpler implementation and better compatibility with many-core hardware accelerators like graphical computing units (GPU). However, we noticed that the virtual tester ran too slow at a strain rate that matched actual mechanical testing. In this case, simulating a microstructure till an assumed failure strain of 0.02 required 1.62$\times10^8$ time steps, taking almost one hour to finish on an NVIDIA Tesla V100 GPU under maximum throughput. The running time is impractical to our computational pipeline because the simulation budget will be too tight to allow for a proper exploration of the combinatorial design space. As a workaround for this challenge, we used a much larger strain rate (equal to 20) to speed up the simulation while introducing system-wide numerical damping to model material viscosity. Once the velocity field was updated in every time step, nodal velocities $v_{i,j}$ were smoothed by a 3$\times$3 filter defined as
\begin{linenomath}
\begin{ceqn}
\begin{equation}
    v_{i,j}' = (1 - \alpha)v_{i,j} + \frac{\alpha}{9} \sum_{k=-1}^1\sum_{l=-1}^1 v_{i+k,j+l},
\end{equation}
\end{ceqn}
\end{linenomath}
where $\alpha = \gamma \Delta t$ is the product of the time step $\Delta t$ and a constant damping coefficient $\gamma$. Under the assumption that all nodes have identical mass, this step effectively interpolates between the momentum of each node and the mean momentum in a 3$\times$3 neighborhood. A larger $\gamma$ leads to stronger damping effects and potentially accounts for more material viscoelasticity. Our observations suggested that larger $\gamma$ values work better with stiffer microstructures in suppressing oscillation. As there is no trivial way to determine the value of $\gamma$ for all microstructures, we optimized $\gamma$ in system identification instead.

The virtual tester for Young's modulus has a much smaller grid of 80$\times$40 quadrilateral elements, consisting of 2$\times$1 microstructure units (Fig.~\ref{fig:simulation_config}B). The material assignment to elements was computed using the same image upsampling method and no notch or triangular cut was added. This configuration qualitatively matches the gauge area of an ASTM D638 Type 1 specimen. A constant displacement boundary condition was enforced in the y-direction where the global strain remained fixed at $10^{-4}$. Nodal forces were computed from the aforementioned Neo-Hookean model. Explicit time integration was used to update velocities and displacements. Negative exponential damping was applied to the velocity field as follows
\begin{linenomath}
\begin{ceqn}
\begin{equation}
    v_{i,j}' = v_{i,j}e^{-\gamma \Delta t},
\end{equation}
\end{ceqn}
\end{linenomath}
where $\Delta t \approx 10^{-7}$ s was used as the time step and the damping coefficient $\gamma$ was optimized through system identification. The solver was allowed to reach a static equilibrium within $10^4$ time steps, after which Young’s modulus was derived from the gauge stress. Although an implicit linear elasticity solver is usually used to virtually measure Young's modulus, we directly reused the explicit solver in the toughness simulator for simplicity. Note that the Young's modulus virtual tester is not the bottleneck of the whole simulation.

In pursuit of maximal computation efficiency, our customized simulator was implemented using the Taichi programming language ({\it 53}), 
an emerging high-performance programming language tailored to computer graphics applications including physics-based simulation. Taichi compiles our source code written in Python syntax into optimized machine programs that, when executed, exploit the massive parallel computing power of GPUs. However, we found that simulating microstructures sequentially resulted in low GPU utilization due to inadequate workload. Our solution was adding a batching mechanism to enable concurrent simulation of multiple microstructures, where each microstructure can be evaluated using a unique set of optimizable parameters. This allows for maximum GPU usage regardless of its power, resulting in a much higher throughput.

\begin{figure}[!ht]
    \centering
    \includegraphics[width=0.7\textwidth]{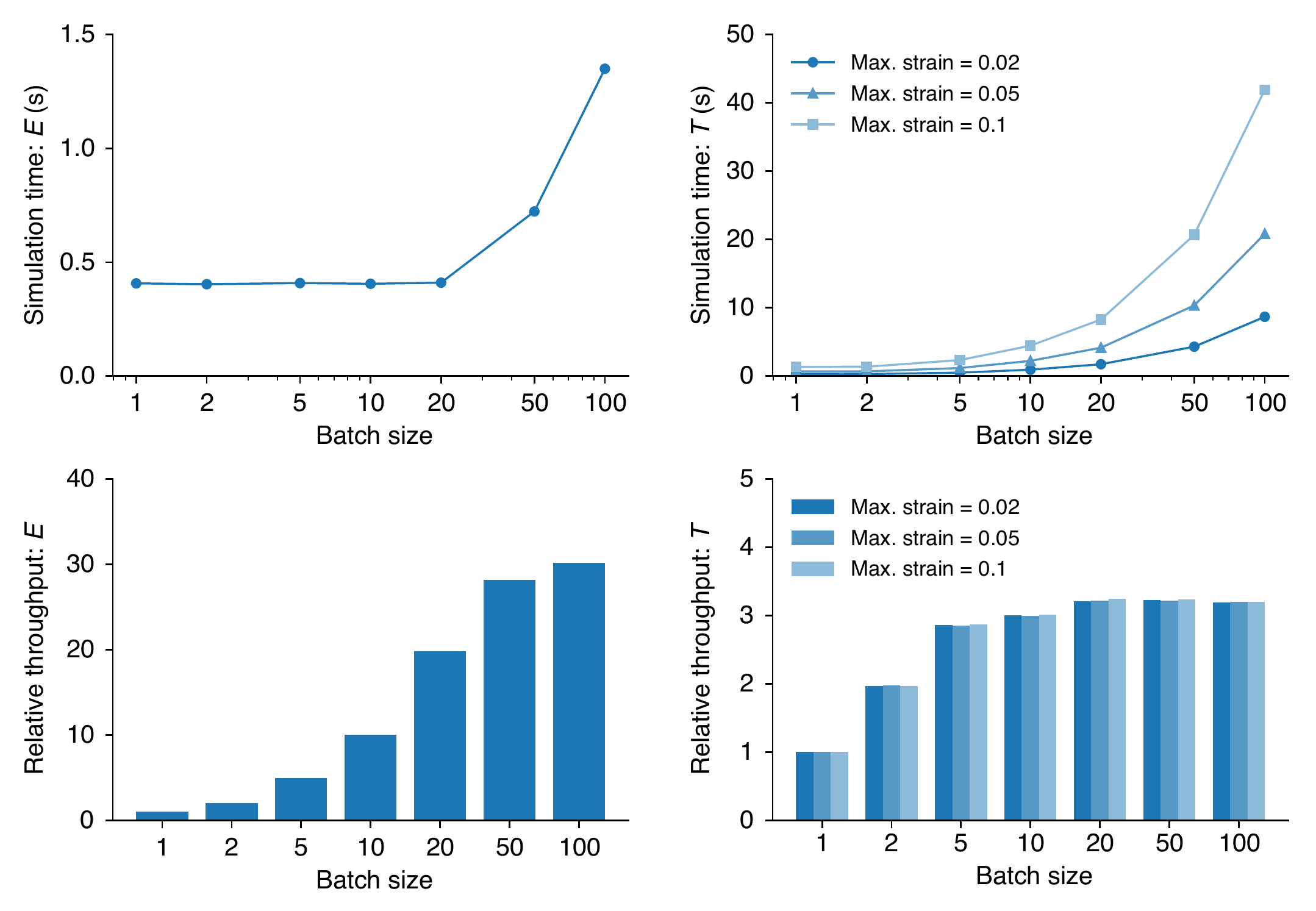}
    \caption{{\bf Time consumption and throughput of the simulator under various input batch sizes.} The column on the left contains results for Young's modulus and the column on the right shows results for toughness. The toughness virtual tester was evaluated under three strain limits: 0.02, 0.05, and 0.1.}
    \label{fig:simulation_time}
\end{figure}

We evaluated simulation speed by timing the simulator under different batch sizes and strain limits. All performance data was collected on a machine with 8 CPU cores and an NVIDIA Tesla V100 GPU averaged across 5 consecutive runs. Results are shown in Fig.~\ref{fig:simulation_time}. For Young's modulus, simulation only becomes slower when at least 20 microstructures are batched together. While increasing the batch size beyond this point leads to a much longer simulation time, the throughput continues to improve, implying that GPU utilization is still approaching its maximum. A batch size of 100 is required to reach a peak throughput of less than 0.015 s per sample. For toughness, the highest throughput occurs at a batch size of 20, where it takes less than 0.2 s to stretch a sample to $\varepsilon = 0.05$. Relative to the sequential setting, our batching mechanism yields a 30$\times$ boost in throughput for Young’s modulus and more than 3$\times$ for toughness regardless of the strain limit.

\subsection{System Identificaton}\label{sec:sid}

\begin{figure}[!ht]
    \centering
    \includegraphics[width=0.7\textwidth]{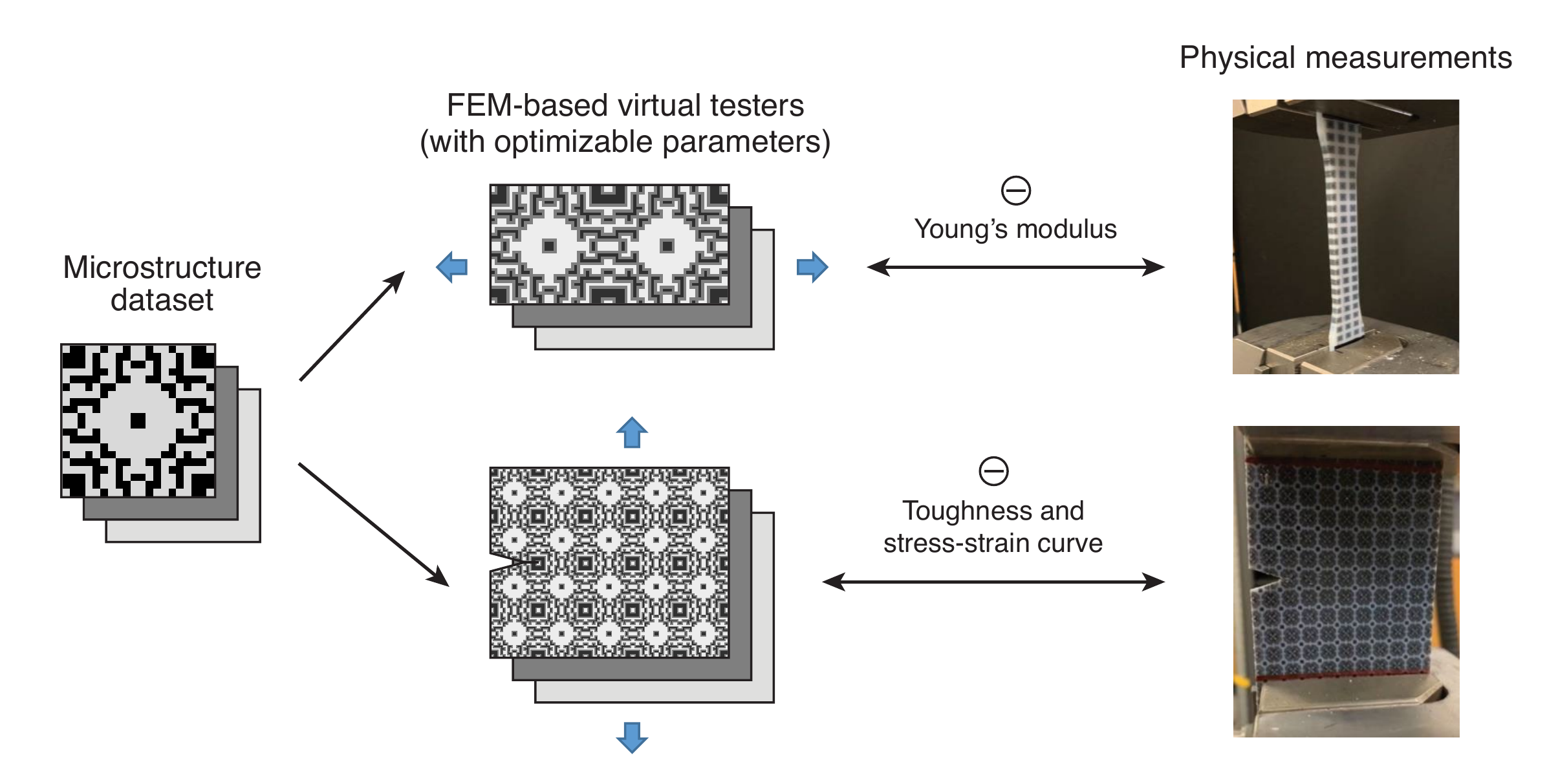}
    \caption{{\bf Schematics of the system identification problem.} Each virtual tester has a separate set of optimizable parameters. '$\ominus$' denotes error minimization via parameter optimization.}
    \label{fig:system_identification}
\end{figure}

System identification finds the optimal parameter setting of the simulator to match simulation results with physical measurements (Fig.~\ref{fig:system_identification}). We first define some mathematical notations before providing a formal description of the problem. For any microstructure $x$, let $f(x, \theta): X \times P \rightarrow \mathbb{R}$ denote the virtual tester for Young's modulus with a parameter setting $\theta$, where $X \subset \{0, 1\}^{20\times20}$ is the design space of microstructures and $P \subset \mathbb{R}^7$ is the feasible set of $\theta$. The estimated Young's modulus of $x$ is directly obtained from $\hat{E} = f(x, \theta)$. Similarly, let $g(x, \eta): X \times Q \rightarrow C[0, \infty)$ denote the virtual tester for toughness with a separate parameter setting $\eta$, where $Q \subset \mathbb{R}^{10}$ is the feasible set of $\eta$. The output $g$ is not a single number but a stress-strain curve $\hat{\sigma}(\varepsilon)$ defined as a continuous function over non-negative strains. Thus, the toughness of $x$ is estimated by computing the following integral
\begin{linenomath}
\begin{ceqn}
\begin{equation}
    \hat{T} = I[\hat{\sigma}] = \int_0^\infty \hat{\sigma}(\varepsilon)\,d\varepsilon.
\end{equation}
\end{ceqn}
\end{linenomath}
Both $\theta$ and $\eta$ are tuples of optimizable simulation parameters, including
\begin{itemize}
    \item Reference Young's moduli of base materials (denoted as TB and VW) and the interface material (denotes as IF): $E_\text{TB}, E_\text{VW}, E_\text{IF}$;
    \item Reference Poisson's ratios: $\nu_\text{TB}, \nu_\text{VW}, \nu_\text{IF}$;
    \item Failure thresholds of strain energy density: $W_\text{TB}, W_\text{VW}, W_\text{IF}$;
    \item A system-wide damping coefficient: $\gamma$.
\end{itemize}
Since fracture simulation is not involved in predicting Young's modulus, we have
\begin{linenomath}
\begin{ceqn}
\begin{align}
\begin{split}
    \theta &= (E_\text{TB}, E_\text{VW}, E_\text{IF}, \nu_\text{TB}, \nu_\text{VW}, \nu_\text{IF}, \gamma) \\
    \eta &= (E_\text{TB}, E_\text{VW}, E_\text{IF}, \nu_\text{TB}, \nu_\text{VW}, \nu_\text{IF}, W_\text{TB}, W_\text{VW}, W_\text{IF}, \gamma).
\end{split}
\end{align}
\end{ceqn}
\end{linenomath}

Based on optimizable parameters, the prediction errors of virtual testers were minimized over a group of microstructures $X_c = \left\{x_i\,|\,i = 1, 2, \cdots, N \right\}$, called the system identification dataset. The optimization process solved the following problems
\begin{linenomath}
\begin{ceqn}
\begin{align}
\begin{split}
    \theta^* &= \underset{\theta}{\operatorname{argmin}}\,\sum_{i=1}^N w_i \frac{\left|\hat{E}_i - E_i\right|}{E_i} \\
    \eta^* &= \underset{\eta}{\operatorname{argmin}}\,\sum_{i=1}^N w_i \frac{\left|\hat{T}_i - T_i\right| + \Tilde{\lambda} I\left[\left|\hat{\sigma}_i - \sigma_i\right|\right]}{T_i},
\end{split}
\end{align}
\end{ceqn}
\end{linenomath}
where $\hat{E}_i, \hat{T}_i, \hat{\sigma}_i$ are the predicted Young's modulus, toughness, and stress-strain curve of each microstructure $x_i$; $E_i, T_i, \sigma_i$ are the corresponding physical measurements. For Young's modulus, the objective is simply the average relative prediction error over the dataset. For toughness, the objective combines errors in both toughness values and stress-strain curves. Since stress-strain curves are functions, the error between two curves is calculated by integrating their absolute difference. Geometrically, this represents the symmetric difference of the areas under the curves. The two error terms are balanced by a multiplier $\Tilde{\lambda} = 1$.
\begin{figure}[!ht]
    \centering
    \includegraphics[width=0.9\textwidth]{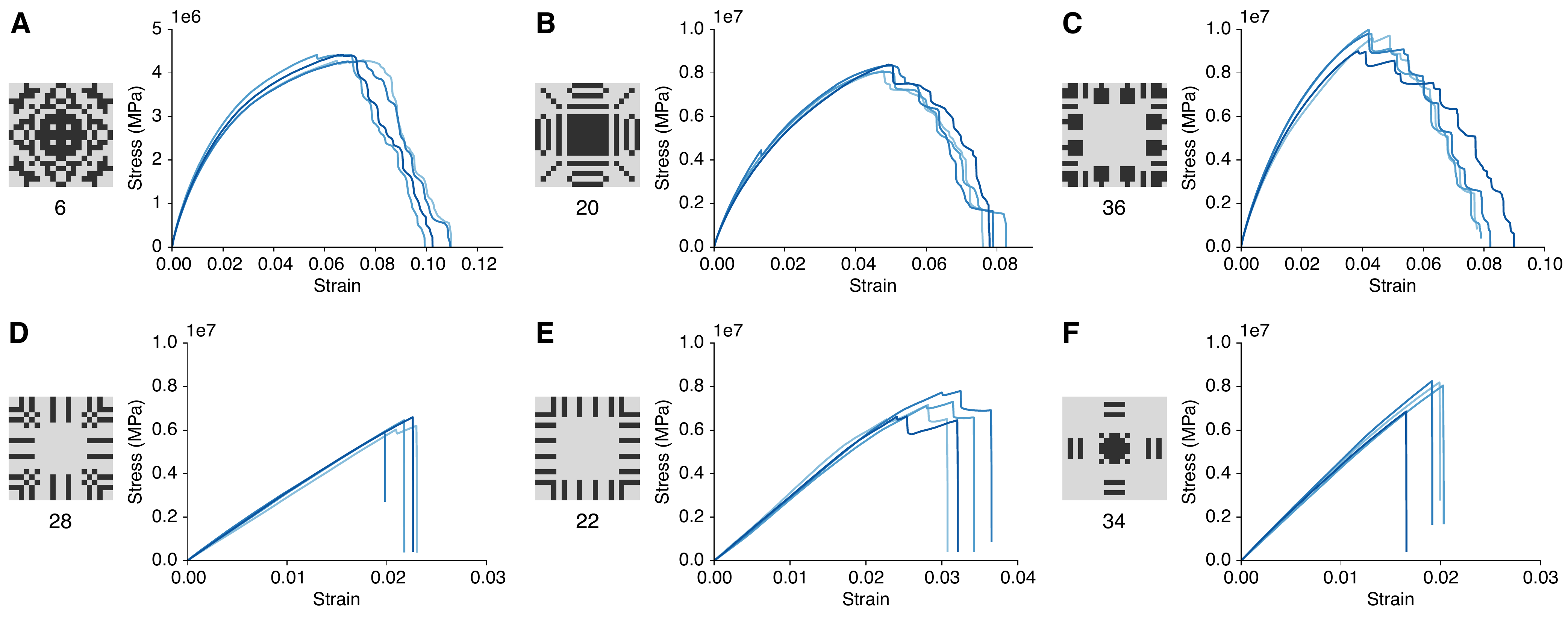}
    \caption{{\bf Measured stress-strain curves from the toughness test of six microstructures discovered by our pipeline.} The microstructures ({\bf A}-{\bf F}) are selected from random stratified sampling to cover a wide range of $\phi$. We showcase the stress-strain curves of all tested specimens (distinguished by color) for each design.}
    \label{fig:raw_data}
\end{figure}
\begin{figure}[!ht]
    \centering
    \includegraphics[width=0.75\textwidth]{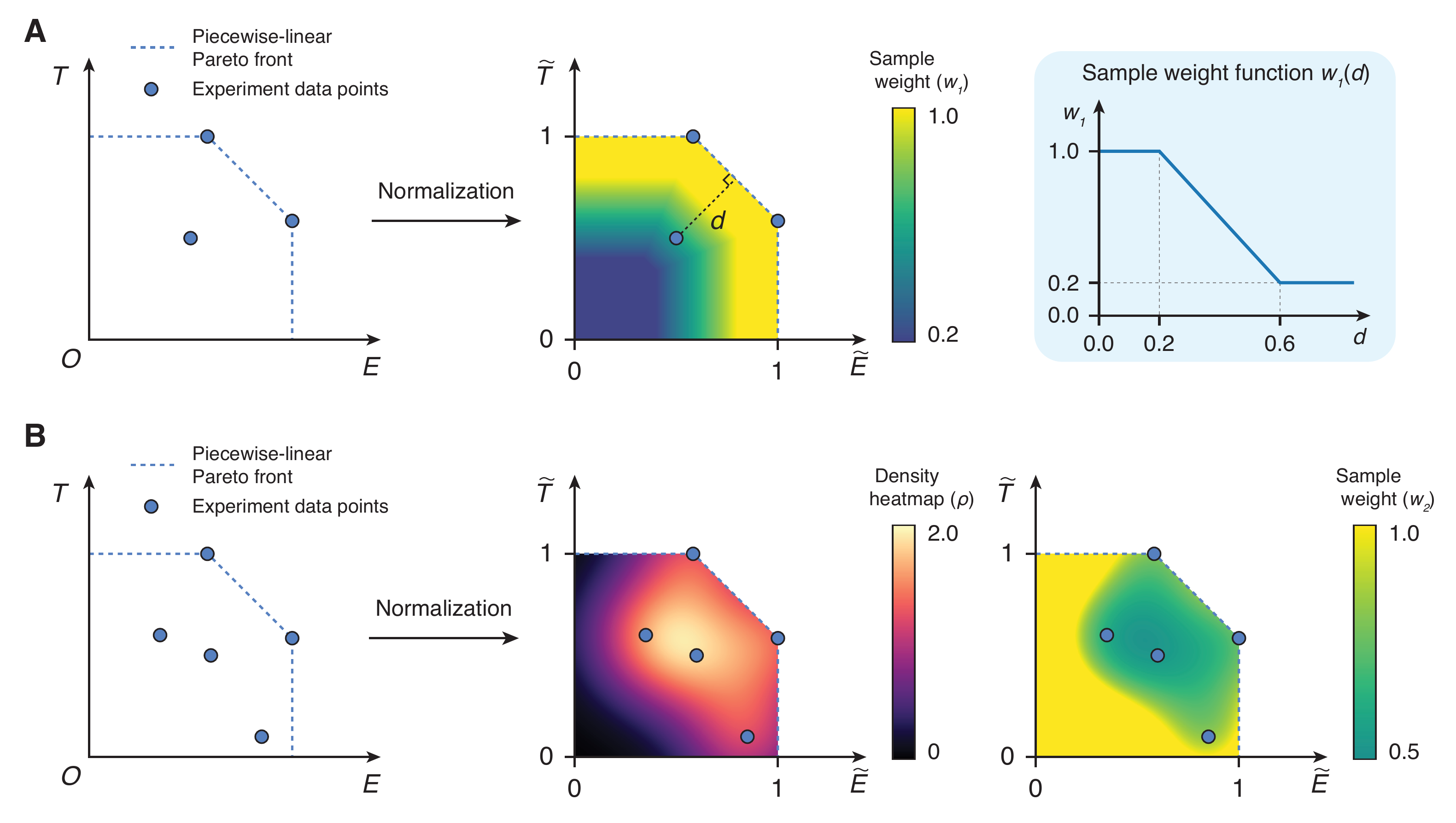}
    \caption{{\bf Calculation of sample weights in system identification.} ({\bf A}) An example of deriving the first sample weight component, $w_1$, in system identification using three data points. ({\bf B}) An example of deriving the second sample weight component, $w_2$, using five data points.}
    \label{fig:sid_weight}
\end{figure}

Additionally, as there are fewer optimizable parameters than microstructures in the dataset, we postulate that the simulation model might have an insufficient capacity to capture the variation of experiment data. Thus, each microstructure $x_i$ is weighted by $w_i \in [0, 1]$ to indicate priority. The weight is defined as the product of two components $w_i = w_{1i}w_{2i}$ that reflect two intuitive principles. First, microstructures with better mechanical properties are prioritized for accuracy since Pareto-optimal microstructures are our focus. To that end, the physical measurements $y = (E, T)$ of microstructures in the dataset were normalized into $\Tilde{y} = (\Tilde{E}, \Tilde{T})$ in a unit square, and a piecewise-linear Pareto front was computed in the normalized performance space. Then, the weight component $w_{1i}$ of microstructure $x_i$ was derived from the Euclidean distance between $\Tilde{y}_i$ and the Pareto front, denoted by $d_i$, via the following function (Fig.~\ref{fig:sid_weight}A)
\begin{linenomath}
\begin{ceqn}
\begin{equation}
    w_1(d) = \begin{cases}
        1 &\quad 0 \leq d < 0.2 \\
        1.4 - 2d &\quad 0.2 \leq d < 0.6 \\
        0.2 &\quad d \geq 0.6
    \end{cases}.
\end{equation}
\end{ceqn}
\end{linenomath}
Second, to prevent any bias from an uneven sample distribution in the performance space, we introduce $w_{2i}$ as a penalty term for excessively densely distributed microstructures. The distribution is estimated using a density heatmap $\rho(\Tilde{y})$ over the normalized performance space (Fig.~\ref{fig:sid_weight}B). The density heatmap is derived from a Gaussian mixture model
\begin{linenomath}
\begin{ceqn}
\begin{equation}
    \rho(\Tilde{y}) = \sum_{i=1}^N \mathrm{exp}(-\gamma_d \left\Vert\Tilde{y}_i - \Tilde{y}\right\Vert_2^2)
\end{equation}
\end{ceqn}
\end{linenomath}
where $\gamma_d$ is a parameter that controls the spread of each Gaussian function. Quantitatively, $\rho(\Tilde{y})$ estimates the number of microstructures in the vicinity of any position $\Tilde{y}$, and variates smoothly between adjacent samples. Then, $w_{2i}$ is defined as a weight discount for every sample $x_i$ whose density value exceeds a certain threshold, $\rho_0$.
\begin{linenomath}
\begin{ceqn}
\begin{equation}
    w_2(\Tilde{y}) = \text{min}\left\{\frac{\rho_0}{\rho(\Tilde{y})}, 1\right\}
\end{equation}
\end{ceqn}
\end{linenomath}
In this study, we found $\gamma_d = 200$ and $\rho_0 = 2$ to be a desirable parameter setting. Noteworthy, the design principles of sample weights generalize to other problem domains. Proper modifications can be made when our approach is repurposed to other computational design problems.

We used batched Bayesian Optimization (BO) as the optimizer for both $\theta$ and $\eta$. The BO algorithm outputs the best solution after 200 iterations with 3 evaluations per iteration, hence performing $600N$ simulations at most. The manually specified feasible ranges of simulation parameters are listed in Table~\ref{tab:simulation_parameter_ranges} and \ref{tab:simulation_parameter_ranges_2}. The algorithm was repeated with five different random seeds and the solution that had the smallest loss value was chosen. To harness the high throughput of the simulator, we implemented the algorithm using the GPyOpt library ({\it 63}) 
and batched all $3N$ simulations in a BO iteration into a single simulator run.

\begin{table}[!ht]
    \centering
    \begin{tabular}{c|OOc|OOc}
        \toprule
        \multirow{2}{*}{\textbf{Sample}} & \multicolumn{2}{c}{\textbf{$E$ (GPa)}} & \textbf{Err.} & \multicolumn{2}{c}{\textbf{$T$ (MPa)}} & \textbf{Err.} \\
        & Exp. & Sim. & \textbf{(\%)} & Exp. & Sim. & \textbf{(\%)} \\
        \midrule
        TB & 1.0094e-3 & 1.0091e-3 & 0.031 & 5.4335e-2 & 5.4271e-2 & 0.118 \\
        VW & 2.6613 & 2.6613 & 0.0002 & 7.4217e-2 & 7.4484e-2 & 0.360 \\
        IF & 0.5642 & 0.5642 & 0.003 & 0.8274 & 0.8193 & 0.984 \\
        \bottomrule
    \end{tabular}
    \caption{{\bf Simulation accuracy after system identification on each base material separately.} The accuracy is measured by relative errors (Err.) of simulation predictions (Sim.) from physical measurements (Exp.)}
    \label{tab:simulation_base_materials}
\end{table}

\begin{table}[!ht]
    \centering
    \begin{tabular}{cc|RRRR|QQRQ}
        \toprule
        \multirow{3}{*}{\textbf{Round}} & \multirow{3}{*}{$N$} & \multicolumn{4}{c|}{\textbf{Rel. error: $E$ (\%)}} & \multicolumn{4}{c}{\textbf{Rel. error: $T$ (\%)}} \\
        & & \multicolumn{2}{c}{\textbf{Reference}} & \multicolumn{2}{c|}{\textbf{After}} & \multicolumn{2}{c}{\textbf{Reference}} & \multicolumn{2}{c}{\textbf{After}} \\
        & & Avg. & Max. & Avg. & Max. & Avg. & Max. & Avg. & Max. \\
        \midrule
        1 & 18 & 31.09 & 93.15 & \textbf{6.53} & 22.95 & 135.59 & 767.44 & \textbf{22.85} & 62.23 \\
        2 & 26 & 25.19 & 93.15 & \textbf{6.29} & 20.40 & 108.26 & 767.44 & \textbf{23.92} & 100.53 \\
        3 & 34 & 21.11 & 93.15 & \textbf{6.32} & 22.75 & 93.09 & 767.44 & \textbf{24.68} & 84.30 \\
        4 & 42 & 18.95 & 93.15 & \textbf{6.50} & 25.40 & 82.38 & 767.44 & \textbf{27.10} & 88.87 \\
        \bottomrule
    \end{tabular}
    \caption{{\bf Simulation accuracy after system identification in each outer loop round since Round 1.} Simulation errors of the updated model after system identification are compared against the reference model (in Round 0). $N$ denotes the number of microstructures in the system identification dataset. Different from Fig.~3F, the error in each round is computed using microstructures in the dataset only.}
    \label{tab:simulation_errors}
\end{table}

In Round 0, we fitted the simulation to the base materials using system identification, obtaining a reference simulation model. It achieved almost perfect accuracy, where Young's modulus predictions are within 0.1\% off from measurements and toughness prediction errors are below 1\% (Table~\ref{tab:simulation_base_materials}). However, after the system identification in each round, the updated model had much smaller errors in predicting the performance of microstructures than the reference model (Table~\ref{tab:simulation_errors}). From a theoretical perspective, a possible explanation is that the localized mechanical performance of base materials might differ from bulk properties due to the microstructured placement. Thus, modeling the base materials in microstructures using physical measurements on bulk specimens could result in larger simulation errors.

For a complete reference, Table~\ref{tab:simulation_e} and \ref{tab:simulation_t} list the physical measurements, simulation predictions, and relative errors of all microstructures discovered by our computational approach in every round of the outer loop.

\subsection{NMO - the Inner Loop}\label{sec:inner-loop}

Microstructures with optimal trade-offs between simulated Young's modulus and toughness were explored following the system identification step. Here, a multi-objective optimization (MOO) problem was solved in the microstructure design space. As pixel-wise material assignments are discrete variables, the MOO problem can be tackled using genetic algorithms like NSGA-II ({\it 31}). 
However, a naive NSGA-II implementation often suffers from low sample efficiency attributed to uninformative random mutations and crossovers. Alternatively, topology optimization (TO) methods ({\it 16, 28, 64, 65}) 
formulate the MOO problem as continuous optimization where the base material properties of finite elements are parameterized by continuous material assignment variables. While the continuous formulation allows for gradient derivation with respect to material assignments, TO algorithms are prone to local optima and do not easily scale up to multiple objectives. Particularly, if a full Pareto front is desired rather than a single Pareto-optimal solution, TO has to harness many starting points and weighted combinations of objectives to properly explore the design space. Another potential option is multi-objective Bayesian optimization (MOBO) which performs a more guided exploration of the design space ({\it 29, 30, 66}) 
than NSGA-II.
Most MOBO algorithms leverage Gaussian-process-based (GP-based) surrogate models that make fast and approximative performance predictions, thus greatly reducing the number of simulations required. However, the surrogate models are heavily constrained by the Gaussian prior and can struggle in approximating sophisticated discrete mappings. Furthermore, since the computational cost of fitting Gaussian processes is cubic to the number of data points, the dataset is often limited to a few hundred in practice, which is far from adequate to model the complex relationship between microstructure patterns and toughness.

Inspired by MOBO, we developed a neural network-accelerated multi-objective optimization algorithm (NMO) that is specially tailored to the design space of microstructure patterns, harnessing the impressive capabilities of deep learning. The NMO workflow is illustrated in Fig.~2a. In the beginning, a set of 10 microstructures were randomly sampled from the design space and simulated to obtain their performance, constituting an initial dataset. In every iteration, the dataset was used to train the predictor, which acts as a surrogate model for the simulator. Ten predicted Pareto-optimal samples were then computed by solving a surrogate multi-objective optimization problem where simulation was replaced by the surrogate model. As the surrogate model is much cheaper to evaluate, we applied NSGA-II to solve the surrogate problem, which ran for 200 generations at a population size of 100. Then, the predicted best microstructures were verified in simulation before being added to the dataset for the next iteration. The algorithm lasted for 500 iterations as dictated by a pre-specified budget of 5,000 simulations. Due to its iterative nature, NMO transforms into the inner loop of our pipeline, where simulation data and proposed designs alternately flow between the simulator and the predictor.

\begin{figure}[ht]
    \centering
    \includegraphics[width=\textwidth]{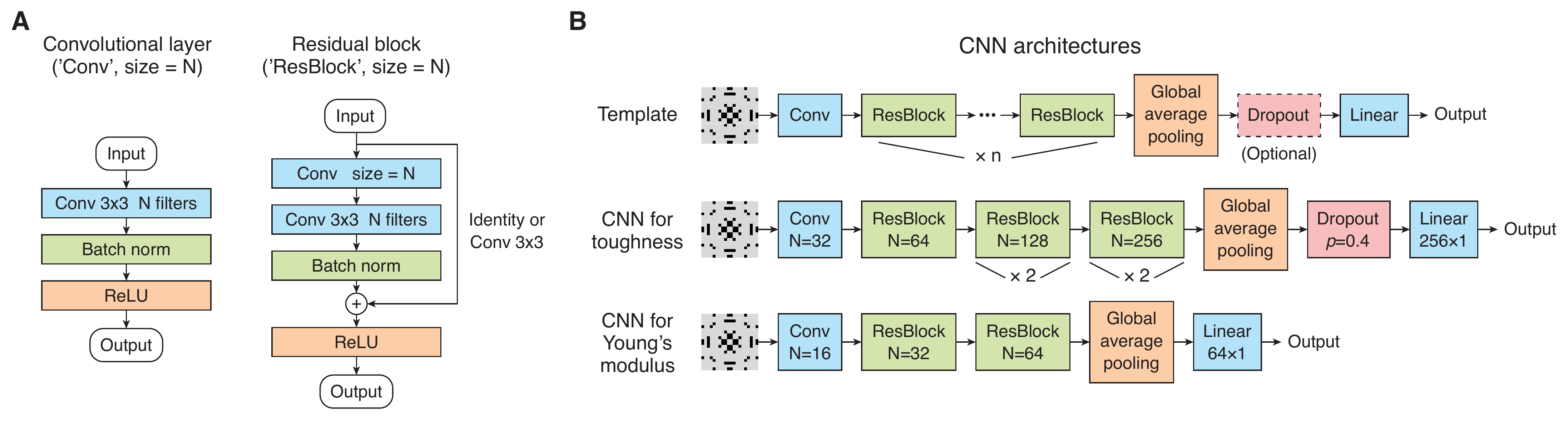}
    \caption{{\bf CNN architectures used in the predictor.} ({\bf A}) Definitions of the architectural units of CNNs. ({\bf B}) Prediction networks of Young's modulus and toughness constructed from an architecture template.}
    \label{fig:nn_architecture}
\end{figure}

NMO has several major improvements over MOBO. Most importantly, the surrogate model is based on convolutional neural networks (CNN), which specializes in image-based reasoning, instead of Gaussian processes. We model toughness and Young’s modulus using two separate residual networks ({\it 54}). 
Both networks were constructed from the same architecture template demonstrated in Fig.~\ref{fig:nn_architecture}. The template is a concatenation of one stem convolution layer, one or more residual blocks, a global average pooling layer, an optional dropout layer, and a fully connected layer. The network for toughness has a relatively larger capacity, with 5 residual blocks and a penultimate dropout layer for regularization. The network for Young’s modulus is much smaller, with only 2 residual blocks. The networks were progressively trained on the dataset of microstructures discovered by the algorithm, where an Adam optimizer ({\it 67}) 
was used at a learning rate of 1e-3. The network for toughness underwent 300 training epochs whereas the network for Young’s modulus went through 100 epochs. The maximum batch size was 2,048. To alleviate overfitting on small datasets, the networks were penalized by weight decay coefficients of 1e-2 and 1e-6, respectively. \edit{On a side note, the user may employ more complex network architectures, e.g., Vision Transformers (ViTs) ({\it 68}), 
to accelerate the inner loop when applying our computational pipeline to alternative design problems. Our work uses CNNs as they typically perform better than ViTs in data-efficient scenarios and entail minimal expert knowledge of advanced neural network models for microstructure patterns.}

The next improvement is a customized mutation operator in NSGA-II specially designed for microstructure patterns. Normally, NSGA-II strikes a balance between exploitation and exploration of the design space using a mutation operator and a crossover operator. In our case, we omitted the crossover step but augmented the mutation operator to enforce exploration. The customized mutation operator carried out one of the following options probabilistically.
\begin{itemize}
    \item With $p_1 = 0.86$, flip a random pixel;
    \item With $p_2 = 0.11$, rewrite a rasterized line of pixels with a unified random value of 0 or 1, where the starting and ending positions are randomly chosen;
    \item With $p_3 = 0.03$, select a random pixel as the top-left corner, then rewrite a square region of pixels (the size of the square is randomly designated) with the complement of the top-left pixel.
\end{itemize}
After that, symmetry was restored by mirroring the changes across the entire pattern. Note that the mutated pattern might violate the structural constraints on the design space. Thus, potential violations were resolved through the procedure below
\begin{enumerate}
    \item If the resulting $\phi$ falls below 0.5 after mutation, the pattern is flipped, and the four corner pixels are reassigned with the stiff material.
    \item If other violations still exist, the operator continues applying new mutation attempts without undoing any changes until the violation is cleared.
\end{enumerate}
Our implementation of the mutation operator takes into account that the toughness of a microstructure is directly affected by structural features in the pattern, such as pixels, beams, and chunks of the soft material. It balances between exploitation and exploration by modifying these structural features at multiple scales probabilistically.

The last improvement is a deduplication mechanism using hash tables. It prevents the algorithm from rediscovering previous samples in later iterations and allows the algorithm to escape local optima during exploration. Normally, single-pixel mutations create limited variation and easily result in duplicate patterns. With a hash table, however, the mutation operator automatically switches to larger perturbations once the hash table entries of neighboring patterns are filled. The exploration then continues in other regions of the design space. In our definition, the hash value of a microstructure pattern (viewed as a flattened array of binaries $x = (b_1, b_2, \cdots, b_{400})$) was computed as
\begin{linenomath}
\begin{ceqn}
\begin{equation}
    \text{hash}(x) = \left(\sum_{i=1}^{400}b_i k^{i-1}\right) \text{mod}\;m
\end{equation}
\end{ceqn}
\end{linenomath}
where $k = 99,997$ and $m = 100,000,007$ are prime numbers. Each hash table had $m$ entries. NMO had a global hash table, while each run of the surrogate problem solver used a separate table. After the modified NSGA-II algorithm finished, the predicted Pareto-optimal microstructures were ordered by non-dominated sorting and checked for hash collision inside the global hash table. The first 10 collision-free samples were validated by the simulator.

The exploration efficiency of NMO was evaluated using Pareto hypervolume growth within a budget of 5,000 simulations. Our modified NSGA-II algorithm was also evaluated. Compared with NMO, it still benefits from the custom mutation operator and the hash table albeit not having any surrogate model. For reference, we included a baseline NSGA-II implementation that performs single-pixel mutations and one-point crossovers at a rate of 0.1 and 0.8, respectively. Any crossover or mutation attempt that violated design constraints was reverted before trying the next one. Hence, the NSGA-II baseline is generally more conservative in exploration.

For TO, we chose bidirectional evolutionary structure optimization (BESO) ({\it 26, 27}) 
and solid isotropic material with penalization (SIMP) ({\it 28}) 
as representatives since previous works have tackled tough composite design using both strategies ({\it 69, 70}). 
Considering that they only optimize a single objective by definition, we extended them to finding the Pareto front by randomly sampling many weighted combinations of normalized Young's modulus and toughness as meta-objectives. To enforce exploration, we run BESO using multiple random initial solutions and set multiple random target volume fractions for SIMP. This way, both algorithms balance exploitation and exploration without leveraging any expert knowledge of Young's modulus or toughness, making a fair comparison with our method. Implementation-wise, we augmented the FEM simulator to support gradient calculation through auto-differentiation ({\it 71}). 
We adopted the extended BESO method from ({\it 72}) 
and solved SIMP using method of moving asymptotes (MMA). For MOBO algorithms, we evaluated DGEMO ({\it 29}) 
and TSEMO ({\it 30}) 
as baselines. DGEMO has state-of-the-art performance in several synthetic MOO benchmarks, while TSEMO is more commonly adopted.

Here, we additionally introduce the detailed optimization setting for TO-SIMP, where the MMA solver maximizes a weighted combination of normalized Young's modulus and toughness. An active constraint is imposed on the volume fraction of the rigid base material ($\phi$). Below we provide the optimization statement in mathematical notations. Let $\rho \in [0, 1]^{20\times 20}$ denote the density field of a microstructure unit. SIMP defines the element-wise Young's modulus $E_e$ as
\begin{linenomath}
\begin{ceqn}
\begin{equation}
E_e(\rho_e) = \rho_e^p E_\text{rigid} + (1 - \rho_e^p)E_\text{soft}
\end{equation}
\end{ceqn}
\end{linenomath}
where $\rho_e$ refers to the density of the element $e$; $p$ denotes the SIMP penalization exponent; $E_\text{rigid}$ and $E_\text{soft}$ represent the Young's moduli of rigid and soft base materials. Let $\Tilde{E}(\rho)$ be the normalized Young's modulus given by the Young's modulus simulator, $\Tilde{T}(\rho)$ be the normalized toughness given by the toughness simulator. The following problem is solved using MMA:
\begin{linenomath}
\begin{ceqn}
\begin{equation}
\begin{aligned}
\max_{\rho} \quad & w_1 \Tilde{E}(\rho) + w_2 \Tilde{T}(\rho) \\
s.t. \quad & \int_{\Omega} \rho\,d\Omega \le \phi_\text{max} \\
& \rho\text{ obeys }p4m\text{ symmetry}
\end{aligned}
\end{equation}
\end{ceqn}
\end{linenomath}
Since the optimizer only converges to one microstructure design at a time, we repeat this process using randomly sampled $w_1$, $w_2$, and $\phi_\text{max}$ to explore the design space in different directions. We use a constant SIMP penalization $p = 8$ in all cases without continuation.

Regarding the evaluation methodology, ten microstructures were simulated per iteration for all algorithms except TO which does not break down to generations. NMO and NSGA-II were run for 500 iterations. MOBO algorithms were capped by 200 iterations for an acceptable time consumption as fitting Gaussian processes became extremely slow with hundreds of samples. For TO algorithms, we paired random meta-objectives and initial designs (or target volume fractions) into optimization tasks and allowed the algorithm to execute as many optimization tasks as possible before exhausting the evaluation budget. All algorithms were repeated on the same five random seeds, which led to identical starting samples. The resulting curves of Pareto hypervolume growth were averaged.

\begin{figure}[ht]
    \centering
    \includegraphics[width=0.6\textwidth]{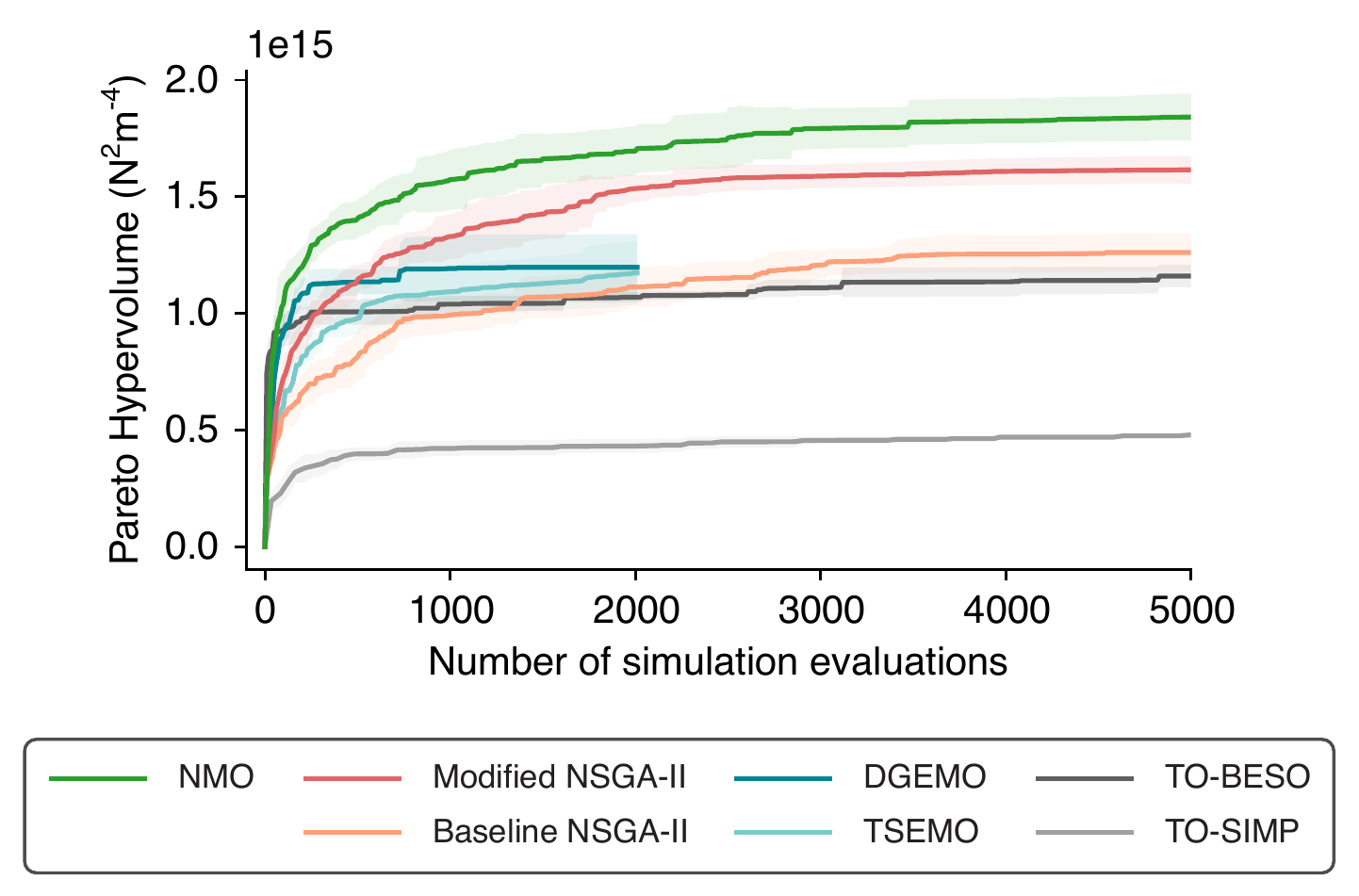}
    \caption{{\bf Pareto hypervolume growth of the gamut using various optimization algorithms in Round 0.} Candidates are NMO, NSGA-II, TO, and MOBO algorithms. Solid curves are averages. Colored regions indicate standard deviations.}
    \label{fig:comp_exploration_hv}
\end{figure}

\begin{table}[!ht]
    \centering
    \begin{tabular}{c|OQc|c|OQc}
        \toprule
        \multirow{2}{*}{\textbf{Round}} & \multicolumn{2}{c}{\textbf{Pareto hypervolume}} & \textbf{Improv.} & \multirow{2}{*}{$V_\text{target}$} & \multicolumn{2}{c}{\textbf{\#Samples till }$V_\text{target}$} & \textbf{Improv.} \\
        & NSGA-II & Ours & \textbf{(\%)} & & NSGA-II & Ours & \textbf{($\times$)} \\
        \midrule
        0 & 1.6154 & \textbf{1.8412} & 13.98 & 1.7202 & 20000 & \textbf{2200} & 9.10 \\
        1 & 1.6012 & \textbf{1.8835} & 17.63 & 1.8225 & 20000 & \textbf{4190} & 4.78 \\
        2 & 0.9956 & \textbf{1.0904} & 9.52 & 1.0776 & 19760 & \textbf{3240} & 6.10 \\
        3 & 1.0406 & \textbf{1.2016} & 15.47 & 1.0911 & 19950 & \textbf{1830} & 10.90 \\
        4 & 0.9457 & \textbf{1.0078} & 6.57 & 1.0078 & 17140 & \textbf{4990} & 3.44 \\
        \bottomrule
    \end{tabular}
    \caption{{\bf Comparison in exploration efficiency between NMO and our modified NSGA-II in each outer loop round.} Exploration efficiency is measured by: (1) the discovered Pareto hypervolume within 5,000 simulations (unit: $10^{15}\,\text{N}^2\text{m}^{-4}$); (2) the number of simulated samples before reaching a target Pareto hypervolume, denoted by $V_\text{target}$. $V_\text{target}$ is equal to the Pareto hypervolume of NSGA-II at 20,000 simulations by default. If NMO does not reach that hypervolume within 5,000 simulations, we change $V_\text{target}$ into NMO's hypervolume at 5,000 simulations instead.}
    \label{tab:comp_exploration_efficiency}
\end{table}

Fig.~\ref{fig:comp_exploration_hv} demonstrates the Pareto hypervolume growth of evaluated algorithms in Round 0. NMO secured victory with the fastest growth and the largest final hypervolume. The modified NSGA-II algorithm also pulled ahead of baseline NSGA-II considerably in hypervolume growth. In contrast, the TO variants failed to cover a decent Pareto hypervolume despite completing dozens of optimization tasks. They not only suffered from local optima but had to consume a majority of the budget on intermediate solutions along optimization trajectories, which crippled their sample efficiency. Both MOBO algorithms severely underperformed as well due to the aforementioned weaknesses of Gaussian processes. Furthermore, even when compared against our modified NSGA-II, NMO took up to an order of magnitude fewer simulations to reach the same Pareto hypervolume (Table~\ref{tab:comp_exploration_efficiency}). These results showcase NMO's evidently superior sample efficiency over other state-of-the-art structural optimization or generic MOO algorithms. In addition, we would like to note that NMO is compatible with any MOO algorithm when solving the surrogate problem. It is recommended to choose an algorithm that adequately fits the design task for ideal exploration efficiency in practice.

\begin{figure}[!ht]
    \centering
    \includegraphics[width=0.75\textwidth]{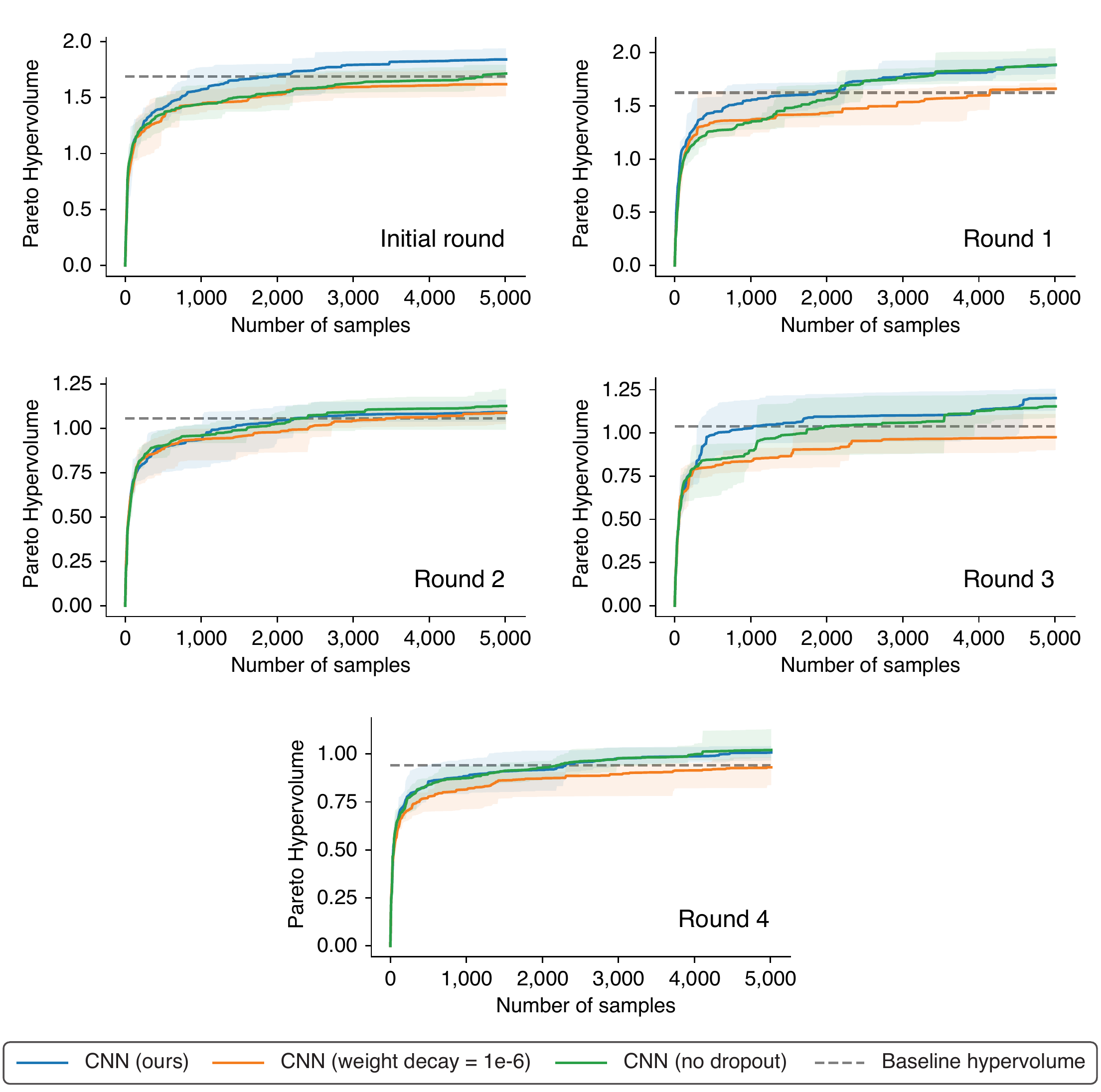}
    \caption{{\bf Impact of network regularization on the Pareto hypervolume growth of NMO in all outer loop rounds.} "Ours" means our hyperparameter setting including a large weight decay of 1e-2 plus a penultimate dropout layer. The unit of the hypervolume is $10^{15}\,\text{N}^2\text{m}^{-4}$.}
    \label{fig:cnn_tuning}
\end{figure}

By tuning the CNNs' hyperparameters, we noticed that network regularization plays an important role in accelerating hypervolume growth. In this experiment, our CNNs were compared against two less regularized variants, one using a smaller weight decay (1e-6) and the other excluding dropout before the fully-connected layer. As shown in Fig.~\ref{fig:cnn_tuning}, our networks outperformed less regularized versions and led to better hypervolume growth in most test cases. While the no-dropout variant had negligible wins in Rounds 2 and 4, the small-weight-decay variant kept trailing behind. Considering the relatively small amount of training data, a reasonable explanation is that regularization prevents the networks from overfitting and improves generalization to unseen samples. Thus, properly regularized networks can be better at discovering high-performing samples in simulation.

\begin{figure}[!ht]
    \centering
    \includegraphics[width=0.65\textwidth]{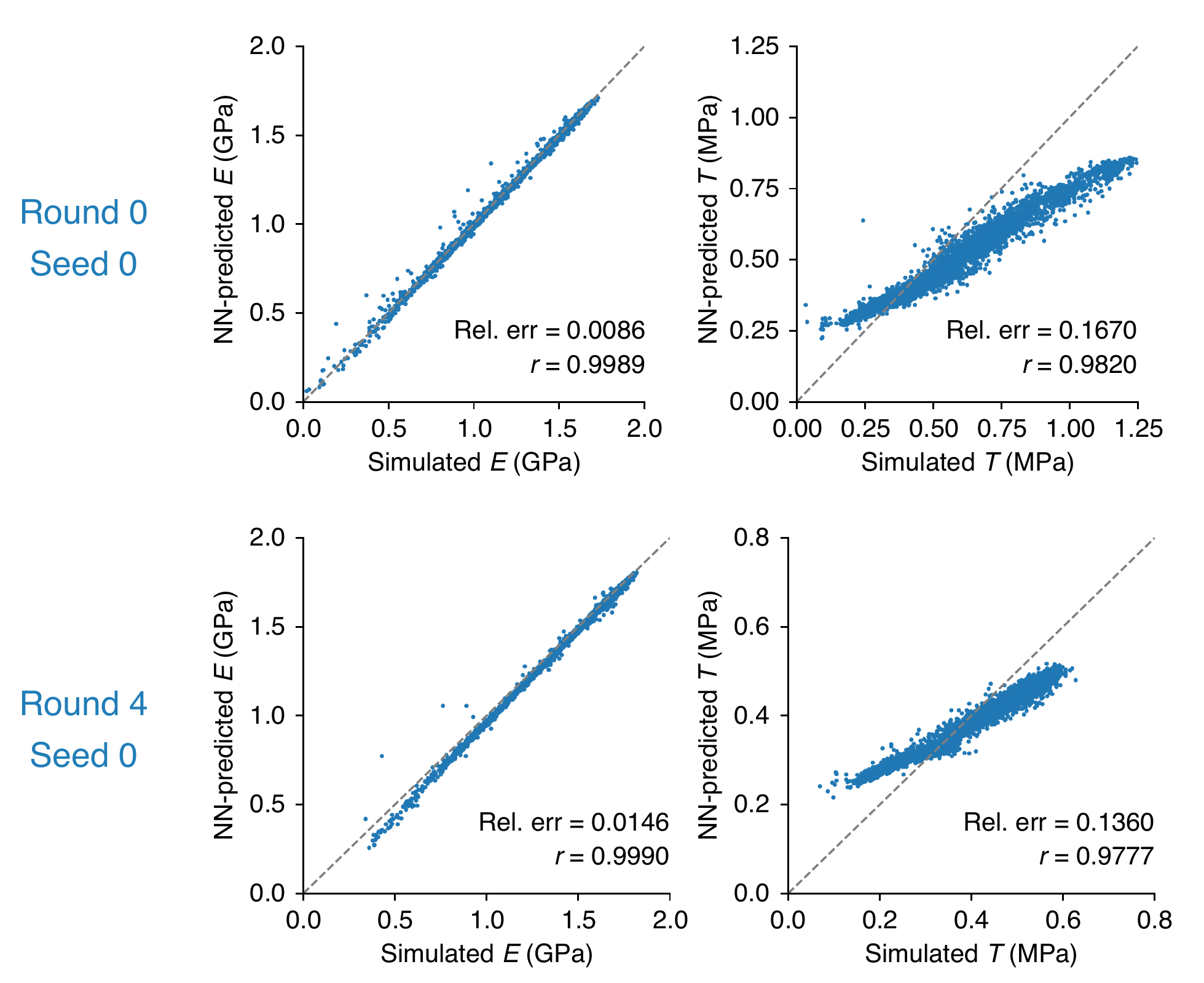}
    \caption{{\bf CNN prediction accuracy in the last iteration of NMO.} The examples shown are Round 0 and Round 4 with a random seed of 0. Simulation results are regarded as the ground truth. $r$ means correlation coefficient.}
    \label{fig:nn_accuracy}
\end{figure}

As an additional test, we evaluated the prediction accuracy of CNNs on training datasets in the last iteration of NMO (Fig.~\ref{fig:nn_accuracy}). The network for Young's modulus achieved very high accuracy with an average relative error of less than 2\%. The network for toughness had larger errors in comparison, at 16.7\% and 13.6\%, respectively. Nonetheless, perfect accuracy is unnecessary in this case since the microstructures for simulation validation in the modified NSGA-II were selected by Pareto dominance rather than performance values. Therefore, it is sufficient to predict correct property rankings, which is indicated by the correlation coefficient ($r$) between predictions and the ground truth. Here, our network for toughness managed to get $r > 0.97$ and demonstrated much faster hypervolume growth than other evolutionary algorithms. As a side note, searching for the best network architecture or hyperparameter setting is beyond the scope of this work and thereby left for future research.

\subsection{Selection of Validation Samples}\label{sec:selection}

\begin{figure}[!ht]
    \centering
    \includegraphics[width=0.75\textwidth]{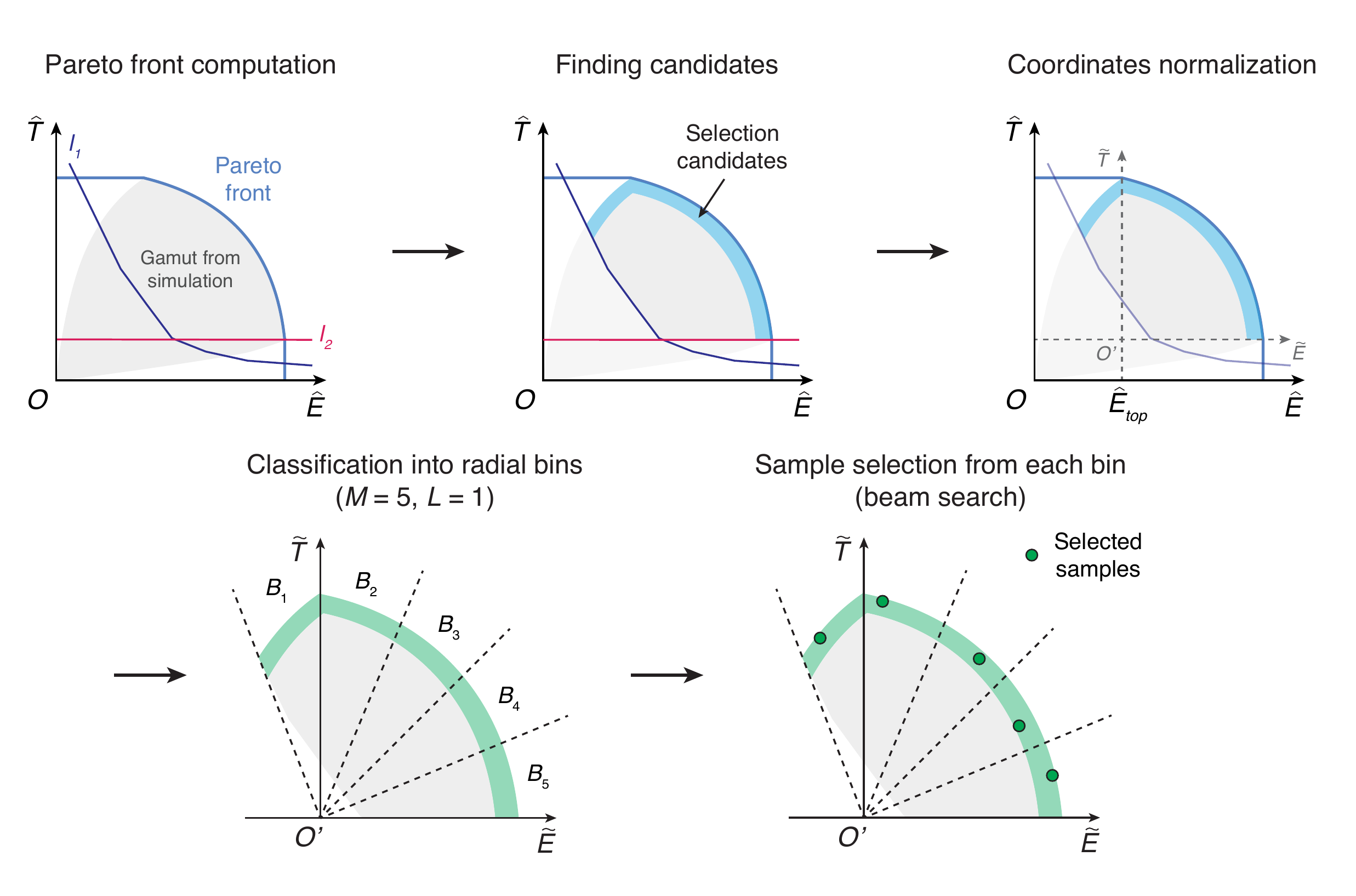}
    \caption{{\bf The selection process of validation microstructures near the Pareto front.} $M = 5$ and $L = 1$ are used in this illustration for simplicity.}
    \label{fig:sample_selection}
\end{figure}

The simulated gamut discovered by NMO typically contains a large pool of microstructures and a few samples are selected for physical validation due to a limited experiment budget. Overall, the selected samples should cover a wide performance range and be distinguished from existing samples in appearance. To fulfill these requirements, we propose a two-stage selection approach visualized in Fig.~\ref{fig:sample_selection}.

The first stage extracted samples closest to the gamut boundary. This step can be done in various ways depending on the shape of the gamut. We employed a simple method where samples in the gamut were ranked in the order of Pareto dominance. By definition, Pareto-optimal samples of the entire gamut are in the 1st rank, and the $i$-th rank contains microstructures on the Pareto front of a partial gamut where those in higher ranks are removed. We regarded samples up to the 3rd rank as near the boundary. Meanwhile, the gamut was intersected by two benchmark lines $l_1$ and $l_2$. $l_1$ is a polyline that connects the physical measurements of homogeneous composites. $l_2$ is a horizontal line going through the sample with the largest Young’s modulus in simulation. Samples that were both near the boundary and above the benchmark lines were deemed as candidates for selection.

In the next stage, the candidate samples, $S \subset X$, were classified into two groups of radial bins $B_1, B_2, \cdots, B_L$ and $B_{L+1}, B_{L+2}, \cdots, B_M$. $M$ denotes the total number of samples to select, and $L$ denotes the number of samples to select from the first group and that $0 \leq L < M$. To that end, we mapped the properties of samples in $S$ to a normalized performance space in $[-1, 1] \times [0, 1]$. While both Young's modulus and toughness went through linear mappings, there are two circumstances for Young's modulus specifically. As shown in Fig.~\ref{fig:sample_selection}, the Young's modulus of the toughest sample $\hat{E}_{top}$ is mapped to zero by default. However, the resulting image will not fit into $[-1, 1]$ if $\hat{E}_{top}$ exceeds the midpoint of the Young's modulus range, hence the midpoint is mapped to zero in that case. Within the normalized performance space, the radial bins are outlined by $M + 1$ rays shooting from the origin, dubbed as $r_1, r_2, \cdots, r_{M+1}$. $r_{L+1}$ and $r_{M+1}$ always align with the coordinate axes regardless of $L$. If $L > 0$, $r_1$ passes through the sample in $S$ that has the largest orientation angle from the positive horizontal axis, $\varphi_{max}$. The $i$-th radial bin $B_i$ refers to the area between $r_i$ and $r_{i+1}$, and we let $S_i$ denote the subset of $S$ classified into $B_i$. $B_1, B_2, \cdots, B_L$ are equally sized and located in the second quadrant, while the other bins are equally sized and located in the first quadrant. The value of $L$ is determined by $\varphi_{max}$ using the rules below
\begin{linenomath}
\begin{ceqn}
\begin{equation}
    L = \begin{cases}
        0 &\quad \varphi_{max} < \dfrac{\pi}{2} + \varphi_{th} \\
        \left\lceil\left(1 - \dfrac{\pi}{2\varphi_{max}}\right)M\right\rceil &\quad \varphi_{max} \geq \dfrac{\pi}{2} + \varphi_{th}
    \end{cases},
\end{equation}
\end{ceqn}
\end{linenomath}
where $\varphi_{th} = \pi / 36$ (i.e., $5^\circ$). Practically, when $\varphi_{max}$ is at least $95^\circ$, $L$ is a minimum number such that the bins in the second quadrant are no larger than those in the first quadrant.

The following combinatorial optimization problem was subsequently solved to search for a group of microstructure patterns that maximally differentiate from physically measured ones. Let $R$ denote the set of microstructures that have been discovered by the computational pipeline, and $\Omega = (x_1, x_2, \cdots, x_M)$ (where $x_i \in S_i$ for $i = 1, 2, \cdots, M$) denote a possible selection. The objective is to maximize the average L1 distance of each $x_i$ to the rest of the samples in $R \cup \Omega$.
\begin{linenomath}
\begin{ceqn}
\begin{equation}
    \Omega^* = \underset{\Omega}{\operatorname{argmax}}\,\frac{1}{M} \sum_{i=1}^M \min_{u \in R \cup \Omega \backslash \{x_i\}} \left\|x_i - u\right\|_1.
\end{equation}
\end{ceqn}
\end{linenomath}
As the problem is likely intractable by enumeration when $S$ contains over 100 samples, we used a beam search algorithm to find an approximate solution. The beam size was set to 100 to preserve the quality of the solution at an acceptable computational cost.

\subsection{The Outer Loop}\label{sec:outer-loop}

\begin{table}[ht]
    \centering
    \begin{tabular}{c|c}
        \toprule
        \textbf{$\phi$ interval} & \textbf{Patterns} \\
        \midrule
        $0.5 \leq \phi < 0.6$ & \quad\includegraphics[width=0.225\textwidth]{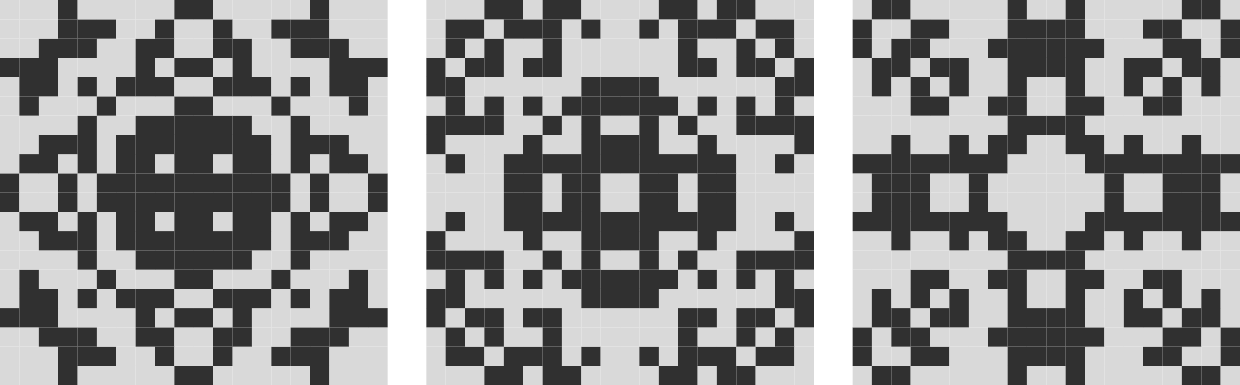}\quad \\
        \midrule
        $0.6 \leq \phi < 0.7$ & \quad\includegraphics[width=0.225\textwidth]{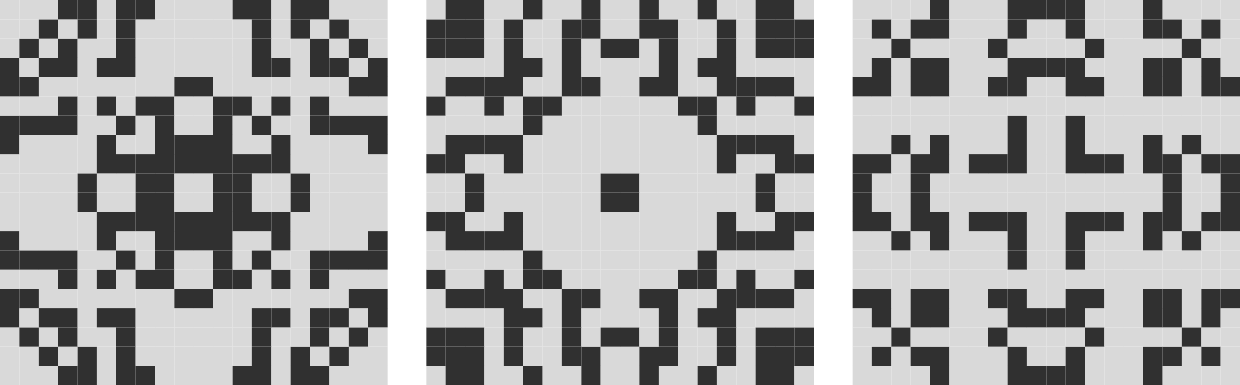}\quad \\
        \midrule
        $0.7 \leq \phi < 0.8$ & \quad\includegraphics[width=0.15\textwidth]{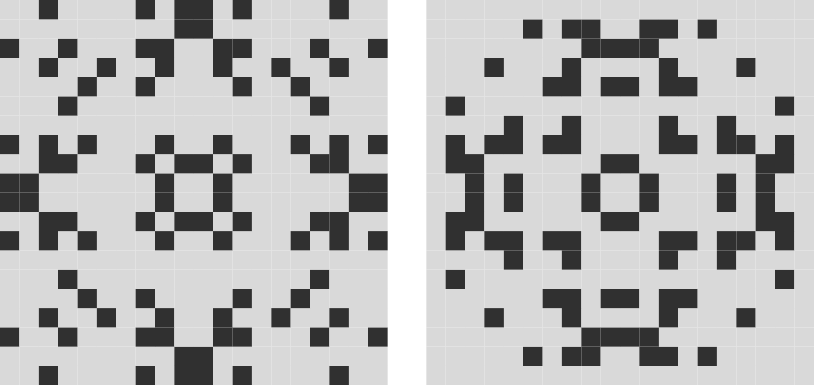}\quad \\
        \midrule
        $0.8 \leq \phi \leq 0.85$ & \quad\includegraphics[width=0.15\textwidth]{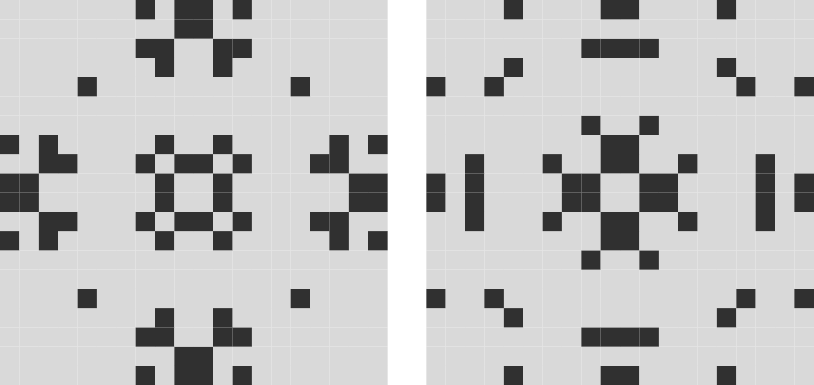}\quad \\
        \bottomrule
    \end{tabular}
    \caption{{\bf Random microstructure designs in the system identification dataset categorized by $\phi$.}}
    \label{tab:random_microstructures}
\end{table}

\begin{figure}[!ht]
    \centering
    \includegraphics[width=0.9\textwidth]{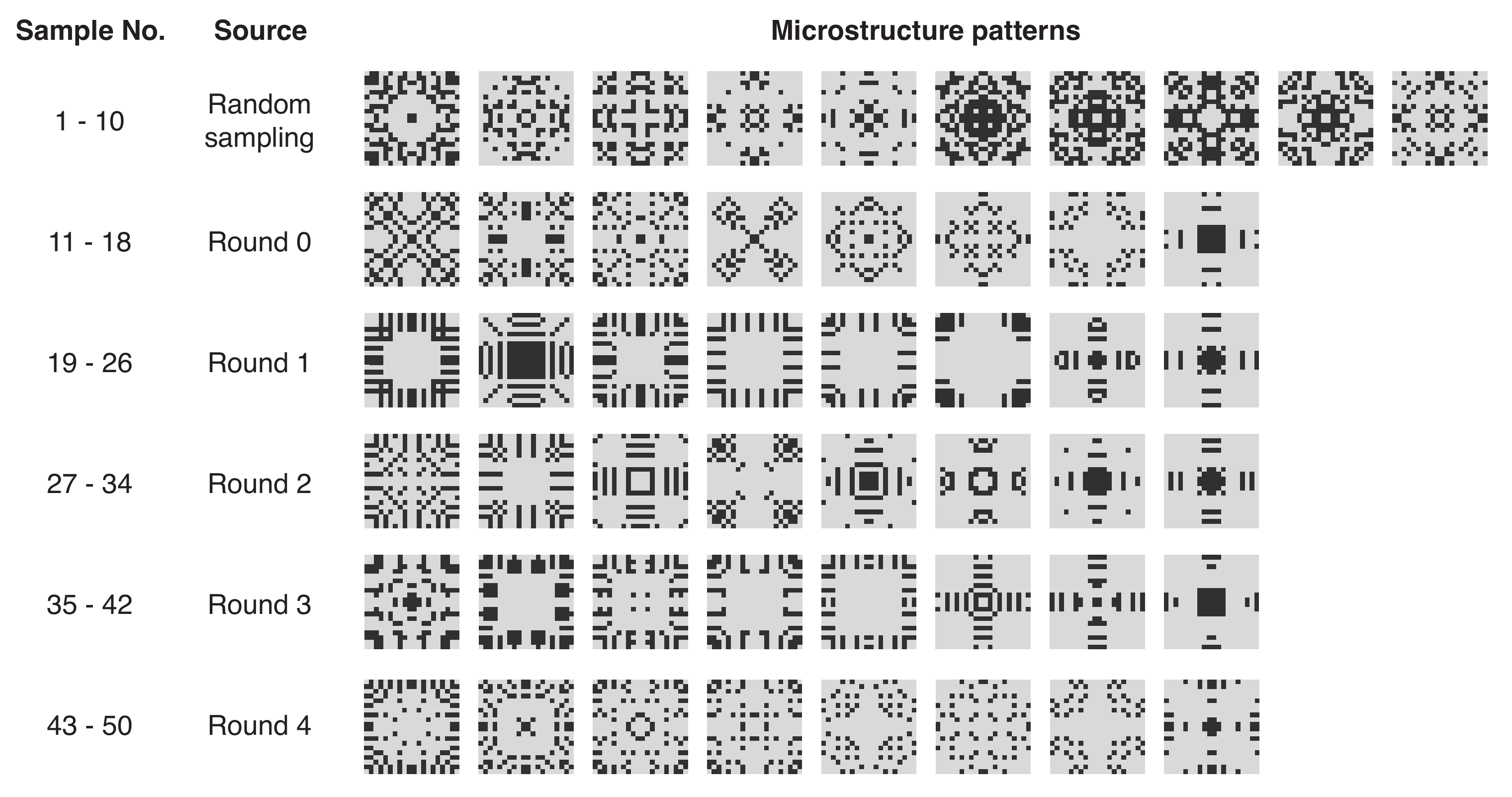}
    \caption{{\bf Microstructures discovered by the computational pipeline.} The first row comprises 10 random patterns in the system identification dataset. The following rows are patterns proposed in outer loop rounds.}
    \label{fig:microstructures}
\end{figure}

\begin{table}[!ht]
    \centering
    \begin{tabular}{cc|PP|QQ|c}
        \toprule
        \multirow{3}{*}{\textbf{Sample}} & \multirow{3}{*}{$\phi$} & \multicolumn{2}{c|}{\multirow{2}{*}{\textbf{Properties}}} & \multicolumn{2}{c|}{\textbf{Improvement}} & \textbf{Improv.} \\
        & & & & \multicolumn{2}{c|}{\textbf{(homogeneous)}} & \textbf{(base)} \\
        & & $E$ (GPa) & $T$ (MPa) & $E$ ($\times$) & $T$ ($\times$) & $T$ ($\times$) \\
        \midrule
        1 & 0.79 & 1.402 & 0.491 & 1.35 & 2.00 & \textbf{7.01} \\
        2 & 0.85 & 1.810 & 0.473 & \textbf{1.61} & \textbf{2.76} & 6.75 \\
        3 & 0.85 & 1.764 & 0.458 & 1.54 & 2.48 & 6.53 \\
        4 & 0.84 & 1.790 & 0.355 & 1.27 & 1.76 & 5.07 \\
        5 & 0.85 & 1.807 & 0.304 & 1.28 & 1.59 & 4.33 \\
        \bottomrule
    \end{tabular}
    \caption{{\bf Performance of five best microstructures discovered by our approach.} Their improvements are calculated relative to homogeneous composites and base materials. For homogeneous composites, the improvement in one property is calculated assuming that the other property is identical. The largest improvement factors are marked in bold.}
    \label{tab:best_microstructures}
\end{table}

The outer loop was held between the mechanical tester and the simulator as follows. In Round 0, the mechanical tester conducted physical measurements on the base materials (the rigid, soft, plus the interface material). Our system then matched simulation and experiment using a system identification. Each material was involved in a separate system identification process that only optimized simulation parameters pertaining to that material, including Young’s modulus, Poisson’s ratio, energy density threshold, and damping coefficient. This yielded three optimized values of the damping coefficient where the median was chosen. Using the calibrated simulator, our system ran the inner loop to find microstructures with the optimal trade-offs between Young's modulus and toughness in simulation. As previously noted, we used the modified NSGA-II algorithm to compute the simulation Pareto front and left the NMO algorithm to a separate comparison with other optimization algorithms. The NSGA-II was run in five different random seeds under a simulation budget of 2$\times10^4$ samples. The resulting gamuts were merged into one gamut with approximately $10^5$ samples, after which 8 validation microstructures were selected near the Pareto front and transferred back to the mechanical tester for validation.

Starting from Round 1, our system maintained a dataset of physically measured microstructures, starting from 18 initial designs. Besides the proposed designs from Round 0, the dataset also had 10 random microstructures drawn from stratified sampling according to $\phi$ (Table~\ref{tab:random_microstructures}). Microstructure patterns in each interval of $\phi$ were generated by randomly specifying a $\phi$ value and filling a zero matrix with a corresponding number of 1's. These random microstructures help regularize system identification, especially in early rounds of the outer loop when the dataset is small. As shown in Fig.~3b, the mechanical tester provided the simulator with physical measurements of all microstructures in the dataset to improve its accuracy through system identification. Different from Round 0, system identification was conducted using experiment data from microstructures exclusively. After that, the simulator computed its gamut and proposed 8 Pareto-optimal designs back to the mechanical tester for validation. The proposed designs were then added to the microstructure dataset. Thus, the system identification dataset had $8n + 18$ microstructures at the end of Round $n$ ($n \ge 0$). The outer loop stopped at $n = 4$.

Microstructured composites discovered during the outer loop (Fig.~\ref{fig:microstructures}) were compared against homogeneous composites. Among the microstructures that showed better performance trade-offs than homogeneous counterparts, we picked five best samples near the Pareto front and with more than 1.5$\times$ improvement in toughness from the benchmark line (Table~\ref{tab:best_microstructures}). Sample 2 has the maximum improvement in both Young's modulus and toughness at 1.61$\times$ and 2.76$\times$, respectively. Samples 1 and 3 also show more than 2$\times$ improvement in toughness. These results sufficiently prove that computationally designed microstructured composites present much better trade-offs between Young's modulus and toughness than homogeneous composites. We also calculated the improvement of the best samples over base materials in toughness (VW+ has a higher toughness of 0.070 MPa than TB+), whereas Sample 1 has 7.01$\times$ higher toughness than VW+.

\section{Analysis on Pareto-Optimal Microstructures}

\subsection{Discovery of Microstructure Families}

Microstructure families were discovered from microstructures with near-optimal trade-offs between Young's modulus and toughness after the nested-loop pipeline. Here, we grouped 11 near Pareto-optimal microstructures into 4 families by patterns and mechanical performance (Fig.~4b). They were referred to as seed microstructures and leveraged to generate pattern variations in each family. Then, we validated the discovered families by confirming that the generated microstructures had similar performance to seed microstructures. The procedure is detailed as follows.

\begin{table}[!ht]
    \centering
    \begin{tabular}{cc|RRRR|RRRR}
        \toprule
        \multirow{3}{*}{\textbf{Family}} & \multirow{3}{*}{$N$} & \multicolumn{4}{c|}{\textbf{Rel. error: $E$ (\%)}} & \multicolumn{4}{c}{\textbf{Rel. error: $T$ (\%)}} \\
        & & \multicolumn{2}{c}{\textbf{R4}} & \multicolumn{2}{c|}{\textbf{FS}} & \multicolumn{2}{c}{\textbf{R4}} & \multicolumn{2}{c}{\textbf{FS}} \\
        & & Avg. & Max. & Avg. & Max. & Avg. & Max. & Avg. & Max. \\
        \midrule
        1 & 3 & 10.77 & 19.98 & \textbf{6.25} & 12.56 & 40.17 & 48.25 & \textbf{4.55} & 8.28 \\
        2 & 4 & 2.99 & 5.45 & \textbf{1.22} & 3.45 & 41.61 & 56.81 & \textbf{11.32} & 21.29 \\
        3 & 2 & 7.69 & 9.64 & \textbf{2.13} & 4.22 & 47.42 & 62.99 & \textbf{2.69} & 5.35 \\
        4 & 2 & 3.96 & 5.76 & \textbf{2.16} & 4.29 & 33.03 & 43.64 & \textbf{0.45} & 0.73 \\
        \bottomrule
    \end{tabular}
    \caption{{\bf Prediction errors of the simulator after system identification in each microstructure family.} The family-specific simulators (FS) are compared with the simulator in Round 4 (R4) using the performance of seed microstructures in each family. $N$ denotes the number of samples. Better values are marked in bold. Physical measurements and corresponding predictions are provided in Table~\ref{tab:simulation_family_e} and \ref{tab:simulation_family_t}.}
    \label{tab:simulation_error_family}
\end{table}

\begin{figure}[!ht]
    \centering
    \includegraphics[width=0.82\textwidth]{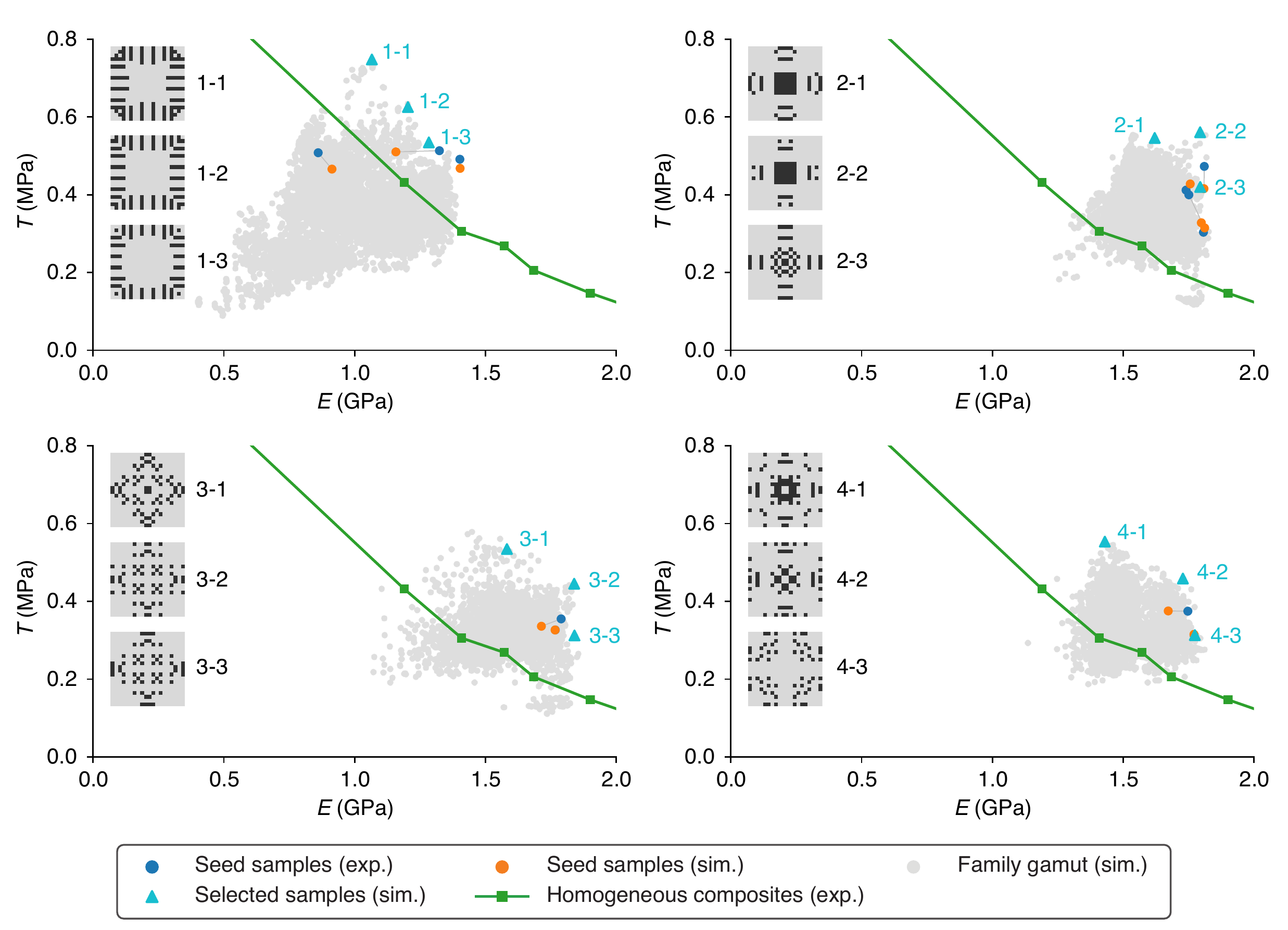}
    \caption{{\bf Gamut of each microstructure family after dense sampling along with selected samples near the Pareto front for validation.} Each family contains around 7,000 to 9,000 samples. 'Exp.' is short for physical measurements and 'sim.' for simulation predictions. The measurements and predictions of a seed microstructure are connected to indicate error.}
    \label{fig:families_gamut}
\end{figure}

To start with, we ran system identification in each family to obtain a family-specific simulator. The system identification involved all near Pareto-optimal samples but assigned them with different weights. Samples in the family had a constant weight of 1, while others were weighted by 0.1 for regularization purposes. Such family-specific system identification leads to a better fitting quality within the family, as illustrated in Table~\ref{tab:simulation_error_family}. Since only seed microstructures and other Pareto-optimal microstructures are used, there is a closer match between simulation predictions and physical measurements than the simulator in Round 4. The contrast in prediction error is particularly obvious in toughness predictions where the family-specific (FS) simulator has an order of magnitude smaller relative errors in three families.

To generate variations in microstructure patterns, we performed dense evolutionary sampling around the seed microstructures. Similar to NSGA-II, the sampling algorithm began with the seed microstructures in the family and generated other microstructures iteratively using our modified mutation operator. However, there was no tournament selection of parents or competition among the population, and any mutation attempt was simply rejected if it led to a violation of design constraints. In every iteration, the algorithm kept generating mutated microstructures until 10 non-existing ones were found, after which they were evaluated using the family-specific simulator. The mutated microstructures were restricted within a maximum of 16-pixel difference in pattern (which is 4\% of all pixels) from the closest seed microstructure. The hash table in Section~\ref{sec:inner-loop} was adopted for deduplication. As the hash table entries were rapidly filled, it became increasingly difficult to generate new microstructures over time. Thus, the algorithm was limited by a maximum number of $3\times10^5$ proposed microstructures per iteration and terminated if it failed to complete an iteration within the limit. Our algorithm generated local gamuts for all four families as shown in Fig.~\ref{fig:families_gamut}.

Finally, we conducted physical measurements on three microstructures in the local gamut with near-Pareto-optimal predicted performance. The samples were selected using the algorithm in Section~\ref{sec:selection} (Fig.~\ref{fig:families_gamut}). Their experiment results revealed comparable performance to the seed microstructures (Table~\ref{tab:simulation_validation}), hence verifying that the performance similarity among microstructures in the same family was preserved after dense sampling.

\subsection{Dimensionality Reduction in Subfamilies}

\begin{figure}[!ht]
    \centering
    \includegraphics[width=0.8\textwidth]{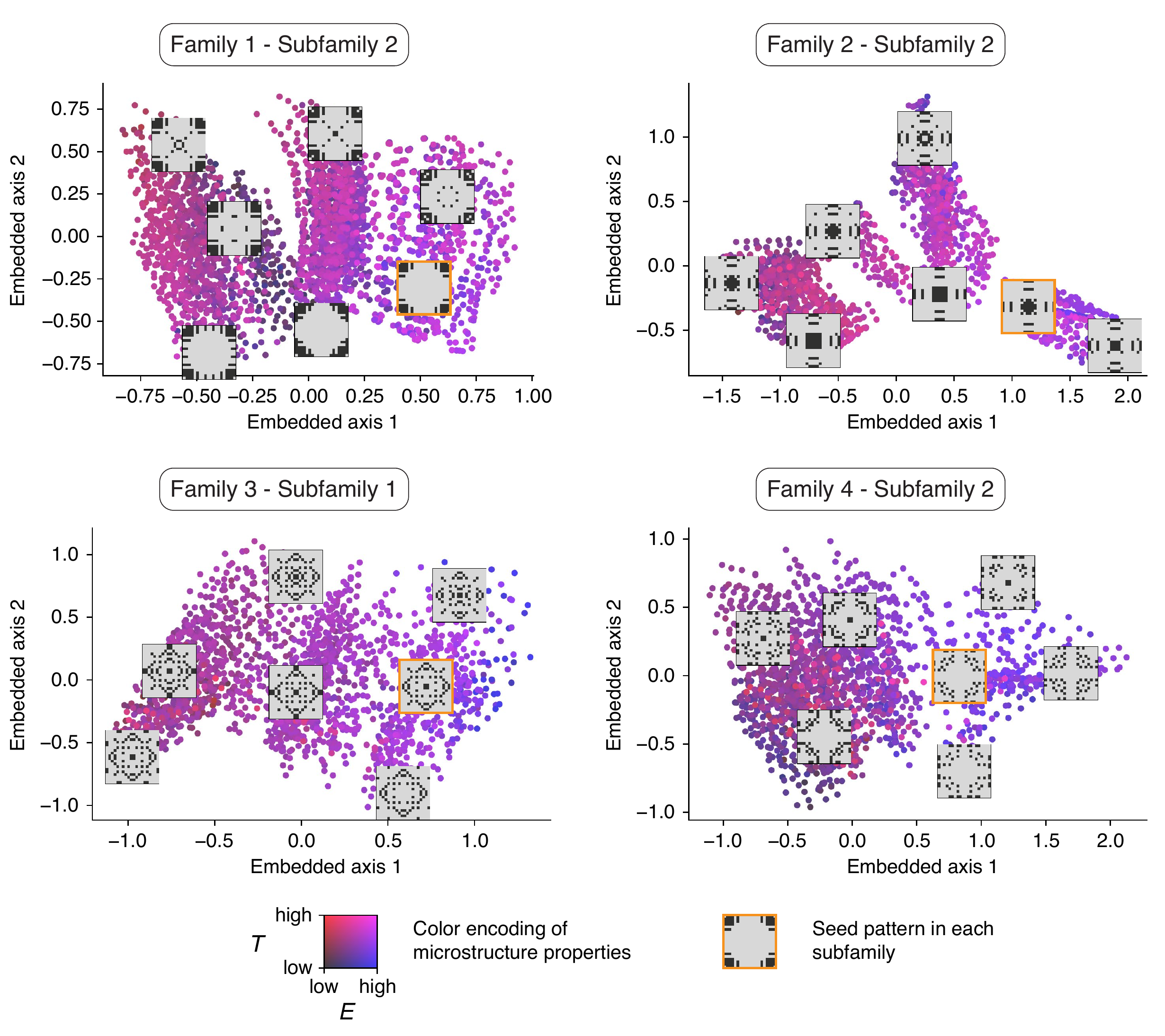}
    \caption{{\bf Isomap embedding spaces of four subfamilies annotated with example patterns.} Each subfamily contains around 2,000 to 3,000 microstructures and a few of them are visualized. Mechanical performance is encoded in colors ($E$: the blue channel; $T$: the red channel). Seed patterns are highlighted by orange boxes.}
    \label{fig:emd_isomap_example}
\end{figure}

For more in-depth analysis, each microstructure family was divided into subfamilies according to similarity with seed microstructures, where a subfamily consists of a seed microstructure and others within a 16-pixel radius as specified in dense sampling. We ran Isomap to reduce 400D microstructure patterns into 2D embedding spaces (with a neighborhood size of 10) in each subfamily, shown in Fig.~\ref{fig:emd_isomap_example}. Microstructure properties are color-coded in the figures. A higher intensity of the red channel indicates better toughness, whereas the blue channel represents Young's modulus. Our dense sampling algorithm was able to create a diversity of variants from the seed microstructures. New shape features can be seen in these variants, such as individual pixels, beams, arcs, and small chunks of soft material.

\begin{figure}[!ht]
    \centering
    \includegraphics[width=0.8\textwidth]{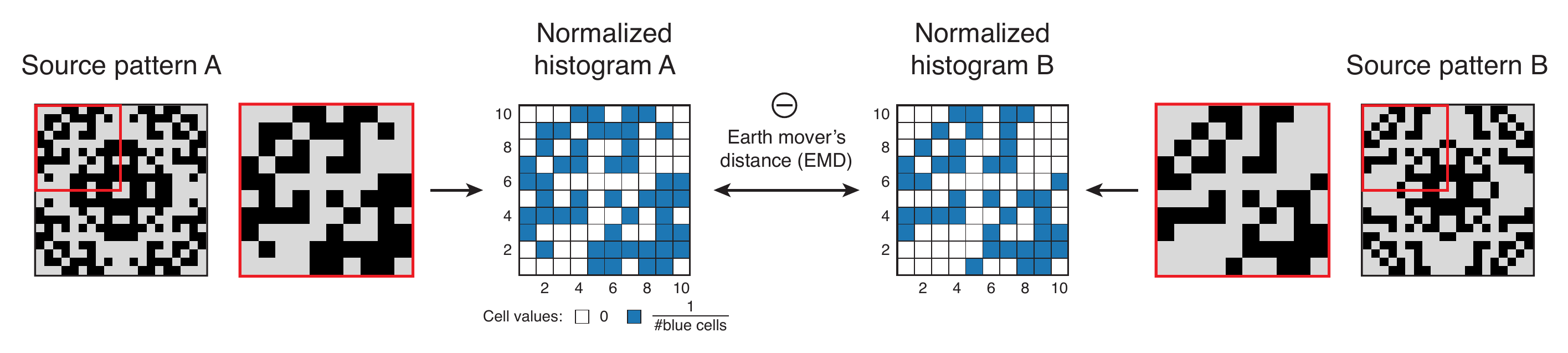}
    \caption{{\bf Calculation of the earth mover’s distance (EMD) between two microstructure patterns.}}
    \label{fig:emd}
\end{figure}

Regarding the distance metric for Isomap, we used the earth mover’s distance (EMD) rather than the Euclidean distance. EMD, or the 1st Wasserstein distance ({\it 55}), 
is a metric between probability distributions originally defined in optimal transportation theory ({\it 57}). 
It intuitively measures the minimal work (defined as mass times distance) required to redistribute a pile of earth into another. Compared with the Euclidean distance, EMD better reflects the shape difference between microstructure patterns by taking into account both pixel values and locations. Since the original definition of EMD is not directly compatible with microstructure patterns, we converted the patterns into quarter-sized normalized histograms that represent discrete probability distributions. The conversion rules, illustrated in Fig.~\ref{fig:emd}, treated pixels of the soft material as shape features and assigned non-zero values to the corresponding cells of the output histogram. Conversely, the other cells associated with the rigid material were set to zero. Therefore, the EMD between two microstructures $x_a$ and $x_b$ is calculated using the 1st Wasserstein distance between their normalized histograms $h_a$ and $h_b$ ($h_a, h_b \in [0, 1]^{100}$).
\begin{linenomath}
\begin{ceqn}
\begin{align}
\begin{split}
    \text{EMD}(x_a, x_b) &= W_1(h_a, h_b) \\
    &= \min_{\pi \in \Pi(h_a, h_b)} \pi \otimes D,
\end{split}
\end{align}
\end{ceqn}
\end{linenomath}
where $\Pi(h_a, h_b)$ is the collection of transportation plans, $\pi$, from $h_a$ to $h_b$, given by
\begin{linenomath}
\begin{ceqn}
\begin{equation}
    \Pi(h_a, h_b) = \left\{\pi \in \mathbb{R}^{100\times100}\,\middle|\,\pi \geq 0,\,\pi\mathbf{1} = h_a,\,\text{and }\pi^\text{T}\textbf{1} = h_b\right\}.
\end{equation}
\end{ceqn}
\end{linenomath}
In the earth-moving analogy, $\pi$ tabulates the amount of earth to transfer from one cell to another. $D$ is the pairwise Euclidean distance matrix of histogram cells. $A \otimes B = \langle A, B\rangle_\text{F} = \text{tr}(A^\text{T}B)$ is the Frobenius inner product of two identically shaped matrices.

\begin{figure}[!ht]
    \centering
    \includegraphics[width=0.85\textwidth]{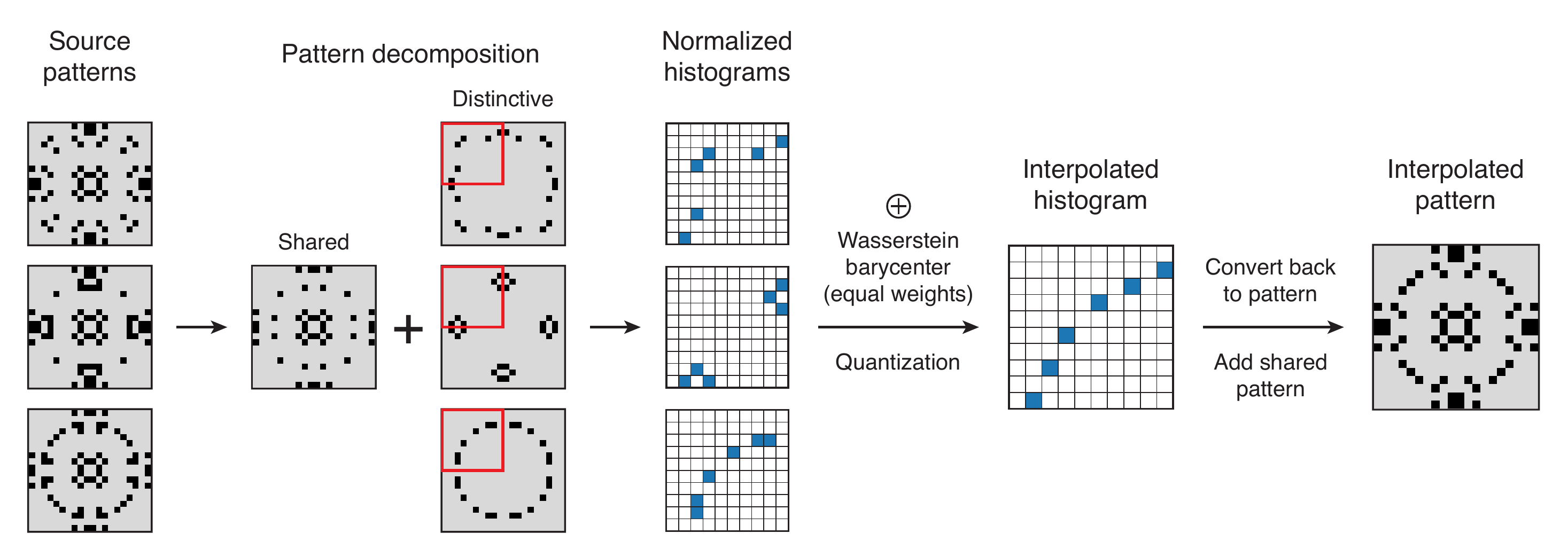}
    \caption{{\bf Interpolation of neighboring microstructure patterns in an Isomap embedding space based on Wasserstein barycenters.}}
    \label{fig:interp}
\end{figure}

\begin{figure}[!ht]
    \centering
    \includegraphics[width=0.7\textwidth]{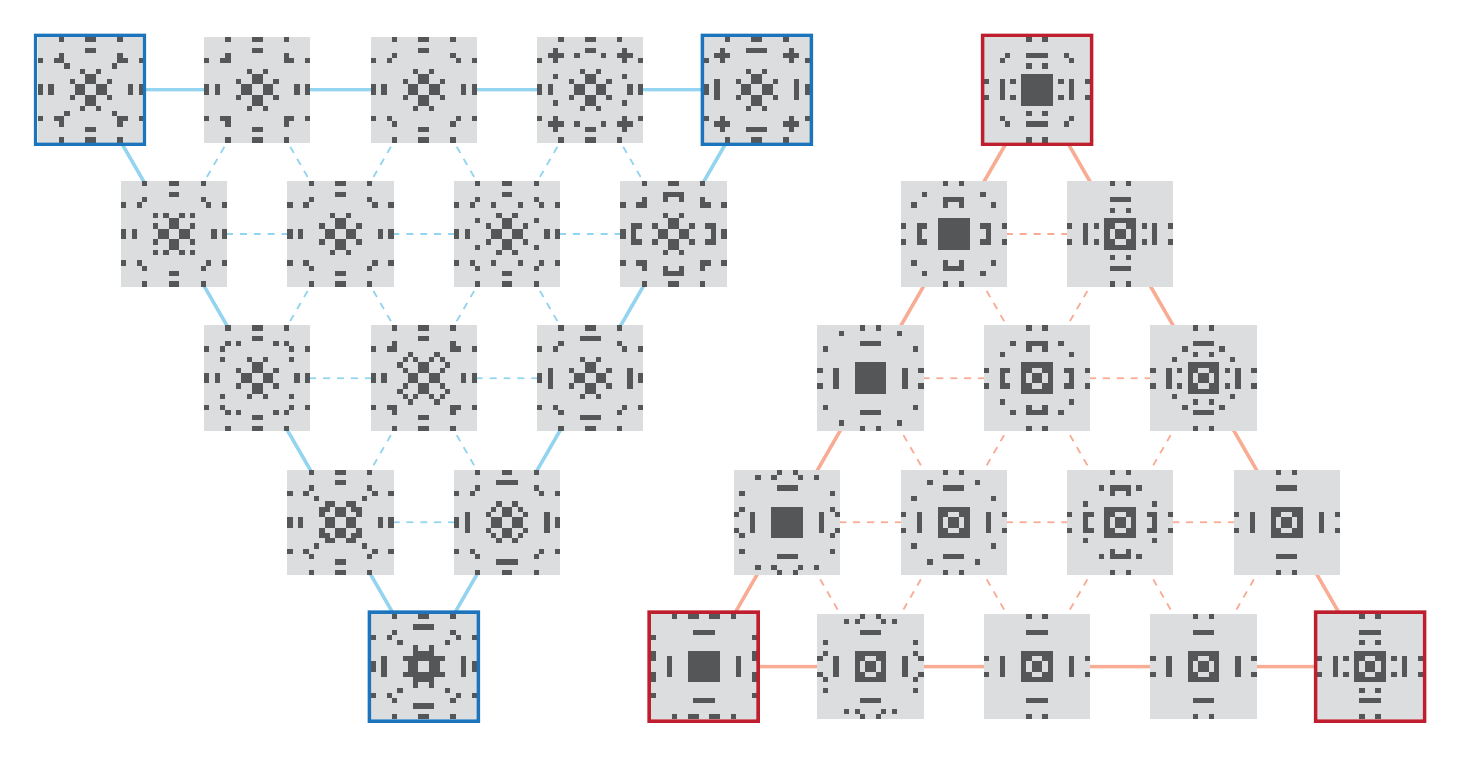}
    \caption{{\bf Examples of interpolation between three microstructures in the same subfamily.}}
    \label{fig:interp_example}
\end{figure}

\begin{figure}[!ht]
    \centering
    \includegraphics[width=0.85\textwidth]{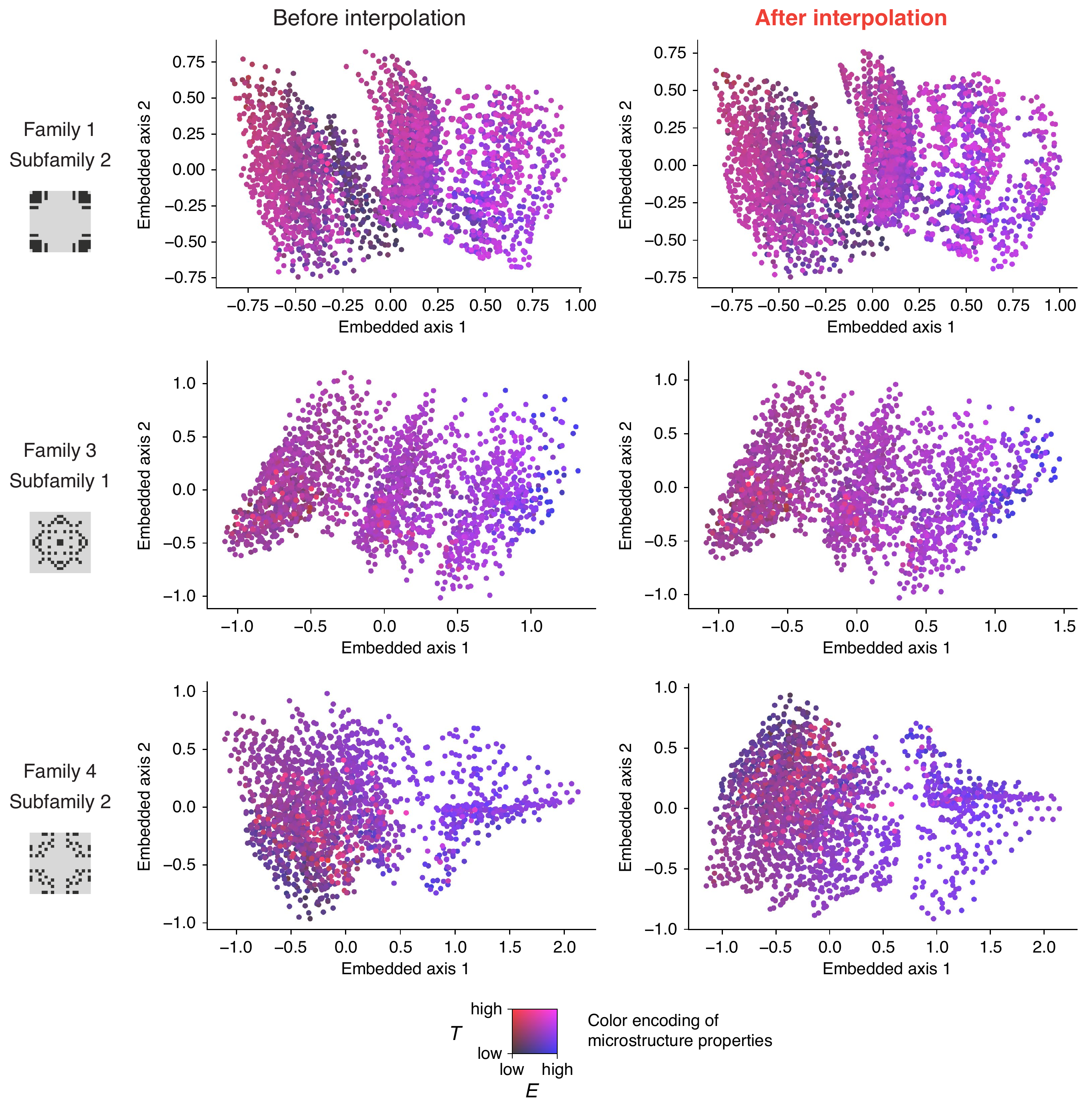}
    \caption{{\bf Isomap embedding spaces of three subfamilies before and after refinement using interpolation.} Microstructure properties are encoded in colors.}
    \label{fig:emd_interp_example}
\end{figure}

The Isomap embedding space of a subfamily can be further refined by samples not covered in dense sampling. To generate these samples, we built an interpolation model for $K$ neighboring microstructure patterns $x_1, x_2, \cdots, x_K$ in the embedding space based on Wasserstein barycenters ({\it 56}), 
which is widely used in shape interpolation ({\it 58}). 
The interpolation algorithm is illustrated in Fig.~\ref{fig:interp}. Each source pattern was decomposed into a common part containing shared shape features and an exclusive part containing distinctive features. The exclusive parts were converted into quarter-sized normalized histograms denoted by $h_1, h_2, \cdots, h_K$. Our model interpolated among the histograms by computing their Wasserstein barycenter $\Bar{h}$ under the weight assignment $w_1, w_2, \cdots, w_K$, defined as
\begin{linenomath}
\begin{ceqn}
\begin{equation}
    \Bar{h} = \underset{h}{\operatorname{argmin}}\,\sum_{i=1}^K w_i W_2(h_i, h).
\end{equation}
\end{ceqn}
\end{linenomath}
$W_2(\cdot, \cdot)$ refers to the 2nd Wasserstein distance,
\begin{linenomath}
\begin{ceqn}
\begin{equation}
    W_2(h_a, h_b) = \min_{\pi \in \Pi(h_a, h_b)} \pi \otimes D_2,
\end{equation}
\end{ceqn}
\end{linenomath}
where $D_2 = D \circ D$ is the pairwise squared distance matrix ($\circ$ refers to the Hadamard product). The minimization problem was tackled by solving the following equivalent linear programming problem ({\it 56}) 
for the optimal transportation plans $\pi_i^*$ from each $h_i$ to $\Bar{h}$ ($i = 1, 2, \cdots, K$).
\begin{linenomath}
\begin{ceqn}
\begin{align}
\begin{split}
    \min_{\{\pi_i\}}\;\;& w_i \left(\pi_i \otimes D_2\right) \\
    s.t. \quad & \pi_i \geq 0, \quad \forall i = 1, 2, \cdots, K \\
    & \pi_i\textbf{1} = h_i, \quad \forall i = 1, 2, \cdots, K \\
    & \pi_i^\text{T}\textbf{1} = \pi_1^\text{T}\textbf{1}, \quad \forall i = 2, 3, \cdots, K
\end{split}
\end{align}
\end{ceqn}
\end{linenomath}
After that, $\Bar{h}$ was obtained by $\Bar{h} = \left(\pi_1^*\right)^\text{T}\textbf{1}$. To convert the interpolated histogram back to a microstructure pattern, $\Bar{h}$ was quantized into a binary matrix that indicates the material assignments of the upper-left quarter of the resulting pattern. Let $c_i$ be the unified value of all non-zero cells in $h_i$ and $c = \left(\sum_{i=1}^K c_i\right)/K$. A cell in $\Bar{h}$ translated into the soft material if no less than $c$, and the rigid material otherwise. Examples of microstructure pattern interpolation are shown in Fig.~\ref{fig:interp_example}.

\begin{table}[!ht]
    \centering
    \begin{adjustbox}{max width=\textwidth}
    \begin{tabular}{c|ccc|cccc|cc|cc|c}
        \toprule
        \textbf{Family} & \multicolumn{3}{c|}{1} & \multicolumn{4}{c|}{2} & \multicolumn{2}{c|}{3} & \multicolumn{2}{c|}{4} & \multirow{2}{*}[-3pt]{\textbf{Total}} \\
        \cmidrule{1-12}
        \textbf{Subfamily} & 1 & 2 & 3 & 1 & 2 & 3 & 4 & 1 & 2 & 1 & 2 & \\
        \midrule
        Before & 3,251 & 2,959 & 3,313 & 1,862 & 1,861 & 1,917 & 1,859 & 1,860 & 1,862 & 1,997 & 1,915 & 24,656 \\
        After & 3,328 & 3,164 & 3,371 & 1,939 & 1,904 & 2,054 & 1,903 & 1,988 & 1,964 & 2,112 & 2,044 & 25,771 \\
        \midrule
        \textbf{Inc. (\%)} & 2.37 & 6.97 & 1.75 & 4.14 & 2.31 & 7.15 & 2.37 & 6.88 & 5.48 & 5.76 & 6.74 & 4.52 \\
        \bottomrule
    \end{tabular}
    \end{adjustbox}
    \caption{{\bf Increases in the size of subfamilies from interpolation.}}
    \label{tab:subfamily_size}
\end{table}

In practice, a Delaunay triangulation was computed in the 2D embedding space to determine neighboring microstructures, namely $K = 3$. The neighborhoods whose longest edge was between 4\% and 40\% of the diameter of the family (i.e., the distance between the two furthest points in the embedding space) were marked as candidates. New microstructures were then generated from uniform grid interpolation in the candidate neighborhoods and added to the subfamily if they were within a 16-pixel radius from the seed microstructure and did not violate any structural constraint. Finally, the refined embedding space was recomputed using Isomap. Our interpolation algorithm generated about 5\% more samples in each subfamily on average (Table~\ref{tab:subfamily_size}). In Fig.~\ref{fig:emd_interp_example}, we make a side-by-side comparison between the Isomap embedding spaces before and after interpolation.

\subsection{Toughening Mechanisms}

\begin{figure}[!ht]
    \centering
    \includegraphics[width=0.8\textwidth]{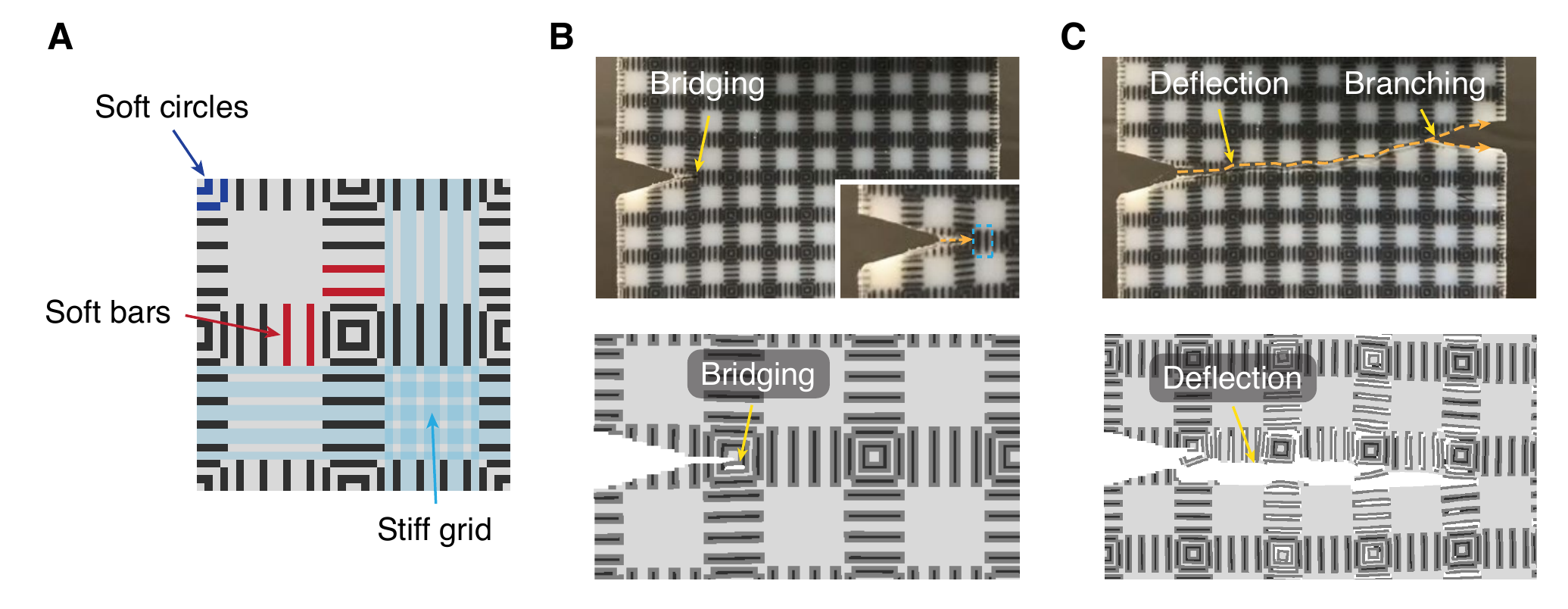}
    \caption{{\bf Major shape features (A) and toughening mechanisms (B, C) of microstructures in Family 1.}}
    \label{fig:mechanisms_family_1}
\end{figure}

\begin{figure}[!ht]
    \centering
    \includegraphics[width=0.8\textwidth]{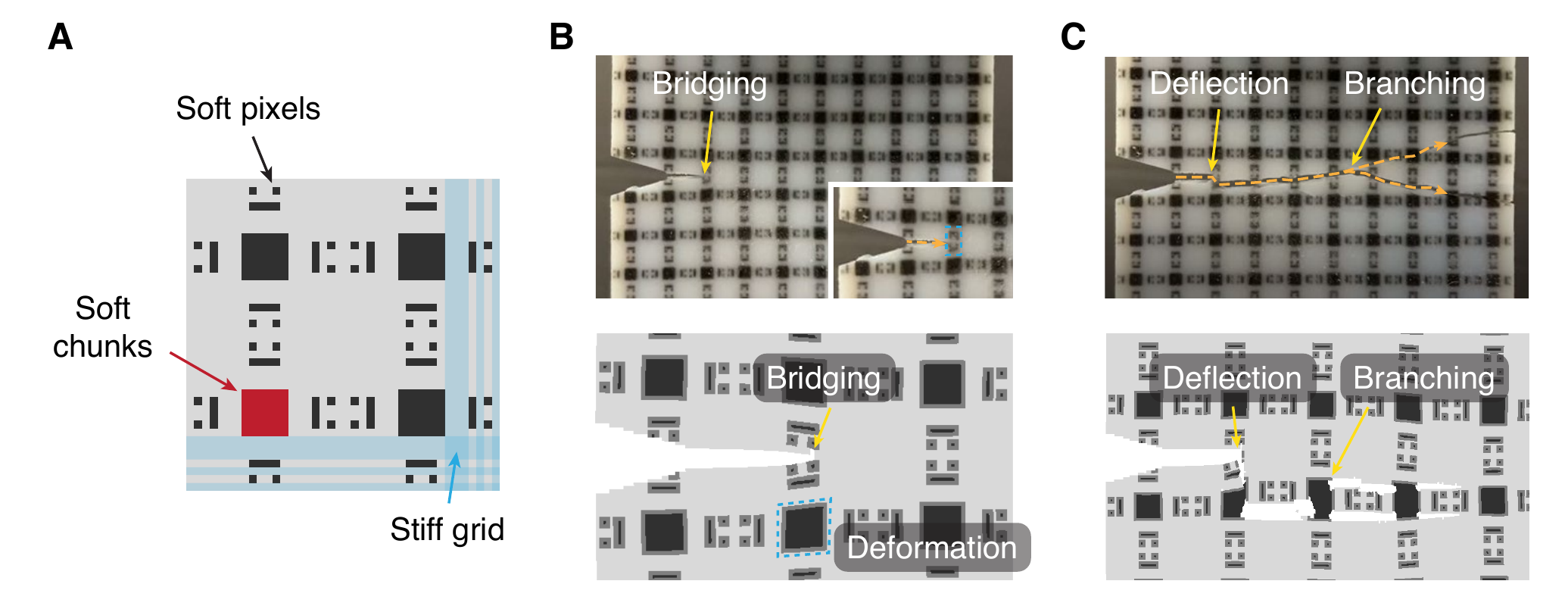}
    \caption{{\bf Major shape features (A) and toughening mechanisms (B, C) of microstructures in Family 2.}}
    \label{fig:mechanisms_family_2}
\end{figure}

Intrinsic toughening mechanisms of each microstructure family were analyzed from video recordings of validation samples during mechanical testing and simulation. We summarize the most representative mechanisms of each family and associate them with shape features in microstructure patterns as follows.

Family 1 maintains a relatively high fraction of soft pixels in the shape of bars or chunks as exemplified in Fig.~\ref{fig:mechanisms_family_1}A. This family is located on the left side of the performance gamut where Young’s modulus is relatively small. To maintain proper \edit{stiffness}, the continuous areas of rigid material form horizontal and vertical grids (stiff grids in Fig.~\ref{fig:mechanisms_family_1}A). A distinctive feature appearing in this family is parallel bars of soft material. Their benefits are twofold. When perpendicular to the pulling direction, they introduce additional stress resistance through elastic deformation. Alternatively, when parallel to the pulling direction, or directly facing crack propagation, they effectively absorb energy and slow down the propagation. Bridging and deflection effects are the most prominent toughening mechanisms (Fig.~\ref{fig:mechanisms_family_1}B and \ref{fig:mechanisms_family_1}C). A zig-zag crack path is observed during fracture which results in high toughness. The propagation energy accumulates under the trapping effect of soft material and dissipates via branched cracks ({\it 42}). 

Different from Family 1, Family 2 concentrates soft pixels in a cross-like band in the center of the unit (Fig.~\ref{fig:mechanisms_family_2}A). Microstructures withstand a larger strain via the elastic deformation of soft material chunks, as indicated by the red square in Fig.~\ref{fig:mechanisms_family_2}B. It is noteworthy that such deformation is accurately reflected in simulation. Stiff grids are also present as a means of improving Young's modulus. Bridging, deflection, and branching (Fig.~\ref{fig:mechanisms_family_2}B and \ref{fig:mechanisms_family_2}C) are essential anti-cracking mechanisms in the family.

\begin{figure}[!ht]
    \centering
    \includegraphics[width=0.8\textwidth]{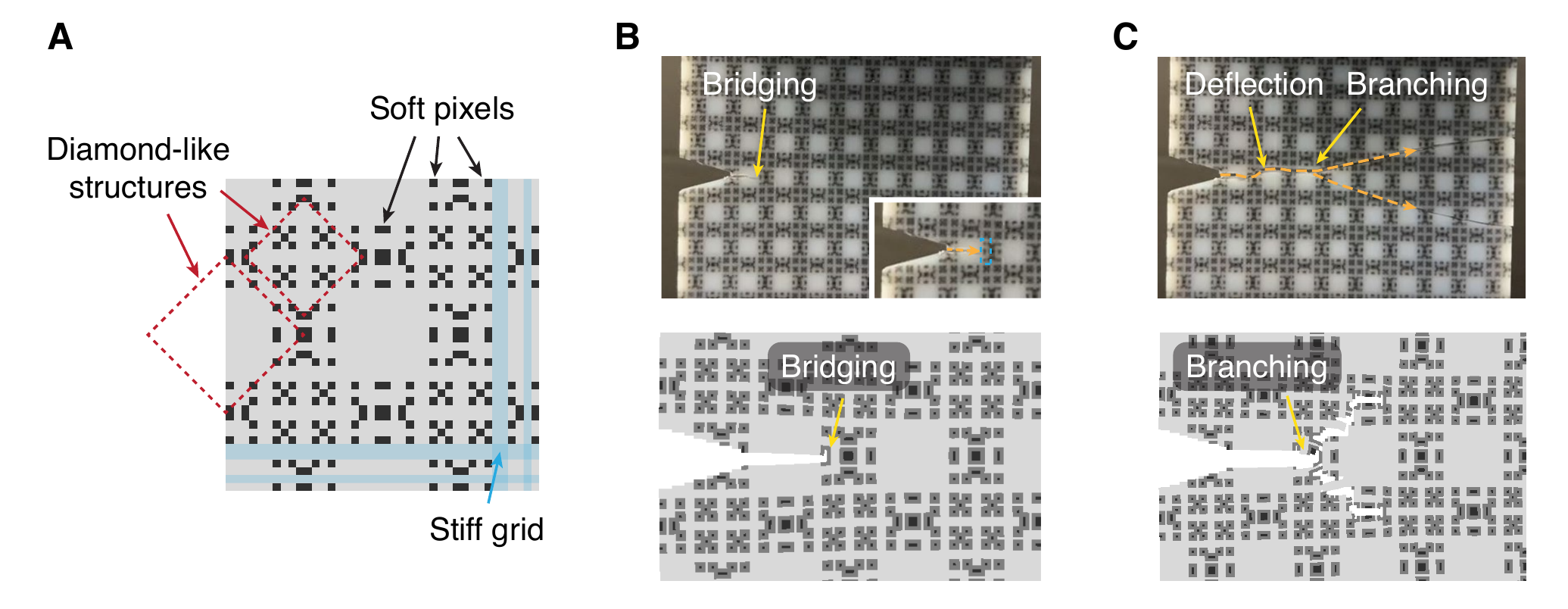}
    \caption{{\bf Major shape features (A) and toughening mechanisms (B, C) of microstructures in Family 3.}}
    \label{fig:mechanisms_family_3}
\end{figure}

\begin{figure}[!ht]
    \centering
    \includegraphics[width=0.8\textwidth]{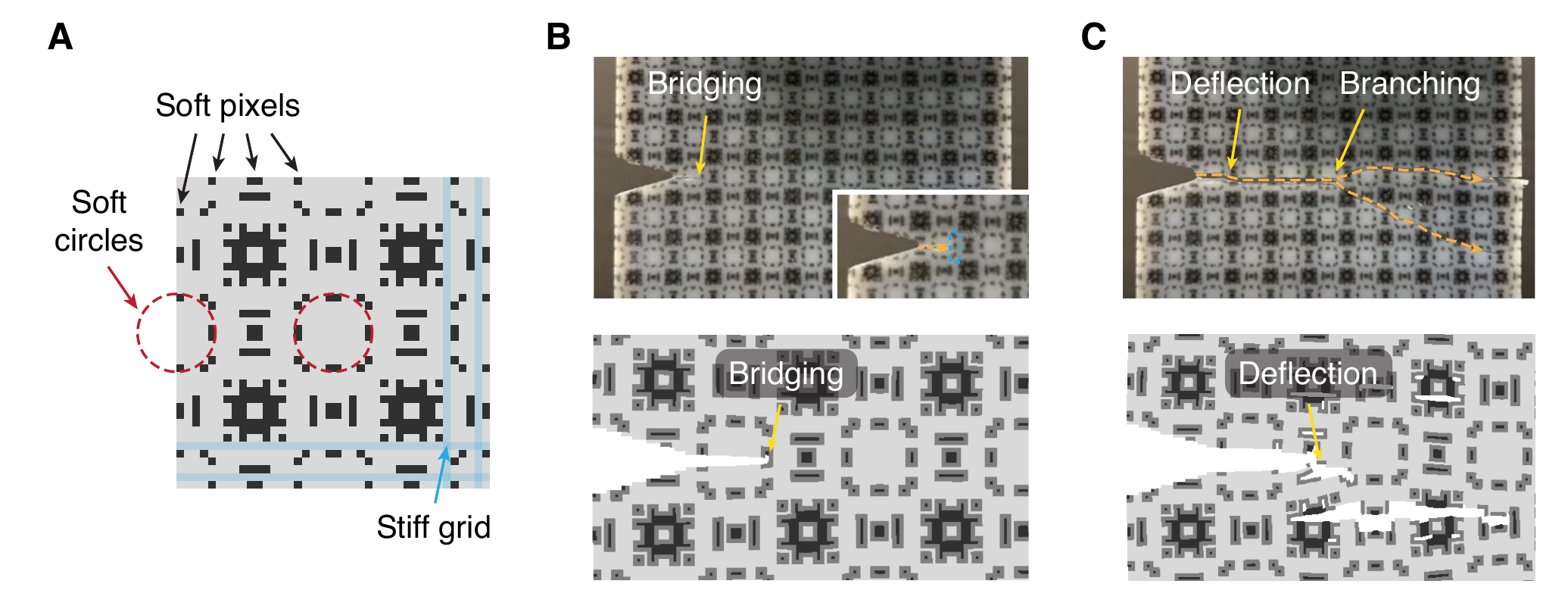}
    \caption{{\bf Major shape features (A) and toughening mechanisms (B, C) of microstructures in Family 4.}}
    \label{fig:mechanisms_family_4}
\end{figure}

While soft pixels are more concentrated in the two families above, they are mostly dispersed in Family 3. Discrete soft pixels form a quarter of a diamond-like structure at corners and another diamond structure in the center (Fig.~\ref{fig:mechanisms_family_3}A). Meanwhile, some soft pixels are located at the edges and ahead of the notch tip. These pixels are typically on the path of crack propagation and act as an energy dissipation zone that delays crack propagation. Since the interface is stronger than the soft material, the soft pixels work as defects that guide crack propagation. Such a deflection effect can increase the crack length dramatically and dissipate much more energy. Additional soft pixels introduced during dense sampling not only enrich the diversity of the family but also increase the overall stress resistance.

As for Family 4, the discrete soft pixels around corners roughly form a quarter of a circular outline (Fig.~\ref{fig:mechanisms_family_4}A). Similar to Family 3, some soft pixels are placed on the edges, which retards crack propagation by dampening the cracking velocity and dissipating energy through elastic deformation. Basic features like stiff grids are preserved to maintain Young's modulus. The same anti-cracking mechanisms, such as bridging, deflection, and branching, are observed in this family (Fig.~\ref{fig:mechanisms_family_4}B and \ref{fig:mechanisms_family_4}C). As Family 3 and 4 bear some resemblance in structure, we observed that they show comparable mechanical performance. Due to the dominance of rigid material at $0.8 \leq \phi \leq 0.85$, microstructures in both families tend to have higher Young's moduli but break at relatively smaller strains.

\section{Case Study: Stiffness-Porosity Trade-off}
With carefully designed geometries, porous structures enable strong materials with light weights. In many engineering applications, it is desirable to design microstructures that are both stiff (large Young's modulus) and lightweight (high porosity). However, porosity and stiffness are often contradictory since materials with higher porosity are generally weaker due to less material supporting the external force. Due to its crucial impact on engineering, the stiffness-porosity trade-off has been extensively studied in the field of solid mechanics and topology optimization. In this section, we directly apply our optimization pipeline to identify the Pareto front between stiffness and porosity for porous microstructure designs. We demonstrate that our pipeline not only remains effective in Pareto front discovery but also has adequate scalability to efficiently explore even larger combinatorial design spaces.

\subsection{Problem setting}
\begin{figure}[!ht]
    \centering
    \includegraphics[width=0.85\textwidth]{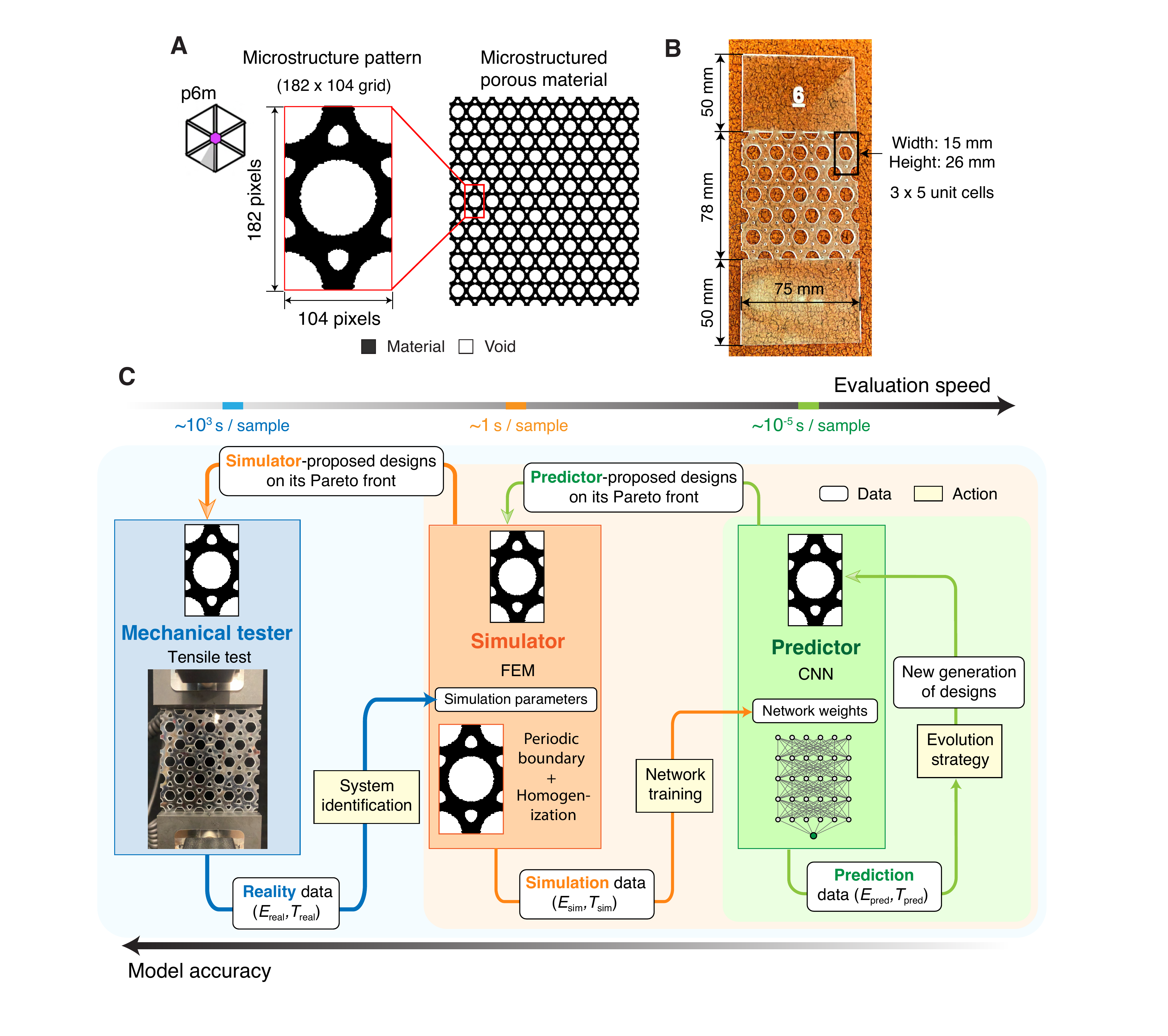}
    \caption{{\bf Application of our approach to discover porous microstructures with optimal stiffness-porosity trade-offs.} ({\bf A}) The porous microstructures consist of periodic unit cells defined on a $182\times104$ regular grid, where each element represents either material or void. We enforce p6m symmetry on microstructure designs for isotropy. ({\bf B}) Sample photograph of a manufactured porous microstructure. The sample is fabricated by laser-cutting a clear acrylic plate. ({\bf C}) Workflow of our pipeline to discover the stiffness-porosity Pareto front.}
    \label{fig:YM_pipeline}
\end{figure}

\textbf{Design space.} We consider single-material, periodic porous structures where each unit is characterized by a regular grid of $182\times104$ elements (Fig.~\ref{fig:YM_pipeline}A). An element can be either void or filled with solid material. This constitutes a much larger design space than the $20\times20$ patterns we considered for the \edit{stiffness}-toughness trade-off. However, a valid microstructure design must contain a single-connected solid phase to exhibit non-zero stiffness. Therefore, we adopt the topology generation algorithm introduced in ({\it 73}) 
to create valid random designs. Meanwhile, we focus on a porosity range from 0.1 to 0.75 to exclude impractical designs with excessive or insufficient material. To further ensure that the generated designs are isotropic, we require all microstructure patterns to have \textit{p6m} symmetry.

\begin{figure}[!ht]
    \centering
    \includegraphics[width=0.6\textwidth]{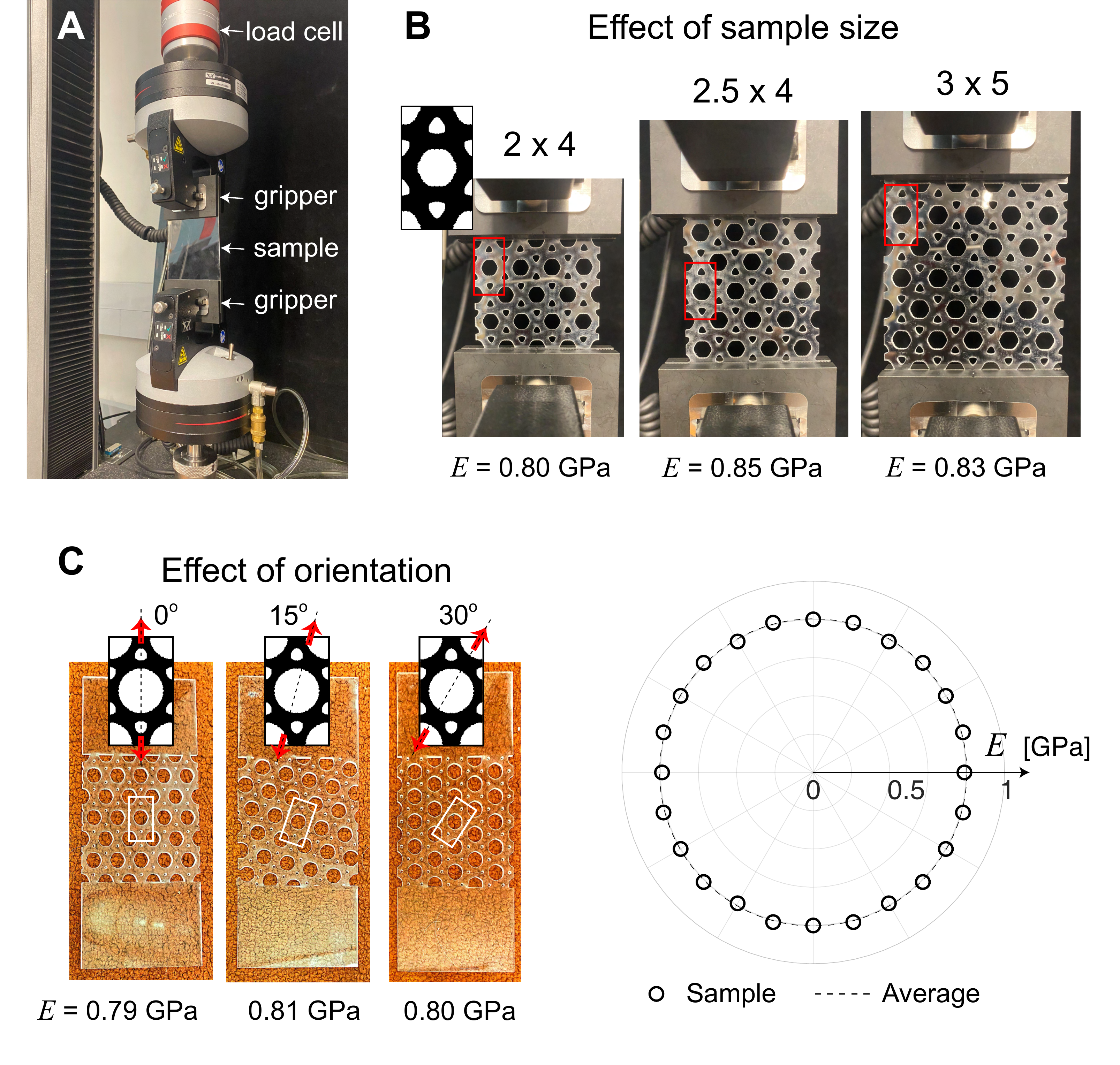}
    \caption{{\bf Experimental measurement of Young's modulus.} ({\bf A}) The sample is gripped by the Instron on both ends outside the microstructure area. The Instron then stretches the sample by 0.1 mm at a rate of 1 mm/minute. The effective Young's modulus is derived from the slope of the measured force-displacement curve. ({\bf B}) Measured effective Young's moduli for samples comprising different numbers of units. The Young's modulus barely varies as we increase the sample size, indicating that the boundary effect of a $3\times5$ sample is sufficiently small. ({\bf C}) Measured effective Young's moduli along different directions, obtained from samples with rotated microstructures. Since the microstructure pattern obeys \textit{p6m} symmetry, we only need to rotate the microstructures by up to $30^\circ$. The results prove that the elastic response of the measured design is isotropic.}
    \label{fig:YM_exp}
\end{figure}

\noindent\textbf{Fabrication and measurement.} Test samples are fabricated on acrylic plates with a thickness of 1.6 mm (McMaster-Carr 8560K172). Microstructure patterns are carved onto the acrylic plate via laser cutting (Universal Laser Systems VLS6.75). More specifically, the laser is programmed to cut along the interface between material and void to produce the designed porous structure (Fig.~\ref{fig:YM_pipeline}B). The dimensions of each unit cell are 15 mm in width and 26 mm in height, resulting in a pixel size of 0.14 mm. Since the resolution of the laser cutter is below 0.05 mm, this fabrication method is sufficiently accurate. Each sample consists of 3$\times$5 unit cells, with a total size of 75 mm in width and 78 in height. In addition, two rectangular gripping areas with a height of 50 mm each are concatenated to the microstructure.

We measure the effective Young's modulus of a microstructure design on a tensile machine (Instron with 500N load cell, see Fig.~\ref{fig:YM_exp}A). The test sample is fixed onto the Instron using two pneumatic grippers (the pressure is 100 pounds per square inch). During the tensile test, the Instron stretches each sample to 0.1 mm at a rate of 1 mm/minute. The effective Young's modulus of the sample is then calculated based on the slope of the measured force-displacement curve and sample size. To prove that the boundary effect is negligible for a sample with 3$\times$5 unit cells, we measured the effective Young's modulus using samples of various sizes: 2$\times$4, 2.5$\times$4, and 3$\times$5 units. According to Fig.~\ref{fig:YM_exp}B, Young's modulus converges at our designated sample size, indicating that the 3$\times$5 size already suffices to minimize the boundary effect. Moreover, we measured several Young's moduli of the same microstructure along multiple directions to verify its isotropy. This is realized by fabricating samples with tilted microstructures. As shown in Fig.~\ref{fig:YM_exp}C, the microstructure design exhibits almost identical Young's moduli in different directions. In all experiments, we measure each test sample 4 times and report the average Young's modulus and standard deviation. These measurements are guaranteed to be consistent since the laser cutter has a sufficiently high resolution and the acrylic plates are relatively thin (3 mm). Nonetheless, we made an extra effort to fabricate three samples for a fixed microstructure design, and the measured Young's moduli differ by less than 1\%. Therefore, to save time and energy, we only fabricate one sample per microstructure design in this case study.

\vspace{1em}
\noindent\textbf{Simulation.} We developed an FEM simulator to calculate the homogenized Young's modulus of a porous microstructure specifically for this study. The FEM grid has identical dimensions to a microstructure unit (i.e., $180 \times 104$ elements). Since stiffness calculation is a linear problem, we use linear elasticity to model the base material. The model is characterized by two exposed parameters: Young's modulus $E_\text{base}$ and Poisson's ratio $\nu_\text{base}$ of the base material. Each finite element is parameterized by material density using a continuous variable between 0 and 1. A density equal to 1 represents the base material whereas a zero density corresponds to void. Based on the homogenization theory, we computed the homogenized stiffness tensor of the microstructure using periodic boundary conditions, from which we derived homogenized Young's moduli in multiple directions. The minimal Young's modulus was chosen as the objective for structural optimization to ensure material isotropy. Moreover, we implemented gradient calculation and SIMP in the simulator to compare our pipeline against existing topology optimization algorithms that utilize sensitivity information.

\subsection{Results}

In the initial round, the exposed parameters of the simulator, $E_\text{base}$ and $\nu_\text{base}$, were determined by the Young's modulus and Poisson's ratio of an acrylic plate. Since the vendor only provides a flexural modulus of 2.76 GPa and leaves out the value of its Poisson's ratio, we made an initial guess of $E_\text{base} = 2.76$ GPa and $\nu_\text{base} = 0.33$.

\begin{figure}[!ht]
    \centering
    \includegraphics[width=0.95\textwidth]{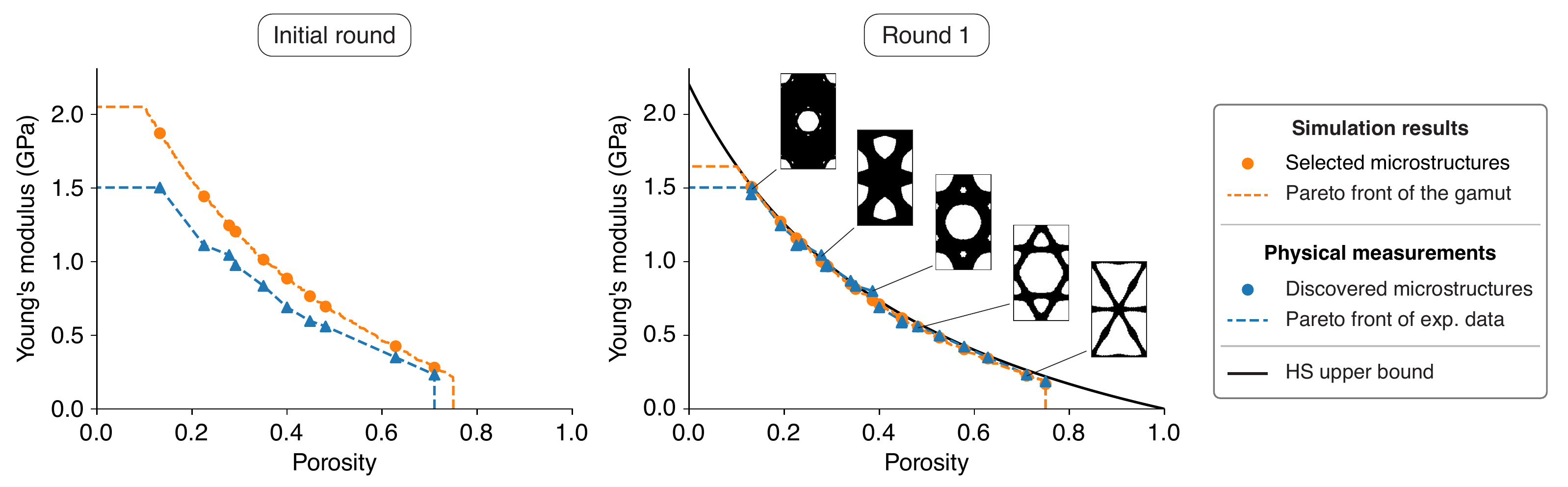}
    \caption{{\bf Evolution of experimental and simulation Pareto fronts during the outer loop.} The sim-to-real gap closes after one round. The discovered microstructures are validated against the HS upper bound (black line). We provide snapshots of several representative designs on the HS upper bound.}
    \label{fig:YM_results}
\end{figure}

We ran the inner loop of our pipeline (NMO) to find the Pareto front of the simulator, from which we selected 10 designs for experimental verification (see the orange dashed line and markers in Fig.~\ref{fig:YM_results}). As expected, there is a noticeable gap between the experimental Pareto front (blue) and the simulation Pareto front (orange), since the exact properties of the base material are unknown. At the beginning of Round 1, our pipeline conducted system identification on discovered designs by adjusting the values of $E_\text{base}$ and $\nu_\text{base}$. Next, we ran the inner loop (NMO) again and identified 10 additional designs on the simulation Pareto front. After physically measuring their Young's moduli, we noticed that the measurements were almost identical to simulation predictions (see Round 1 in Fig.~\ref{fig:YM_results}). In contrast to the \edit{stiffness}-toughness problem, our pipeline is able to close the sim-to-real gap after merely one round. The reasons for the fast progress include: 1) stiffness is a linear property; 2) the fabrication process of test samples is precise; 3) experimental behaviors of the samples are very consistent; 4) the simulator only has two exposed parameters.


\vspace{1em}
\noindent\textbf{Hashin-Shtrikman upper bound.} Once our pipeline finished, we verified that the discovered designs on the experimental Pareto front are theoretically optimal solutions. In 1963, Hashin and Shtrikman derived the theoretical upper bound of the Young's modulus of an isotropic material under different porosities as ({\it 43, 44, 73}), 
\begin{linenomath}
\begin{ceqn}
\begin{equation}\label{eq:hs}
    E_\text{HS} = E_\text{base} (1-\phi)/(1+2\phi),
\end{equation}
\end{ceqn}
\end{linenomath}
where $E_\text{base}$ is the Young’s modulus of the base material, $\phi$ is porosity, and $E_\text{base}$ gives the Hashin-Shrikman (HS) upper bound. Therefore, for any porous material with porosity $\phi$, its isotropic Young's modulus $E\leq E_\text{HS}(\phi)$.
Equation \eqref{eq:hs} defines the theoretical upper bound for the Pareto front of Young’s modulus and porosity, providing a benchmark for structural optimization algorithms. In Fig.~\ref{fig:YM_results}, we compare experimental and simulation results of Pareto-optimal designs discovered by our pipeline with the HS upper bound (black line). We find that our results closely align with the theoretical upper bound. This proves that our pipeline is able to find theoretically optimal solutions to the stiffness-porosity problem.

\subsection{Sample efficiency}

In this section, we compare NMO with modified NSGA-II and TO in sample efficiency when discovering the simulation Pareto front. Here, the modified NSGA-II algorithm uses a carefully designed mutation operator. The mutation operator probabilistically adds/removes a group of randomly selected boundary pixels, adds/removes ellipses with random shapes and orientations, or adds a line with random thickness and orientation. Each random number controlling the magnitude of a mutation is drawn from an exponential distribution governed by its rate parameter $\lambda$. The TO algorithm implements SIMP and runs continuous gradient-based optimization using the method of moving asymptotes (MMA). Echoing the valid porosity range of the design space, we set the volume fraction constraint ($V_\mathrm{max} = 1 - \phi$) between 0.25 and 0.9 and maximize the homogenized Young's modulus to discover porous microstructures near the HS upper bound. Considering that TO only yields one design at a time, we conduct a series of optimization tasks under evenly sampled $V_\mathrm{max}$ and arrange their order using a greedy strategy to expedite hypervolume growth. More specifically, we choose each subsequent task to maximize the expected Pareto hypervolume improvement from discovered designs assuming that the HS upper bound will be reached. The resulting design is discarded if it violates the volume fraction constraint after quantization.

\begin{figure}[!ht]
    \centering
    \includegraphics[width=\textwidth]{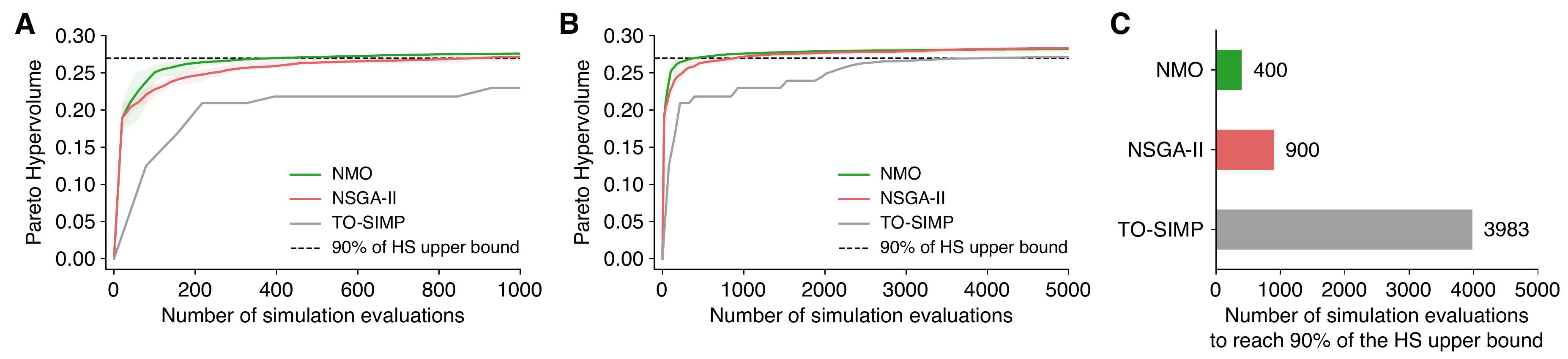}
    \caption{{\bf Comparison between NMO and other multi-objective optimization algorithms in Pareto hypervolume growth.} The comparison is made within a budget of ({\bf A}) 1,000 simulation evaluations and ({\bf B}) 5,000 simulation evaluations. ({\bf C}) Number of evaluations required for NMO and other baseline algorithms to reach 90\% of the theoretical maximal hypervolume.}
    \label{fig:YM_inner_loop}
\end{figure}

Fig.~\ref{fig:YM_inner_loop} illustrates how Pareto hypervolume grows with the number of simulation evaluations for NMO, NSGA-II, and TO-SIMP (with an MMA solver). Considering that TO only yields one optimal design at a time, we conduct a series of optimizations using different weighted combinations of properties and active constraints. Their order is arranged in a greedy strategy that maximizes theoretical hypervolume improvement according to the HS upper bound. Again, NMO shows the highest sample efficiency especially in the early stage of the optimization when fewer than 1,000 samples are evaluated (Fig.~\ref{fig:YM_inner_loop}A). It takes NMO only 400 simulations to reach 90\% hypervolume of the theoretical HS upper bound, 900 simulations for NSGA-II, and almost 4,000 simulations for TO (Fig.~\ref{fig:YM_inner_loop}C). This is because the neural network inside NMO quickly explores the entire design space and proposes valuable designs to the simulator after the optimization starts. Since stiffness measurement is a linear problem, CNNs can learn to predict Young's modulus reasonably well with a small amount of training data. This allows NMO to rapidly expand its Pareto hypervolume. In comparison, TO has to follow a sequence of optimization trajectories to reach the entire HS upper bound, consuming a large number of intermediate evaluations despite the usage of MMA.

We note that NMO is surpassed by NSGA-II as the Pareto hypervolume approaches the HS upper bound (Fig.~\ref{fig:YM_inner_loop}B). This is mainly due to the shrinking room for Pareto hypervolume growth at a later stage of the optimization. Although the neural network initially accelerates the exploration of the design space, giving accurate predictions for remaining designs near the HS upper bound eventually comes in higher priority. Therefore, the hypervolume growth of NMO slows down as the network tries to improve predictions for those top-performing designs. The \edit{stiffness}-toughness trade-off introduced in the manuscript, however, has an extremely non-smooth landscape and an unknown theoretical upper bound. In that case, it is always preferable to explore the design space using NMO to expedite hypervolume growth within a tight simulation budget.


\clearpage
\newpage

\appendix

\section{Tables}

\begin{table}[!ht]
    \centering
    \begin{tabular}{QQQ|QQQ}
        \toprule
        \multicolumn{3}{c|}{\textbf{Young's modulus}} & \multicolumn{3}{c}{\textbf{Toughness}} \\
        Name & Min. & Max. & Name & Min. & Max. \\
        \midrule
        $E_\text{TB}$ & 8.08e5 & 1.21e6 & $E_\text{TB}$ & 1e5 & 2e6 \\
        $E_\text{VW}$ & 2.13e9 & 3.19e9 & $E_\text{VW}$ & 1e7 & 1e9 \\
        $E_\text{IF}$ & 4.51e8 & 6.77e8 & $E_\text{IF}$ & 1e7 & 1e9 \\
        $\nu_\text{TB}$ & 0.4 & 0.5 & $\nu_\text{TB}$ & 0.4 & 0.5 \\
        $\nu_\text{VW}$ & 0.2 & 0.4 & $\nu_\text{VW}$ & 0.2 & 0.4 \\
        $\nu_\text{IF}$ & 0.3 & 0.4 & $\nu_\text{IF}$ & 0.3 & 0.4 \\
        - & - & - & $W_\text{TB}$ & 5e4 & 1e6 \\
        - & - & - & $W_\text{VW}$ & 5e4 & 1e7 \\
        - & - & - & $W_\text{IF}$ & 1e6 & 1e8 \\
        $\gamma$ & 2e4 & 2e6 & $\gamma$ & 2e4 & 4e6 \\
        \bottomrule
    \end{tabular}
    \caption{{\bf Value ranges of simulation parameters used in system identification when fitting to physical measurements of base materials.}}
    \label{tab:simulation_parameter_ranges}
\end{table}

\begin{table}[!ht]
    \centering
    \begin{tabular}{QQQ|QQQ}
        \toprule
        \multicolumn{3}{c|}{\textbf{Young's modulus}} & \multicolumn{3}{c}{\textbf{Toughness}} \\
        Name & Min. & Max. & Name & Min. & Max. \\
        \midrule
        $E_\text{TB}$ & 8e5 & 1.2e6 & $E_\text{TB}$ & 5e5 & 5e8 \\
        $E_\text{VW}$ & 2.08e9 & 3.12e9 & $E_\text{VW}$ & 5e7 & 1e9 \\
        $E_\text{IF}$ & 4.48e8 & 6.72e8 & $E_\text{IF}$ & 1e7 & 5e8 \\
        $\nu_\text{TB}$ & 0.4 & 0.5 & $\nu_\text{TB}$ & 0.4 & 0.5 \\
        $\nu_\text{VW}$ & 0.2 & 0.4 & $\nu_\text{VW}$ & 0.2 & 0.4 \\
        $\nu_\text{IF}$ & 0.3 & 0.4 & $\nu_\text{IF}$ & 0.3 & 0.4 \\
        - & - & - & $W_\text{TB}$ & 5e4 & 1e6 \\
        - & - & - & $W_\text{VW}$ & 5e5 & 1e7 \\
        - & - & - & $W_\text{IF}$ & 1e6 & 1e8 \\
        $\gamma$ & 2e4 & 2e6 & $\gamma$ & 2e4 & 4e6 \\
        \bottomrule
    \end{tabular}
    \caption{{\bf Value ranges of simulation parameters used in system identification when fitting to physical measurements of microstructures.}}
    \label{tab:simulation_parameter_ranges_2}
\end{table}

\clearpage
\newpage

\newcommand{\bl}[1]{\textcolor{blue}{#1}\normalcolor}
\begin{table}[!ht]
    \centering
    \begin{adjustbox}{max width=0.9\textwidth}
    \begin{tabular}{cc|cc|cc|cc|cc|cc|cc}
        \toprule
        \multirow{2}{*}{\textbf{Round}} & \multirow{2}{*}{\textbf{No.}} & \multirow{2}{*}{$\phi$} & \multirow{2}{*}{\textbf{Exp.}} & \textbf{Sim.} & \textbf{Rel.} & \textbf{Sim.} & \textbf{Rel.} & \textbf{Sim.} & \textbf{Rel.} & \textbf{Sim.} & \textbf{Rel.} & \textbf{Sim.} & \textbf{Rel.} \\
        & & & & \textbf{(R0)} & \textbf{err.} & \textbf{(R1)} & \textbf{err.} & \textbf{(R2)} & \textbf{err.} & \textbf{(R3)} & \textbf{err.} & \textbf{(R4)} & \textbf{err.} \\
        \midrule
        \multirow{18}{*}{0} & 1 & 0.61 & 0.640 & 0.402 & 0.372 & 0.653 & 0.020 & 0.633 & 0.011 & 0.632 & 0.013 & 0.637 & 0.004 \\
         & 2 & 0.76 & 1.400 & 0.980 & 0.300 & 1.281 & 0.085 & 1.216 & 0.131 & 1.185 & 0.153 & 1.185 & 0.154 \\
         & 3 & 0.67 & 0.948 & 0.663 & 0.301 & 0.931 & 0.018 & 0.889 & 0.062 & 0.878 & 0.074 & 0.881 & 0.070 \\
         & 4 & 0.83 & 1.541 & 1.333 & 0.135 & 1.646 & 0.068 & 1.559 & 0.012 & 1.510 & 0.020 & 1.505 & 0.023 \\
         & 5 & 0.84 & 1.747 & 1.509 & 0.136 & 1.826 & 0.045 & 1.718 & 0.017 & 1.654 & 0.053 & 1.647 & 0.058 \\
         & 6 & 0.51 & 0.347 & 0.118 & 0.659 & 0.346 & 0.002 & 0.340 & 0.020 & 0.349 & 0.005 & 0.357 & 0.029 \\
         & 7 & 0.51 & 0.301 & 0.133 & 0.560 & 0.370 & 0.229 & 0.363 & 0.204 & 0.370 & 0.228 & 0.378 & 0.254 \\
         & 8 & 0.54 & 0.330 & 0.023 & 0.932 & 0.347 & 0.053 & 0.342 & 0.037 & 0.356 & 0.080 & 0.368 & 0.115 \\
         & 9 & 0.60 & 0.533 & 0.261 & 0.511 & 0.526 & 0.013 & 0.511 & 0.041 & 0.515 & 0.034 & 0.523 & 0.020 \\
         & 10 & 0.75 & 1.178 & 0.986 & 0.163 & 1.231 & 0.045 & 1.180 & 0.002 & 1.155 & 0.020 & 1.153 & 0.021 \\
         & 11 & 0.66 & 0.954 & 0.603 & 0.368 & 0.801 & 0.160 & 0.784 & 0.178 & 0.778 & 0.185 & 0.780 & 0.182 \\
         & 12 & 0.74 & 1.308 & 0.980 & 0.251 & 1.232 & 0.059 & 1.171 & 0.105 & 1.142 & 0.127 & 1.140 & 0.128 \\
         & 13 & 0.76 & 1.375 & 1.075 & 0.218 & 1.330 & 0.033 & 1.264 & 0.080 & 1.231 & 0.104 & 1.228 & 0.107 \\
         & 14 & 0.82 & 1.659 & 1.255 & 0.243 & 1.539 & 0.072 & 1.470 & 0.114 & 1.424 & 0.141 & 1.420 & 0.144 \\
         & 15 & 0.84 & 1.790 & 1.479 & 0.174 & 1.775 & 0.008 & 1.681 & 0.061 & 1.625 & 0.092 & 1.617 & 0.096 \\
         & 16 & 0.85 & 1.767 & 1.523 & 0.138 & 1.827 & 0.034 & 1.729 & 0.021 & 1.674 & 0.053 & 1.666 & 0.057 \\
         & 17 & 0.85 & 1.771 & 1.604 & 0.094 & 1.915 & 0.081 & 1.807 & 0.020 & 1.742 & 0.016 & 1.733 & 0.022 \\
         & 18 & 0.85 & 1.807 & 1.729 & 0.043 & 2.080 & 0.151 & 1.934 & 0.070 & 1.844 & 0.020 & 1.833 & 0.014 \\
        \midrule
        \multirow{8}{*}{1} & 19 & 0.64 & 0.819 & 0.605 & 0.261 & \bl{0.854} & \bl{0.042} & 0.810 & 0.012 & 0.793 & 0.032 & 0.796 & 0.028 \\
         & 20 & 0.58 & 0.860 & 0.736 & 0.144 & \bl{0.919} & \bl{0.068} & 0.876 & 0.018 & 0.854 & 0.007 & 0.853 & 0.008 \\
         & 21 & 0.67 & 0.980 & 0.874 & 0.108 & \bl{1.115} & \bl{0.139} & 1.054 & 0.076 & 1.021 & 0.042 & 1.019 & 0.041 \\
         & 22 & 0.77 & 1.324 & 1.162 & 0.123 & \bl{1.474} & \bl{0.113} & 1.378 & 0.041 & 1.328 & 0.003 & 1.325 & 0.001 \\
         & 23 & 0.79 & 1.572 & 1.356 & 0.138 & \bl{1.678} & \bl{0.067} & 1.566 & 0.004 & 1.502 & 0.045 & 1.496 & 0.049 \\
         & 24 & 0.79 & 1.402 & 1.479 & 0.054 & \bl{1.792} & \bl{0.278} & 1.661 & 0.185 & 1.583 & 0.129 & 1.574 & 0.123 \\
         & 25 & 0.85 & 1.764 & 1.613 & 0.086 & \bl{1.963} & \bl{0.113} & 1.838 & 0.042 & 1.760 & 0.002 & 1.751 & 0.007 \\
         & 26 & 0.85 & 1.810 & 1.738 & 0.040 & \bl{2.089} & \bl{0.154} & 1.945 & 0.074 & 1.856 & 0.025 & 1.845 & 0.019 \\
        \midrule
        \multirow{8}{*}{2} & 27 & 0.68 & 0.987 & 0.774 & 0.215 &  &  & \bl{0.961} & \bl{0.026} & 0.944 & 0.044 & 0.944 & 0.043 \\
         & 28 & 0.74 & 1.227 & 1.056 & 0.140 &  &  & \bl{1.273} & \bl{0.038} & 1.232 & 0.004 & 1.230 & 0.002 \\
         & 29 & 0.75 & 1.277 & 1.166 & 0.087 &  &  & \bl{1.361} & \bl{0.066} & 1.313 & 0.028 & 1.309 & 0.025 \\
         & 30 & 0.79 & 1.308 & 1.281 & 0.021 &  &  & \bl{1.480} & \bl{0.132} & 1.432 & 0.095 & 1.427 & 0.091 \\
         & 31 & 0.80 & 1.450 & 1.355 & 0.065 &  &  & \bl{1.577} & \bl{0.088} & 1.516 & 0.046 & 1.510 & 0.042 \\
         & 32 & 0.85 & 1.630 & 1.541 & 0.055 &  &  & \bl{1.782} & \bl{0.093} & 1.709 & 0.048 & 1.702 & 0.044 \\
         & 33 & 0.85 & 1.690 & 1.723 & 0.019 &  &  & \bl{1.923} & \bl{0.138} & 1.833 & 0.085 & 1.822 & 0.078 \\
         & 34 & 0.85 & 1.679 & 1.724 & 0.027 &  &  & \bl{1.930} & \bl{0.150} & 1.841 & 0.097 & 1.831 & 0.091 \\
        \midrule
        \multirow{8}{*}{3} & 35 & 0.68 & 0.847 & 0.724 & 0.145 &  &  &  &  & \bl{0.890} & \bl{0.051} & 0.891 & 0.052 \\
         & 36 & 0.66 & 0.861 & 0.904 & 0.050 &  &  &  &  & \bl{1.035} & \bl{0.203} & 1.033 & 0.200 \\
         & 37 & 0.74 & 1.317 & 1.033 & 0.216 &  &  &  &  & \bl{1.206} & \bl{0.084} & 1.204 & 0.086 \\
         & 38 & 0.76 & 1.302 & 1.207 & 0.073 &  &  &  &  & \bl{1.362} & \bl{0.047} & 1.358 & 0.043 \\
         & 39 & 0.79 & 1.528 & 1.305 & 0.146 &  &  &  &  & \bl{1.459} & \bl{0.045} & 1.454 & 0.048 \\
         & 40 & 0.82 & 1.629 & 1.443 & 0.114 &  &  &  &  & \bl{1.594} & \bl{0.021} & 1.587 & 0.025 \\
         & 41 & 0.85 & 1.741 & 1.673 & 0.039 &  &  &  &  & \bl{1.805} & \bl{0.037} & 1.796 & 0.032 \\
         & 42 & 0.85 & 1.751 & 1.749 & 0.001 &  &  &  &  & \bl{1.858} & \bl{0.061} & 1.847 & 0.054 \\
        \midrule
        \multirow{8}{*}{4} & 43 & 0.70 & 1.093 & 0.968 & 0.115 &  &  &  &  &  &  & \bl{1.112} & \bl{0.017} \\
         & 44 & 0.74 & 1.164 & 1.015 & 0.128 &  &  &  &  &  &  & \bl{1.167} & \bl{0.003} \\
         & 45 & 0.77 & 1.297 & 1.151 & 0.112 &  &  &  &  &  &  & \bl{1.299} & \bl{0.002} \\
         & 46 & 0.79 & 1.372 & 1.248 & 0.090 &  &  &  &  &  &  & \bl{1.396} & \bl{0.017} \\
         & 47 & 0.84 & 1.683 & 1.505 & 0.105 &  &  &  &  &  &  & \bl{1.639} & \bl{0.026} \\
         & 48 & 0.85 & 1.683 & 1.538 & 0.086 &  &  &  &  &  &  & \bl{1.672} & \bl{0.006} \\
         & 49 & 0.85 & 1.649 & 1.561 & 0.053 &  &  &  &  &  &  & \bl{1.701} & \bl{0.032} \\
         & 50 & 0.85 & 1.593 & 1.680 & 0.055 &  &  &  &  &  &  & \bl{1.801} & \bl{0.130} \\
        \bottomrule
    \end{tabular}
    \end{adjustbox}
    \caption{{\bf Experimental Young's moduli, simulation predictions, and relative errors of all microstructures discovered by our approach.} R$i$ = Round $i$. Text in blue indicates predictions for the next round. The unit of Young's modulus is GPa.}
    \label{tab:simulation_e}
\end{table}

\clearpage
\newpage

\begin{table}[!ht]
    \centering
    \begin{adjustbox}{max width=0.9\textwidth}
    \begin{tabular}{cc|cc|cc|cc|cc|cc|cc}
        \toprule
        \multirow{2}{*}{\textbf{Round}} & \multirow{2}{*}{\textbf{No.}} & \multirow{2}{*}{$\phi$} & \multirow{2}{*}{\textbf{Exp.}} & \textbf{Sim.} & \textbf{Rel.} & \textbf{Sim.} & \textbf{Rel.} & \textbf{Sim.} & \textbf{Rel.} & \textbf{Sim.} & \textbf{Rel.} & \textbf{Sim.} & \textbf{Rel.} \\
        & & & & \textbf{(R0)} & \textbf{err.} & \textbf{(R1)} & \textbf{err.} & \textbf{(R2)} & \textbf{err.} & \textbf{(R3)} & \textbf{err.} & \textbf{(R4)} & \textbf{err.} \\
        \midrule
        \multirow{18}{*}{0} & 1 & 0.61 & 0.323 & 0.528 & 0.633 & 0.361 & 0.116 & 0.254 & 0.214 & 0.239 & 0.261 & 0.309 & 0.045 \\
         & 2 & 0.76 & 0.201 & 0.329 & 0.637 & 0.227 & 0.131 & 0.236 & 0.173 & 0.143 & 0.290 & 0.221 & 0.102 \\
         & 3 & 0.67 & 0.271 & 0.398 & 0.470 & 0.395 & 0.458 & 0.327 & 0.207 & 0.286 & 0.056 & 0.273 & 0.009 \\
         & 4 & 0.83 & 0.276 & 0.307 & 0.114 & 0.274 & 0.007 & 0.246 & 0.109 & 0.190 & 0.310 & 0.203 & 0.264 \\
         & 5 & 0.84 & 0.375 & 0.277 & 0.261 & 0.310 & 0.171 & 0.282 & 0.247 & 0.214 & 0.429 & 0.211 & 0.436 \\
         & 6 & 0.51 & 0.309 & 0.487 & 0.574 & 0.255 & 0.177 & 0.275 & 0.110 & 0.190 & 0.387 & 0.275 & 0.110 \\
         & 7 & 0.51 & 0.245 & 0.231 & 0.057 & 0.270 & 0.099 & 0.132 & 0.461 & 0.141 & 0.425 & 0.221 & 0.098 \\
         & 8 & 0.54 & 0.117 & 0.022 & 0.809 & 0.102 & 0.126 & 0.078 & 0.329 & 0.102 & 0.124 & 0.143 & 0.226 \\
         & 9 & 0.60 & 0.252 & 0.222 & 0.119 & 0.344 & 0.365 & 0.194 & 0.227 & 0.204 & 0.188 & 0.244 & 0.029 \\
         & 10 & 0.75 & 0.211 & 0.482 & 1.283 & 0.343 & 0.622 & 0.263 & 0.248 & 0.227 & 0.073 & 0.343 & 0.624 \\
         & 11 & 0.66 & 0.371 & 1.304 & 2.515 & 0.410 & 0.105 & 0.359 & 0.031 & 0.310 & 0.165 & 0.386 & 0.039 \\
         & 12 & 0.74 & 0.261 & 1.112 & 3.268 & 0.407 & 0.564 & 0.413 & 0.586 & 0.218 & 0.162 & 0.394 & 0.514 \\
         & 13 & 0.76 & 0.391 & 1.091 & 1.791 & 0.398 & 0.018 & 0.294 & 0.247 & 0.303 & 0.225 & 0.420 & 0.074 \\
         & 14 & 0.82 & 0.109 & 0.945 & 7.674 & 0.159 & 0.457 & 0.120 & 0.104 & 0.110 & 0.012 & 0.099 & 0.095 \\
         & 15 & 0.84 & 0.355 & 0.865 & 1.436 & 0.236 & 0.335 & 0.278 & 0.218 & 0.179 & 0.495 & 0.131 & 0.630 \\
         & 16 & 0.85 & 0.326 & 0.786 & 1.407 & 0.342 & 0.048 & 0.315 & 0.035 & 0.276 & 0.154 & 0.222 & 0.319 \\
         & 17 & 0.85 & 0.313 & 0.603 & 0.927 & 0.291 & 0.069 & 0.222 & 0.292 & 0.223 & 0.289 & 0.243 & 0.224 \\
         & 18 & 0.85 & 0.304 & 0.424 & 0.395 & 0.378 & 0.244 & 0.324 & 0.067 & 0.305 & 0.005 & 0.271 & 0.108 \\
        \midrule
        \multirow{8}{*}{1} & 19 & 0.64 & 0.424 & 0.295 & 0.304 & \bl{0.700} & \bl{0.650} & 0.295 & 0.304 & 0.540 & 0.275 & 0.448 & 0.056 \\
         & 20 & 0.58 & 0.426 & 0.553 & 0.300 & \bl{0.892} & \bl{1.095} & 0.355 & 0.166 & 0.345 & 0.189 & 0.458 & 0.075 \\
         & 21 & 0.67 & 0.227 & 0.669 & 1.950 & \bl{0.970} & \bl{3.277} & 0.455 & 1.005 & 0.418 & 0.843 & 0.428 & 0.889 \\
         & 22 & 0.77 & 0.513 & 0.554 & 0.079 & \bl{0.982} & \bl{0.913} & 0.449 & 0.125 & 0.526 & 0.024 & 0.346 & 0.326 \\
         & 23 & 0.79 & 0.426 & 0.452 & 0.062 & \bl{0.865} & \bl{1.033} & 0.452 & 0.063 & 0.494 & 0.161 & 0.343 & 0.193 \\
         & 24 & 0.79 & 0.491 & 0.315 & 0.359 & \bl{0.680} & \bl{0.383} & 0.371 & 0.245 & 0.447 & 0.089 & 0.254 & 0.482 \\
         & 25 & 0.85 & 0.458 & 0.280 & 0.389 & \bl{0.584} & \bl{0.276} & 0.431 & 0.058 & 0.374 & 0.182 & 0.266 & 0.418 \\
         & 26 & 0.85 & 0.473 & 0.314 & 0.336 & \bl{0.391} & \bl{0.172} & 0.309 & 0.346 & 0.315 & 0.333 & 0.206 & 0.564 \\
        \midrule
        \multirow{8}{*}{2} & 27 & 0.68 & 0.315 & 0.612 & 0.943 &  &  & \bl{0.598} & \bl{0.897} & 0.338 & 0.074 & 0.467 & 0.483 \\
         & 28 & 0.74 & 0.331 & 0.395 & 0.195 &  &  & \bl{0.611} & \bl{0.847} & 0.340 & 0.029 & 0.304 & 0.081 \\
         & 29 & 0.75 & 0.493 & 0.423 & 0.141 &  &  & \bl{0.589} & \bl{0.196} & 0.530 & 0.075 & 0.390 & 0.209 \\
         & 30 & 0.79 & 0.381 & 0.364 & 0.045 &  &  & \bl{0.541} & \bl{0.417} & 0.269 & 0.295 & 0.394 & 0.034 \\
         & 31 & 0.80 & 0.106 & 0.304 & 1.865 &  &  & \bl{0.505} & \bl{3.761} & 0.175 & 0.648 & 0.172 & 0.620 \\
         & 32 & 0.85 & 0.369 & 0.349 & 0.053 &  &  & \bl{0.466} & \bl{0.263} & 0.330 & 0.105 & 0.289 & 0.216 \\
         & 33 & 0.85 & 0.303 & 0.266 & 0.122 &  &  & \bl{0.369} & \bl{0.218} & 0.151 & 0.501 & 0.260 & 0.143 \\
         & 34 & 0.85 & 0.360 & 0.310 & 0.139 &  &  & \bl{0.327} & \bl{0.092} & 0.172 & 0.521 & 0.218 & 0.394 \\
        \midrule
        \multirow{8}{*}{3} & 35 & 0.68 & 0.339 & 0.504 & 0.485 &  &  &  &  & \bl{0.641} & \bl{0.888} & 0.291 & 0.142 \\
         & 36 & 0.66 & 0.508 & 0.363 & 0.286 &  &  &  &  & \bl{0.684} & \bl{0.347} & 0.306 & 0.397 \\
         & 37 & 0.74 & 0.276 & 0.389 & 0.412 &  &  &  &  & \bl{0.716} & \bl{1.598} & 0.360 & 0.307 \\
         & 38 & 0.76 & 0.278 & 0.386 & 0.388 &  &  &  &  & \bl{0.667} & \bl{1.401} & 0.308 & 0.108 \\
         & 39 & 0.79 & 0.351 & 0.434 & 0.235 &  &  &  &  & \bl{0.633} & \bl{0.802} & 0.312 & 0.111 \\
         & 40 & 0.82 & 0.329 & 0.479 & 0.453 &  &  &  &  & \bl{0.544} & \bl{0.650} & 0.265 & 0.196 \\
         & 41 & 0.85 & 0.412 & 0.294 & 0.287 &  &  &  &  & \bl{0.424} & \bl{0.030} & 0.237 & 0.424 \\
         & 42 & 0.85 & 0.400 & 0.239 & 0.403 &  &  &  &  & \bl{0.261} & \bl{0.346} & 0.173 & 0.568 \\
        \midrule
        \multirow{8}{*}{4} & 43 & 0.70 & 0.357 & 0.713 & 0.997 &  &  &  &  &  &  & \bl{0.590} & \bl{0.653} \\
         & 44 & 0.74 & 0.247 & 0.640 & 1.593 &  &  &  &  &  &  & \bl{0.564} & \bl{1.287} \\
         & 45 & 0.77 & 0.326 & 0.544 & 0.669 &  &  &  &  &  &  & \bl{0.546} & \bl{0.672} \\
         & 46 & 0.79 & 0.425 & 0.516 & 0.213 &  &  &  &  &  &  & \bl{0.510} & \bl{0.200} \\
         & 47 & 0.84 & 0.263 & 0.361 & 0.374 &  &  &  &  &  &  & \bl{0.450} & \bl{0.712} \\
         & 48 & 0.85 & 0.348 & 0.398 & 0.144 &  &  &  &  &  &  & \bl{0.398} & \bl{0.144} \\
         & 49 & 0.85 & 0.348 & 0.419 & 0.203 &  &  &  &  &  &  & \bl{0.369} & \bl{0.060} \\
         & 50 & 0.85 & 0.394 & 0.302 & 0.233 &  &  &  &  &  &  & \bl{0.256} & \bl{0.350} \\
        \bottomrule
    \end{tabular}
    \end{adjustbox}
    \caption{{\bf Experimental toughness, simulation predictions, and relative errors of all microstructures discovered by our approach.} R$i$ = Round $i$. Text in blue indicates predictions for the next round. The unit of toughness is MPa.}
    \label{tab:simulation_t}
\end{table}

\clearpage
\newpage

\begin{table}[!ht]
    \centering
    \begin{tabular}{c|cP|cP|cP|cP}
        \toprule
        \textbf{Family} & \multicolumn{2}{c|}{1} & \multicolumn{2}{c|}{2} & \multicolumn{2}{c|}{3} & \multicolumn{2}{c}{4} \\
        \midrule
        & No. & $E$ (GPa) & No. & $E$ (GPa) & No. & $E$ (GPa) & No. & $E$ (GPa) \\
        \midrule
        \multirow{4}{3cm}{\centering \textbf{Seed microstructures}} & 22 & 1.324 & 18 & 1.807 & 15 & 1.790 & 5 & 1.747 \\
        & 24 & 1.402 & 26 & 1.810 & 16 & 1.767 & 17 & 1.771 \\
        & 36 & 0.861 & 41 & 1.741 & & & & \\
        & & & 42 & 1.751 & & & & \\
        \midrule
        \multirow{3}{3cm}{\centering \textbf{Validation microstructures}} & 1-1 & 1.253 & 2-1 & 1.541 & 3-1 & 1.470 & 4-1 & 1.447 \\
        & 1-2 & 1.329 & 2-2 & 1.779 & 3-2 & 1.703 & 4-2 & 1.679 \\
        & 1-3 & 1.485 & 2-3 & 1.703 & 3-3 & 1.689 & 4-3 & 1.621 \\
        \bottomrule
    \end{tabular} \\
    \vspace{1em}
    \begin{tabular}{c|cP|cP|cP|cP}
        \toprule
        \textbf{Family} & \multicolumn{2}{c|}{1} & \multicolumn{2}{c|}{2} & \multicolumn{2}{c|}{3} & \multicolumn{2}{c}{4} \\
        \midrule
        & No. & $T$ (MPa) & No. & $T$ (MPa) & No. & $T$ (MPa) & No. & $T$ (MPa) \\
        \midrule
        \multirow{4}{3cm}{\centering \textbf{Seed microstructures}} & 22 & 0.513 & 18 & 0.304 & 15 & 0.355 & 5 & 0.375 \\
        & 24 & 0.491 & 26 & 0.473 & 16 & 0.326 & 17 & 0.313 \\
        & 36 & 0.508 & 41 & 0.412 & & & & \\
        & & & 42 & 0.400 & & & & \\
        \midrule
        \multirow{3}{3cm}{\centering \textbf{Validation microstructures}} & 1-1 & 0.369 & 2-1 & 0.343 & 3-1 & 0.330 & 4-1 & 0.381 \\
        & 1-2 & 0.447 & 2-2 & 0.388 & 3-2 & 0.269 & 4-2 & 0.347 \\
        & 1-3 & 0.420 & 2-3 & 0.458 & 3-3 & 0.350 & 4-3 & 0.433 \\
        \bottomrule
    \end{tabular}
    \caption{{\bf Comparison between the physical measurements of seed microstructures and validation microstructures in each family.} Microstructures in the same family show comparable Young's moduli and toughnesses.}
    \label{tab:simulation_validation}
\end{table}

\clearpage
\newpage

\begin{table}[!ht]
    \centering
    \textbf{Seed microstructures} \\
    \vspace{0.8em}
    \begin{tabular}{cc|cc|cc|cc}
        \toprule
        \multirow{2}{*}{\textbf{Family}} & \multirow{2}{*}{\textbf{No}} & \multirow{2}{*}{$\phi$} & \multirow{2}{*}{\textbf{Exp.}} & \textbf{Sim.} & \textbf{Rel.} & \textbf{Sim.} & \textbf{Rel.} \\
        & & & & \textbf{(R4)} & \textbf{err.} & \textbf{(FS)} & \textbf{err.} \\
        \midrule
        \multirow{3}{*}{1} & 22 & 0.77 & 1.324 & 1.325 & 0.001 & 1.158 & 0.126 \\
         & 24 & 0.79 & 1.402 & 1.574 & 0.123 & 1.403 & 0.001 \\
         & 36 & 0.66 & 0.861 & 1.033 & 0.200 & 0.913 & 0.061 \\
        \midrule
        \multirow{4}{*}{2} & 18 & 0.85 & 1.807 & 1.833 & 0.014 & 1.799 & 0.004 \\
         & 26 & 0.85 & 1.810 & 1.845 & 0.019 & 1.809 & 0.001 \\
         & 41 & 0.85 & 1.741 & 1.796 & 0.032 & 1.756 & 0.009 \\
         & 42 & 0.85 & 1.751 & 1.847 & 0.054 & 1.812 & 0.034 \\
        \midrule
        \multirow{2}{*}{3} & 15 & 0.84 & 1.790 & 1.617 & 0.096 & 1.714 & 0.042 \\
         & 16 & 0.85 & 1.767 & 1.666 & 0.057 & 1.767 & 0.000 \\
        \midrule
        \multirow{2}{*}{4} & 5 & 0.84 & 1.747 & 1.647 & 0.058 & 1.672 & 0.043 \\
         & 17 & 0.85 & 1.771 & 1.733 & 0.022 & 1.771 & 0.000 \\
        \bottomrule
    \end{tabular} \\
    \vspace{1em}
    \textbf{Validation microstructures} \\
    \vspace{0.8em}
    \begin{tabular}{cc|cc|cc|cc}
        \toprule
        \multirow{2}{*}{\textbf{Family}} & \multirow{2}{*}{\textbf{No}} & \multirow{2}{*}{$\phi$} & \multirow{2}{*}{\textbf{Exp.}} & \textbf{Sim.} & \textbf{Rel.} & \textbf{Sim.} & \textbf{Rel.} \\
        & & & & \textbf{(R4)} & \textbf{err.} & \textbf{(FS)} & \textbf{err.} \\
        \midrule
        \multirow{3}{*}{1} & F1-1 & 0.73 & 1.253 & 1.222 & 0.024 & 1.066 & 0.149 \\
         & F1-2 & 0.75 & 1.329 & 1.359 & 0.022 & 1.204 & 0.094 \\
         & F1-3 & 0.79 & 1.485 & 1.451 & 0.022 & 1.284 & 0.135 \\
        \midrule
        \multirow{3}{*}{2} & F2-1 & 0.81 & 1.541 & 1.654 & 0.073 & 1.621 & 0.052 \\
         & F2-2 & 0.85 & 1.779 & 1.825 & 0.026 & 1.794 & 0.009 \\
         & F2-3 & 0.85 & 1.703 & 1.827 & 0.073 & 1.795 & 0.054 \\
        \midrule
        \multirow{3}{*}{3} & F3-1 & 0.83 & 1.470 & 1.510 & 0.027 & 1.582 & 0.076 \\
         & F3-2 & 0.85 & 1.703 & 1.725 & 0.013 & 1.840 & 0.081 \\
         & F3-3 & 0.85 & 1.689 & 1.724 & 0.021 & 1.841 & 0.090 \\
        \midrule
        \multirow{3}{*}{4} & F4-1 & 0.80 & 1.447 & 1.426 & 0.015 & 1.430 & 0.012 \\
         & F4-2 & 0.84 & 1.679 & 1.699 & 0.012 & 1.728 & 0.030 \\
         & F4-3 & 0.85 & 1.621 & 1.736 & 0.071 & 1.774 & 0.094 \\
        \bottomrule
    \end{tabular}
    \caption{{\bf Experimental Young's modulus, simulation predictions, and relative errors of seed microstructures and validation microstructures in each family.} 'R4' refers to the virtual testers in Round 4. 'FS' means family-specific virtual testers. The unit of Young's modulus is GPa.}
    \label{tab:simulation_family_e}
\end{table}

\clearpage
\newpage

\begin{table}[!ht]
    \centering
    \textbf{Seed microstructures} \\
    \vspace{0.8em}
    \begin{tabular}{cc|cc|cc|cc}
        \toprule
        \multirow{2}{*}{\textbf{Family}} & \multirow{2}{*}{\textbf{No}} & \multirow{2}{*}{$\phi$} & \multirow{2}{*}{\textbf{Exp.}} & \textbf{Sim.} & \textbf{Rel.} & \textbf{Sim.} & \textbf{Rel.} \\
        & & & & \textbf{(R4)} & \textbf{err.} & \textbf{(FS)} & \textbf{err.} \\
        \midrule
        \multirow{3}{*}{1} & 22 & 0.77 & 0.513 & 0.346 & 0.326 & 0.510 & 0.006 \\
         & 24 & 0.79 & 0.491 & 0.254 & 0.482 & 0.468 & 0.048 \\
         & 36 & 0.66 & 0.508 & 0.306 & 0.397 & 0.466 & 0.083 \\
        \midrule
        \multirow{4}{*}{2} & 18 & 0.85 & 0.304 & 0.271 & 0.108 & 0.328 & 0.080 \\
         & 26 & 0.85 & 0.473 & 0.206 & 0.564 & 0.416 & 0.121 \\
         & 41 & 0.85 & 0.412 & 0.237 & 0.424 & 0.428 & 0.039 \\
         & 42 & 0.85 & 0.400 & 0.173 & 0.568 & 0.315 & 0.213 \\
        \midrule
        \multirow{2}{*}{3} & 15 & 0.84 & 0.355 & 0.131 & 0.630 & 0.336 & 0.054 \\
         & 16 & 0.85 & 0.326 & 0.222 & 0.319 & 0.326 & 0.000 \\
        \midrule
        \multirow{2}{*}{4} & 5 & 0.84 & 0.375 & 0.211 & 0.436 & 0.375 & 0.002 \\
         & 17 & 0.85 & 0.313 & 0.243 & 0.224 & 0.315 & 0.007 \\
        \bottomrule
    \end{tabular} \\
    \vspace{1em}
    \textbf{Validation microstructures} \\
    \vspace{0.8em}
    \begin{tabular}{cc|cc|cc|cc}
        \toprule
        \multirow{2}{*}{\textbf{Family}} & \multirow{2}{*}{\textbf{No}} & \multirow{2}{*}{$\phi$} & \multirow{2}{*}{\textbf{Exp.}} & \textbf{Sim.} & \textbf{Rel.} & \textbf{Sim.} & \textbf{Rel.} \\
        & & & & \textbf{(R4)} & \textbf{err.} & \textbf{(FS)} & \textbf{err.} \\
        \midrule
        \multirow{3}{*}{1} & F1-1 & 0.73 & 0.369 & 0.375 & 0.017 & 0.748 & 1.028 \\
         & F1-2 & 0.75 & 0.447 & 0.386 & 0.136 & 0.626 & 0.399 \\
         & F1-3 & 0.79 & 0.420 & 0.305 & 0.275 & 0.535 & 0.272 \\
        \midrule
        \multirow{3}{*}{2} & F2-1 & 0.81 & 0.343 & 0.273 & 0.204 & 0.546 & 0.593 \\
         & F2-2 & 0.85 & 0.388 & 0.115 & 0.703 & 0.561 & 0.446 \\
         & F2-3 & 0.85 & 0.458 & 0.209 & 0.544 & 0.420 & 0.083 \\
        \midrule
        \multirow{3}{*}{3} & F3-1 & 0.83 & 0.330 & 0.281 & 0.147 & 0.535 & 0.623 \\
         & F3-2 & 0.85 & 0.269 & 0.197 & 0.271 & 0.445 & 0.653 \\
         & F3-3 & 0.85 & 0.350 & 0.169 & 0.516 & 0.312 & 0.106 \\
        \midrule
        \multirow{3}{*}{4} & F4-1 & 0.80 & 0.381 & 0.258 & 0.321 & 0.554 & 0.455 \\
         & F4-2 & 0.84 & 0.347 & 0.252 & 0.274 & 0.459 & 0.322 \\
         & F4-3 & 0.85 & 0.433 & 0.241 & 0.444 & 0.313 & 0.277 \\
        \bottomrule
    \end{tabular}
    \caption{{\bf Experimental toughness, simulation predictions, and relative errors of seed microstructures and validation microstructures in each family.} 'R4' refers to the virtual testers in Round 4. 'FS' means family-specific virtual testers. The unit of toughness is MPa.}
    \label{tab:simulation_family_t}
\end{table}

\clearpage
\newpage

\section{Figures}

\begin{figure}[!ht]
    \centering
    \includegraphics[width=0.8\textwidth]{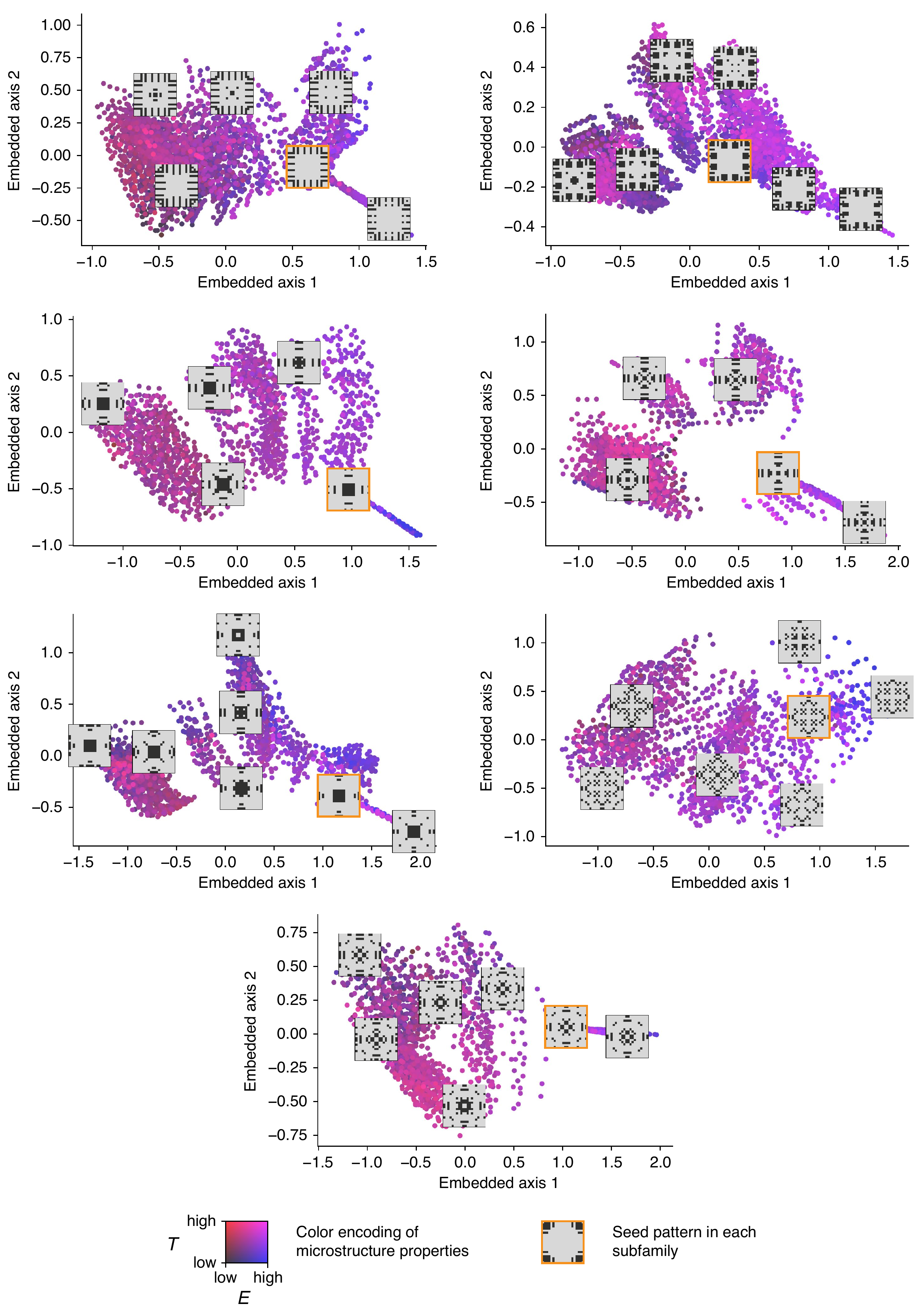}
    \caption{{\bf Isomap embedding spaces of subfamilies not displayed in Fig.~\ref{fig:emd_isomap_example}.} Each embedding space contains around 2,000 to 3,000 samples and a few of them are visualized. Microstructure properties are encoded in colors ($E$: the blue channel; $T$: the red channel). Seed patterns are marked in orange boxes.}
    \label{fig:emd_isomap}
\end{figure}

\clearpage
\newpage

\begin{figure*}[ht]
    \centering
    \includegraphics[width=0.85\textwidth]{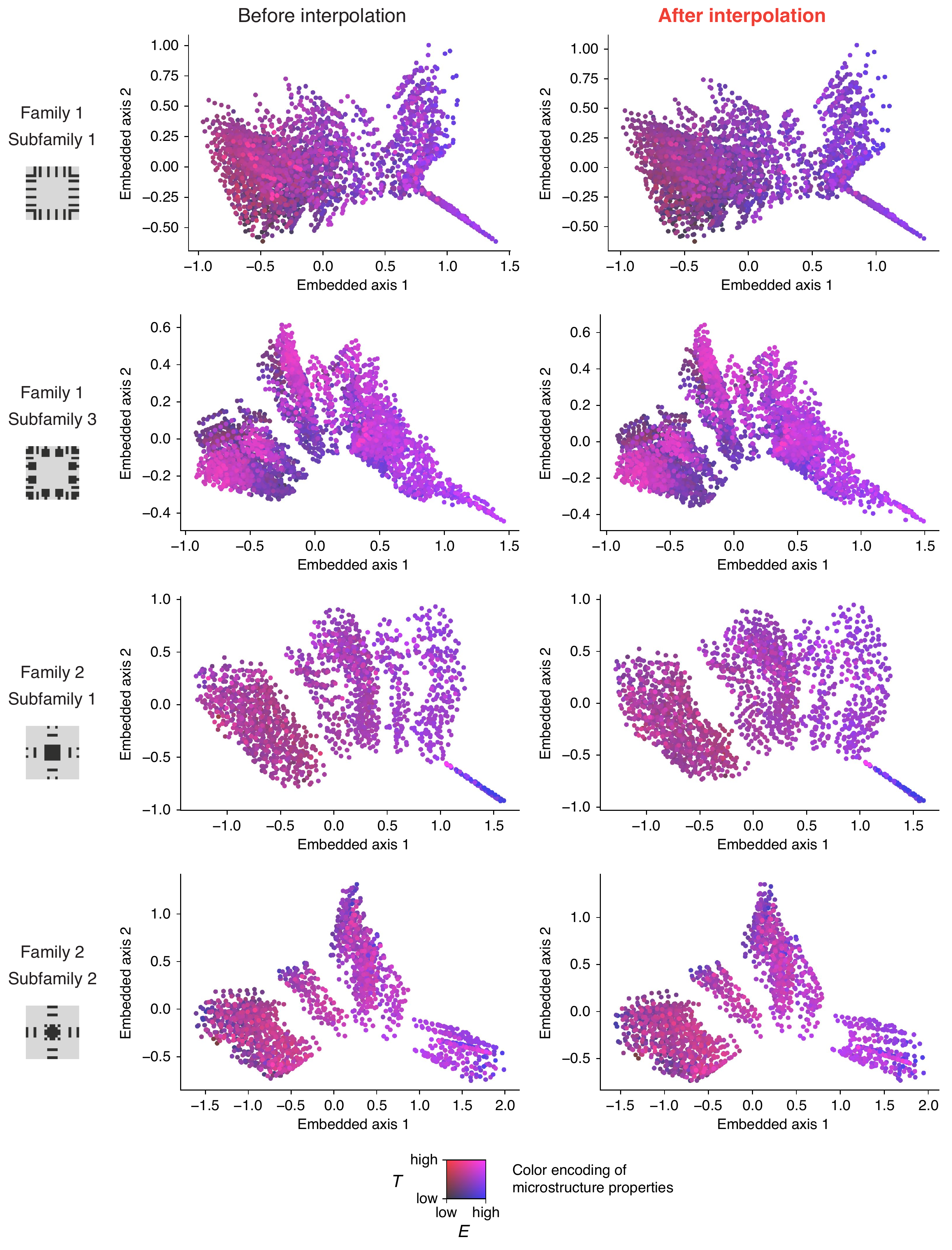}
    \caption{{\bf Isomap embedding spaces of more subfamilies before and after refinement using interpolation (Part 1).} Microstructure properties are encoded in colors.}
    \label{fig:emd_interp}
\end{figure*}

\clearpage
\newpage

\begin{figure}[ht]
    \centering
    \includegraphics[width=0.85\textwidth]{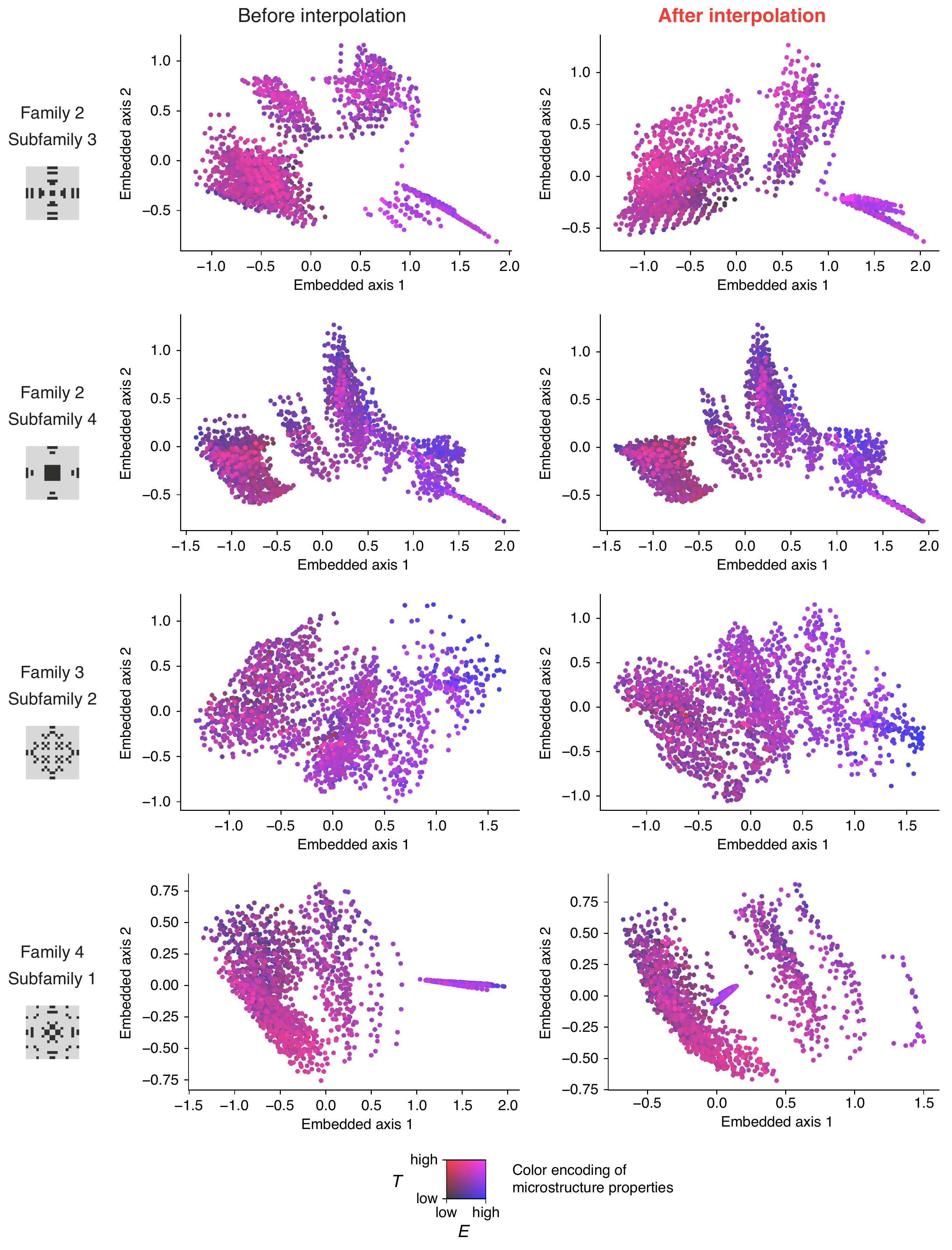}
    \caption{{\bf Isomap embedding spaces of more subfamilies before and after refinement using interpolation (Part 2).} Microstructure properties are encoded in colors.}
    \label{fig:emd_interp_2}
\end{figure}

\clearpage
\newpage

\begin{figure}[ht]
    \centering
    \includegraphics[width=0.9\textwidth]{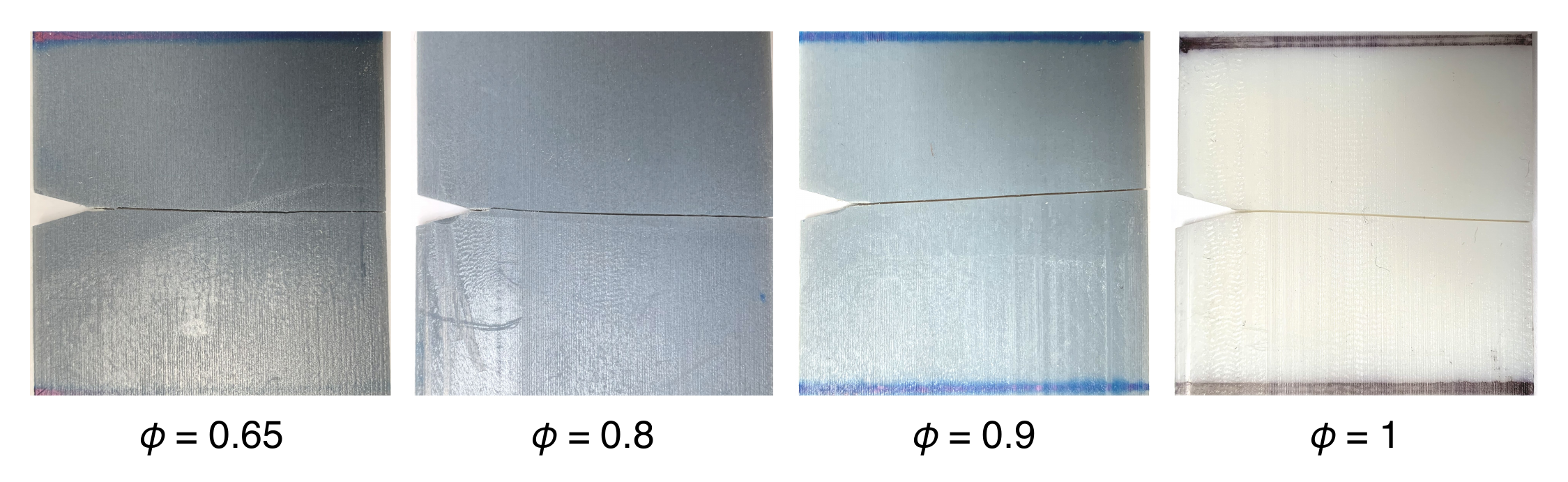}
    \caption{{\bf Clean cleavage is observed after crack propagation in homogeneous composites with various fractions of rigid material.}}
    \label{fig:clean_cleavage}
\end{figure}